\documentclass[twocolumn,trackchanges]{aastex631}
\pdfoutput=1 
\usepackage{amsmath}	
\usepackage{graphicx}	
\usepackage{amssymb}	

\shorttitle{Ranking Supernovae Models with Intracluster gas}
\shortauthors{Batalha et al.}
\graphicspath{{./}{figures/}}

\begin{document}

\title{Ranking Theoretical Supernovae Explosion Models from Observations of the Intracluster Gas}

\correspondingauthor{Rebeca Batalha}
\email{rbatalha.astro@gmail.com}

\author[0000-0002-4949-8351]{Rebeca Batalha}
\affiliation{Observatório Nacional, Rua General José Cristino, 77, São Cristóvão, 20921-400, Rio de Janeiro, Brazil}
\affiliation{Department of Physics and Astronomy, University of Alabama, Box 870324, Tuscaloosa, AL 35487, USA\\}

\author{Renato A. Dupke}
\affiliation{Observatório Nacional, Rua General José Cristino, 77, São Cristóvão, 20921-400, Rio de Janeiro, Brazil}
\affiliation{Department of Physics and Astronomy, University of Alabama, Box 870324, Tuscaloosa, AL 35487, USA\\}
\affiliation{Department of Astronomy, University of Michigan, 930 Dennison Bldg., Ann Arbor, MI 48109-1090, USA}

\author[0000-0002-6090-2853]{Yolanda Jiménez-Teja}
\affiliation{Instituto de Astrofísica de Andalucía–CSIC, Glorieta de la Astronomía s/n, E–18008 Granada, Spain}



\begin{abstract}
The intracluster medium (ICM) is a reservoir of heavy elements synthesized by different supernovae (SNe) types over cosmic history. Different enrichment mechanisms contribute a different relative metal production, predominantly caused by different SNe Type dominance. Using spatially resolved X-ray spectroscopy, one can probe the contribution of each metal enrichment mechanism. However, a large variety of physically feasible supernova explosion models make the analysis of the ICM enrichment history more uncertain. This paper presents a non-parametric PDF analysis to rank different theoretical SNe yields models by comparing their performance against observations. Specifically, we apply this new methodology to rank 7192 combinations of core-collapse SN and Type Ia SN models using 8 abundance ratios from \textit{Suzaku} observations of 18 galaxy systems (clusters and groups) to test their predictions. This novel technique can compare many SN models and maximize spectral information extraction, considering all the individual measurable abundance ratios and their uncertainties. We find that Type II Supernova with nonzero initial metallicity progenitors in general performed better than Pair-Instability SN and Hypernova models and that 3D SNIa models (with the WD progenitor central density of $2.9\times10^9 \mathrm{g\,cm^{-3}}$) performed best among all tested SN model pairs.
\end{abstract}


\section{Introduction} \label{sec:intro}

The discovery of the non-pristine hot intracluster medium (ICM) over 40 years ago \citep{Mitchell_1976, Serlemitsos_1977} established a link between stellar and supernova (SN) nucleosynthesis products in galaxies and metals (i.e., heavy elements) in the ICM (or in the intragroup medium, IGrM). Supernovae events play a dominant role in synthesizing and ejecting heavy elements and can offer clues to processes involved in the ICM chemical enrichment. Core collapse supernovae (SNcc) produce significant amounts of oxygen (O), neon (Ne), magnesium (Mg), silicon (Si), sulfur (S), argon (Ar), and calcium (Ca), as compared to iron (Fe) and its neighboring elements, such as nickel (Ni), mainly produced, instead, primarily by Type Ia supernovae (SNIa). Despite the general picture of the metal production, there are still uncertainties about the amount of synthesized metals (yields) by the different SNe types, related to their explosion mechanisms and the nature of their progenitors.

Type Ia supernovae are generally associated to thermonuclear explosions of carbon and oxygen white dwarfs (C+O WD) in binary star systems. However, the nature of the companion star is still uncertain (see reviews by \citealt{Maeda_2016}, \citealt{Seitenzahl2017}, and \citealt{Livio_2018}). The accreted material can come from a non-degenerate companion star \citep[single degenerate scenario,][]{Whelan_1973} or another white dwarf (WD) \citep[double degenerate scenario,][]{Iben_1984,Webbink1984}. The physics of the propagation burning front is also under debate \citep{Hillebrandt_2000}. 
Central detonation models of near-Chandrasekhar mass ($M_{Ch} \sim 1.4 M_\odot$) WDs \citep{Arnett_1969, Hansen_1969} produce insufficient amounts of intermediate mass elements (IMEs), such as S and Si, needed to explain the observed SNIa spectra. Deflagration models allow the WD to expand to lower density conditions suitable to the synthesis of IMEs but still fall short in reproducing their observed spectra \citep{Fink2014}. Delayed-detonation models such as deflagration-to-detonation transition \citep{Khokhlov1991}, gravitationally confined detonations \citep{Plewa2004,Jordan2008}, and pulsating reverse detonations \citep{Bravo_2009}, arise as a reasonable option to overcome the nucleosynthetic deficiencies to explain normal SNIa. Different mechanisms can also lead to SNIa in the sub-$M_{Ch}$ scenario, involving violent mergers of two C+O WDs \citep[e.g.][]{Pakmor2010,Pakmor2012}, direct collisions of two C+O WDs with or without He-shells \citep[e.g.][]{Papish2016}, double-detonations of WD with He shells \citep[e.g.][]{Leung2020a}, and dynamically driven double-degenerate double-detonations \citep[e.g.][]{Shen2018}.

Core collapse supernovae (SNcc)\footnote{We consider the Types II/Ib/Ic Supernovae, Hypernovae, and pair-instability Supernovae, as the core-collapse Supernovae Type.} are related to the death of massive stars \citep[$\gtrsim 10M_\odot-140M_\odot$ see][for a review]{Nomoto2013}. Their nucleosynthetic products depend on the initial mass and metallicity of the progenitor stars, the mass cut separating the final core mass and the remnant ejected mass, and the final kinetic energy \citep[e.g.][]{Chieffi2004,Nomoto2013}. Moreover, to determine the yields in populations of massive stars, the considered initial mass function (IMF) is also an additional uncertain factor. As a result, all these uncertainties lead to numerous theoretical explosion models for both SNIa and SNcc present in the literature, where competing models often exhibit discrepant heavy element abundance patterns.

The metals in the ICM can be seen as a reservoir of ejecta from SNIa and SNcc explosions. There are many processes involved in producing the chemical enrichment of the ICM, the relative importance of each of them is still under debate \citep[for a review see e.g.][]{Schindler2008,Mernier2018review}. They might include SN II powered galactic winds \citep{Larson1975}, ram-pressure stripping of gas from galaxies \citep{Gunn1972} after cluster formation, or ICM metal enrichment could have happened in the environment before cluster's virialization (pre-enrichment) \citep[e.g.][]{Biffi2018}. The importance of each of these mechanisms can be assessed through comparison to SNe yield models if these models are well constrained. However, as mentioned above, there are still significant uncertainties about the predicted yields for different SN Types, and one can, in principle, invert the problem and use bona fide abundances observed in the ICM to differentiate among various SNe explosion models. This iteration is justified because the ICM is generally optically thin to X-rays and close to collisional ionization equilibrium (see reviews by \citealt{Bohringer2010} and \citealt{Mernier2018review}), making abundance measurements clean and easy to derive.

Abundance measurements from X-ray spectroscopy have proven to be a powerful tool to constrain theoretical supernova yield models \citep{Dupke2000,Finoguenov_2002,Sato_2007,dePlaa2007,Bulbul2012,Mernier2016,Simionescu2019}. This work applies a new technique to improve the power of elemental abundance ratios measured in a selected cluster sample to rank different theoretical SNe explosion models. The sample is composed of 18 carefully selected nearby galaxy systems, such that (i) the internal and external regions can be cleanly separated and (ii) the particle background is low enough, so we do not need to worry about particle background modeling, which would introduce additional uncertainties. Thanks to the very low particle background combined with an excellent spectral resolution, the X-ray Imaging Spectrometer (XIS) on-board of \textit{Suzaku} is the most suitable for measuring individual abundances accurately. Our technique identifies the level of incompatibility with the observations between an extensive set of SNIa and SNcc predicted yield models by individually assessing eight abundance ratios (O/Fe, Ne/Fe, Mg/Fe, Si/Fe, S/Fe, Ar/Fe, Ca/Fe, and Ni/Fe) in a non-parametric analysis based on the Kolmogorov-Smirnov test (KS test, hereafter). We compare them to those predicted by 7192 different combinations SNIa--SNcc pairs among 263 theoretical supernova yield models found in the literature.

This paper is organized as follows. In Section \ref{sec:data}, we describe the data preparation and reduction, as well as the spectra modeling of the sample. The comparison between the predicted yields from SN models and the measured ratio abundances from observations is detailed in Section \ref{sec:SN_models}.
We then implement our methodology to quantify the incompatibility of the theoretical yields with the abundance ratio measurements in Section \ref{sec:KStest} over 263 theoretical supernovae yield models, whose main properties are quickly described in Section \ref{sec:summary_SNmodels}. Our main results are presented in Section \ref{sec:results} and we summarize our main conclusions in Section \ref{sec:conclusion}. We assume cosmological parameters being $H_0=67.8$ km\,s$^{-1}$\,Mpc$^{-1}$, $\Omega_m$=0.308, and $\Omega_\Lambda$=0.692 \citep{Planck_Collab_2016}. Error bars of abundance ratios are given at 68\% significance level. We adopt the metal solar abundance model by \citet[][hereafter ANGR]{Anders_1989}.

\section{Data Preparation} \label{sec:data}
\begin{deluxetable*}{lccccccr}
\tablecaption{Observations of groups and clusters of galaxies.\label{tab:sample}}
\tablehead{
\colhead{Cluster/Group name} & \colhead{RA} & \colhead{DEC} & \colhead{Redshift$^a$} &
\colhead{Cool-core radius} & \colhead{Obs ID} & \colhead{Exposure time} & \colhead{PI Name}\\
\colhead{} & \colhead{(hh:mm:ss)} & \colhead{(dd:mm:ss)} & \colhead{} & 
\colhead{(kpc)} & \colhead{} & \colhead{(ks)} & \colhead{}
}
\decimalcolnumbers
\startdata
NGC 5846   & 15 06 29.62 & +01 36 09.0 & 0.0060 & $\lesssim20$ & 803042010 & 155,9 & M. Loewenstein \\
NGC 4472    & 12 29 46.58	& +08 00 18.0 & 0.0044 & $\lesssim10$ & 801064010 & 121,0 & M. Loewenstein\\
HCG62       & 12 53 05.76 & -09 12 07.9 & 0.0147 & $\lesssim30$ & 800013020 & 114,8 & N. Ota\\
Ophiuchus & 17 12 26.23 & -23 22 44.4 & 0.0280 & $\lesssim40$ & 802046010 & 100,5 & A. Furuzawa\\
NGC 1550    & 04 19 47.71	& +02 24 37.8 & 0.012 & $\lesssim25$ & 803017010 & 83,3  & K. Sato\\
Abell 3581        & 14 07 37.99 & -27 01 11.6 & 0.0226 & $<100$ & 807026010 & 80,4  & K. Sato\\
NGC 507     & 01 23 38.52 & +33 15 35.3 & 0.01902 & $\lesssim40$ & 801017010 & 79,5  & K. Sato\\
Abell 426	  & 03 19 49.30 & +41 30 14.0 & 0.0179 & $\lesssim100$ & 800010010 & 50,4  & A. Fabian\\
Abell 496         & 04 33 39.50 & -13 16 44.8 & 0.0329 & $<90$ & 803073010 & 44,3  & R. Dupke\\
Abell 3571	      & 13 47 26.98 & -32 51 08.6 & 0.0391 & $\lesssim100$ & 808094010 & 38,3  & R. Dupke\\
Abell 262         & 01 52 46.13 & +36 09 32.8 & 0.01641 & $<40$ & 802001010 & 37,2  & K. Matsushita\\
NGC 2300    & 07 31 15.79	& +85 41 47.4 & 0.0070 & $<20$ & 804030010 & 37,1  & K. Matsushita\\
Abell 3526 & 12 48 48.29 & -41 18 47.5 & 0.0104 & $<50$ & 800014010 & 36,5  & Y. Fukazawa\\
MKW4      & 12 04 32.88 & +01 54 50.8 & 0.01975   & $<25$ & 808066010 & 34,6  & Y. Su\\
NGC 5044    & 13 15 24.10 & -16 23 23.6 & 0.0082 & $<40$ & 801046010 & 19,7  & K. Matsushita\\
AWM7      & 02 54 31.44 & +41 35 35.9 & 0.0172 & $<30$ & 801035010 & 19,0  & T. Ohashi \\
NGC 6338    & 17 15 15.10	& +57 25 14.9 & 0.02816 & $<70$ & 809099010 & 15,5  & L. Lovisari\\
UGC 3957    & 07 40 58.39	& +55 25 44.4 & 0.0340 & $\lesssim110$ & 801072010 & 10,8  & C. Scharf\\
\enddata
\tablecomments{The first three columns show the target names and their equatorial coordinates. The $^a$Fourth column (z) shows the corresponding redshifts taken from \citet{https://doi.org/10.26132/ned1} to perform spectral fittings (see Section \ref{subsec:spec_model}). The fifth column (Inner radius) indicates the radius that delimits the inner regions of each group/cluster. The last three columns (ObsID, Exposure time, and PI Name) are related to its observation identification number, exposure time, and the name of the principal investigator on the proposal, respectively.}

\end{deluxetable*}

Our analysis is based on 18 nearby ($z \leq 0.0391$) groups and clusters of galaxies observed with the X-ray Imaging Spectrometer observations on-board the \textit{Suzaku} satellite. Their properties are presented in Table \ref{tab:sample}. Downloaded observations from the HEASARC website were processed as follows. We selected the observations with the longest exposure time available for each object in the sample. We calibrated and screened 4 XIS chips (XIS0, 1, 2 \& 3), when available, via \texttt{aepipeline} tool. We applied standard criteria\footnote{\url{https://heasarc.gsfc.nasa.gov/docs/suzaku/processing/criteria\_xis.html}} using \textsc{HEAsoft} v6.21. Editing mode data 5x5 were converted and combined to 3x3 event files with XSELECT v2.4d. We extracted three regions - inner, outer and total - for each object. The inner region is where one would expect higher contamination of metals from SNIa in the cool-core galaxy groups and clusters (see Section \ref{subsec:regions}). Spectra channels were grouped with $\geq10$ counts per channel. Redistribution matrices and ancillary response files were generated using \texttt{xisrmfgen} and \texttt{xissimarfgen}, respectively. Observations of the North Ecliptic Pole were used as background observations. We applied the same screening criteria described above for the background observation (ID 100018010).

\subsection{Regions selection} \label{subsec:regions}

We used three regions for each cluster, an inner region encompassing (nearly) the entire cool-core, an outer region built to avoid cool-core contamination, and a full field-of-view (FoV) region. We selected cool-core clusters with central Fe abundance gradients. The innermost regions of cool-core clusters have particular characteristics. Fe abundance and surface brightness peak at the center, while temperature falls typically to $\sim2-3$ times lower than the outskirt values. Non-cool-core clusters lack the surface brightness peak and central temperature drop \citep[e.g.][]{DeGrandi_2001}. We exploit the central Fe excess adopting inner region sizes that maximize the Fe abundance excess, encompassing the cool-core domain. The Fe abundance peak in the center of the cool-core galaxy systems may be a result of an extensive period of enrichment by SNIa explosions deposited in or created by the brightest cluster galaxy \citep{DeGrandi2004, Bohringer2004}. 
While early studies suggested a relatively higher contribution of the Fe mass fraction from SNIa ejecta in the central regions of cool core clusters compared to the outer regions \citep[e.g.][]{Dupke2000,Finoguenov2000}, this was not shown to be ubiquitous and also not systematically found across all measured ratios \citep[e.g.][and references therein]{dePlaa2006,Simionescu2009,Bohringer2010,Mernier2017}. However, the presence of such a radial gradient of SNIa/SNcc Fe mass fraction could increase the ranking power of SN explosion model pairs by our method by analyzing the central and outer regions separately. If the gradient is not present, the separation of inner/outer regions would smooth out the ranking ability of specific abundance ratios due to the increased errors associated with having fewer counts. But even so, it should provide similar ranking results. For this reason, we choose to measure the abundances separately from the innermost and outermost regions to potentially maximize the range of variation of SNIa/SNcc yield mass fraction, improving the ranking power of the method.

The redshift spanned by the sample is based on both the point spread function (PSF) and the field-of-view of \textit{Suzaku}. The inner regions are circles, with radii varying from $1\arcmin$ to $4\arcmin$. The variation of the inner region size is chosen to take into account the cool-core radius and the minimization of \textit{Suzaku} PSF contamination. We assessed the cool-core radius from the ACCEPT project \citep{Cavagnolo2009}, except for Abell 3571 \citep{Dupke2000,Eckert2011}, the NGC 6338, NGC 2300, \citep{Rasmussen2007}, NGC 4472 \citep{Kraft2010}, NGC 1550 \citep{Kolokythas2020}, and UGC 3957 \citep{Tholken2016} groups. Outer regions are annuli with a radius $\geq 5\arcmin$ to reduce contamination from the cool-core vicinity\footnote{\url{http://www.astro.isas.jaxa.jp/suzaku/process/caveats/caveats_xrtxis03.html}}. Full FoV regions (i.e., total regions) have $8\arcmin$ of radius.

\subsection{Spectral Models} \label{subsec:spec_model}

We fit each cluster/group to the optically thin thermal plasma cluster/group emission \textit{vapec} model. Elemental abundances of O, Ne, Mg, Si, S, Ar, Ca, Fe, and Ni were free to vary and temperature and normalization. Cluster emission is expected to be absorbed by foreground Galactic gas, characterized by \textit{wabs} model \citep{Morrison1983}. We then adopted a fixed Galactic hydrogen column of $N_H$ collected from the HEASARC $N_H$ tool\footnote{\url{http://heasarc.gsfc.nasa.gov/cgi-bin/Tools/w3nh/w3nh.pl}}. We included a power-law component to model active galaxy nuclei (AGN) emission when needed, based on visual inspection of archival \textit{Chandra} observations (only for inner and total regions). We fixed the power-law index at 1.9, as found by the high-resolution \textit{Hitomi} analysis of NGC 1275 \citep{Hitomi2017}, but its normalization is a free parameter. The energy band range of spectral fits was limited to 0.5--8.0\,keV. We fitted individual spectra for each cluster/group, region, and XIS instrument. For instance, the inner region of the Centaurus Cluster has four abundances of Si (from XIS0, 1, 2 \& 3), and we calculate their error-weighted average to obtain a final value of Si abundance. We repeated this procedure for each object, instrument, region, and elemental abundance measurement. We report the final error-weighted average of temperature and chemical abundances for each region in Table \ref{tab:Inner_Outer_Abunds}. Then, we determined and compared these final abundance ratios of O/Fe, Ne/Fe, Mg/Fe, Si/Fe, S/Fe, Ar/Fe, Ca/Fe, and Ni/Fe with those predicted by SN explosion models. We performed the spectral fittings with XSPEC v12.9.1 \citep{Arnaud_1996} and AtomDB v3.0.7 assuming a $\chi^2$ statistic.

\section{Supernova Model Yields and Observed ICM comparison} \label{sec:SN_models}

Metals in the ICM have origin in different stellar sources. We assume that the heavy elements observed in the ICM/IGrM originate ultimately from ejected material of two SN Types, SNIa and SNcc, the latter representing all core-collapse SNe. The chemical enrichment of the ICM can then be quantified by the linear combination of each SN type. The \textit{observed} number of atoms in the ICM, $N(\mathrm{E})$, of a certain element E, is expressed as

\begin{equation}\label{eq:linear_combination}
     \underbrace{N(\mathrm{E})}_{\text{observed in the ICM}}=\underbrace{n_i N_i^{\text{Ia}}(\mathrm{E}) + n_j N_j^{\text{cc}}(\mathrm{E})}_{\text{predicted by SN models}},
\end{equation}
where $N_i^{\text{Ia}}(\mathrm{E})$ and $N_j^{\text{cc}}(\mathrm{E})$ are the number of atoms predicted by the SNIa model $i$ and SNcc model $j$, respectively, where $i=\{1, 2, ...\,, 232\}$ refers to the 232 SNIa models considered, and $j=\{1, 2, ...\,, 31\}$, indexes the 31 SNcc models. The numerical abundances $A_i^{\text{Ia}}(\mathrm{E})$ and $A_j^{\text{cc}}(\mathrm{E})$ are the number of atoms of a given element E divided by the number of atoms of Hydrogen (H), normalized to the same ratio for the solar value, produced by the SNIa and SNcc models $i$ and $j$, respectively. The coefficients $n_i$ and $n_j$ are the number of SNIa and SNcc that enriched the ICM calculated adopting models $i$ and $j$, respectively. The argument E represents the chemical element considered, where E=\{O, Ne, Mg, Si, S, Ar, Ca, Fe, Ni\}.

To calculate the predicted abundance $A_i^{\text{Ia}}$ (relative to H) of a given element E, defined as 

\begin{equation}\label{eq:A_E}
A_i^{\text{Ia}}=A_i^{\text{Ia}}(\mathrm{E}) \equiv \frac{[N_i^{\text{Ia}}(\mathrm{E})/N(\mathrm{H})]}{[N(\mathrm{E})/N(\mathrm{H})]_{\odot}},\\
\end{equation}

we derive the number of atoms, $N_i^{\text{Ia}}$, of a given element E via the relation below:
\begin{equation}\label{eq:N_E}
N_i^{\text{Ia}}=N_i^{\text{Ia}}(\mathrm{E})=\frac{Y_i^{\text{Ia}}(\mathrm{E})}{\mu(\mathrm{E})},
\end{equation}
where $Y_i^{\text{Ia}}(\mathrm{E})$ is the mass yield of the element E (simulated) by the SNIa model $i$ and $\mu(\mathrm{E})$ is the atomic mass of the element E. $N(\mathrm{E})$ and $N(\mathrm{H})$ are the number of atoms of the elements E and Hydrogen, respectively, where $[N(\mathrm{E})/N(\mathrm{H})]_{\odot}$ represents that ratio for the sun, i.e., solar abundance (in this work we use solar abundances of ANGR). The results obtained in this work are not affected by the choice of the solar abundance table.

Our collection of SNe models encompasses the current models with published yields, $Y_i^{\text{Ia}}$ and  $Y_j^{\text{cc}}$, of a given element E and are listed in Tables \ref{tab:list_SNIa_models} and \ref{tab:list_SNcc_models} in the Appendix \ref{sec:list_SNe_models}.
While $Y_i^{\text{Ia}}$ are taken directly from the yield tables, the SNcc yields are provided only for discrete 
progenitor masses (m), with a given initial progenitor metallicity $Z_{\rm init}$, $Y_j^{\text{cc}}(m, \mathrm{E}, Z_{\rm init})$. We then used a Salpeter IMF \citep{Salpeter1955} with slope $\alpha=-2.35$ to weigh these discrete yields within the mass limits simulated ($m_{j,\text{low}}$ to $m_{j,\text{up}}$) for a common $Z_{\rm init}$. The total average yield mass, $\overline{Y}_j^{\text{cc}}(\mathrm{E}, Z_{\rm init})$, of a given element E weighted over the Salpeter IMF within $m_{j,low}$ and $m_{j,up}$ at a given initial metallicity of the progenitor star, $Z_{\rm init}$, is expressed as 
\begin{equation}\label{eq:IMF_integration}
\overline{Y}_j^{\text{cc}}=\overline{Y}_j^{\text{cc}}(\mathrm{E}, Z_{\rm init})=\frac{\displaystyle \int_{m_{j,\text{low}}}^{m_{j,\text{up}}} Y_j^{\text{cc}}(m, \mathrm{E}, Z_{\rm init}) \ m^{\alpha}\, dm}{\displaystyle \int_{m_{j,\text{low}}}^{m_{j,\text{up}}} m^{\alpha}\ dm} \,,
\end{equation}
where $Y_j^{\text{cc}}(m, \mathrm{E}, Z_{\rm init})$ is the yield predicted by SNcc model $j$ of a given element E produced by an initial progenitor stellar mass $m$ with an initial metallicity $Z_{\rm init}$. 

The individual elemental abundance ratio of a given element E to that of Fe, $R(\mathrm{E})$, using Equations \eqref{eq:linear_combination}--\eqref{eq:N_E}, can be written as:
\begin{equation}\label{eq:linear_combination_R_E}
    \textstyle R(\mathrm{E})=f_i \, R_i^{\text{Ia}}(\mathrm{E}) + f_j \, R_j^{\text{cc}}(\mathrm{E}) \,,
\end{equation}
where
\begin{equation}\label{eq:R_E}
    R_i^{\text{Ia}}=R_i^{\text{Ia}}(\mathrm{E}) \equiv \frac{A_i^{\text{Ia}}(\mathrm{E})}{A_i^{\text{Ia}}(\text{Fe})}=\bigg( \frac{N_i^{\text{Ia}}(\mathrm{E})}{N_i^{\text{Ia}}(\text{Fe})} \bigg) \bigg/
\bigg( \frac{N(\mathrm{E})}{N(\text{Fe})} \bigg)_\odot \,. \\[1mm]
\end{equation}

Here, $R_i^{\text{Ia}}(\mathrm{E})$ and $R_j^{\text{cc}}(\mathrm{E})$ are the abundance ratios of element E predicted by the SNIa model $i$ and SNcc model $j$, respectively. The $f_i$ and $f_j$ represent the iron mass fractions that result from enriching the ICM using the SNIa model $i$ and SNcc model $j$. An analogous expression of Equation \eqref{eq:R_E} can be easily obtained for SNcc using the same formalism denoted by the index $j$ in Equations \eqref{eq:A_E}, \eqref{eq:N_E} and \eqref{eq:R_E}.

The iron mass fraction of the respective SNIa from model $i$ and SNcc from model $j$ that enriches the ICM, $f_i$ and $f_j$, are then expressed as
\begin{small}
\begin{equation} \label{eq:f_i}
f_i = \frac{Y_i^{\text{Ia}}(\mathrm{Fe})\overline{Y}_j^{\text{cc}}(\mathrm{Fe}) - Y_i^{\text{Ia}}(\mathrm{Fe})\overline{Y}_j^{\text{cc}}(\mathrm{E})\frac{A(\mathrm{Fe})}{A(\mathrm{E})}\frac{\mu(\mathrm{Fe})}{\mu(\mathrm{E})}\big( \frac{N(\mathrm{E})}{N(\text{Fe})} \big)_\odot}
{\frac{A(\mathrm{Fe})}{A(\mathrm{E})}\frac{\mu(\mathrm{Fe})}{\mu(\mathrm{E})}\big( \frac{N(\mathrm{E})}{N(\text{Fe})} \big)_\odot \left[ Y_i^{\text{Ia}}(\mathrm{E})\overline{Y}_j^{\text{cc}}(\mathrm{Fe}) - Y_i^{\text{Ia}}(\mathrm{Fe})\overline{Y}_j^{\text{cc}}(\mathrm{E}) \right] } \,,
\end{equation}
\end{small}
where, by definition
\begin{equation} \label{eq:f_j}
f_i + f_j \equiv 1 \,.
\end{equation}

The numerical abundance measured in the ICM of a given element E and Fe are represented by $A(\mathrm{E})$ and $A(\mathrm{Fe})$, respectively. The number of SNIa per SNcc enriching the ICM, $\eta \equiv n_i/n_j$, in terms of fundamental parameters can be derived as
\begin{equation} \label{eq:ni_nj}
\eta \equiv \frac{n_i}{n_j} =  \frac{\overline{Y}_j^{\text{cc}}(\mathrm{E}) \frac{M(\mathrm{Fe})}{M(\mathrm{E})} - \overline{Y}_j^{\text{cc}}(\mathrm{Fe})}
{Y_i^{\text{Ia}}(\mathrm{Fe}) - Y_i^{\text{Ia}}(\mathrm{E})\frac{M(\mathrm{Fe})}{M(\mathrm{E})}} \,,
\end{equation}
where $M(\mathrm{Fe})$ and $M(\mathrm{E})$ are the total gas mass in the ICM originally from Fe and element E, respectively.

In Section \ref{sec:KStest}, we compare these theoretical abundance ratios from a SNIa and SNcc models ($R_i^{\text{Ia}}(\mathrm{E})$ and $R_j^{\text{cc}}(\mathrm{E})$, respectively) to the observed abundance ratio distribution of a given element E.

\section{Evaluation of Supernovae Explosion Models} \label{sec:KStest}

The Kolmogorov-Smirnov test is a non-parametric statistical test that compares the cumulative distribution function (CDF) of two samples under the null hypothesis that these two samples are drawn from the same distribution \citep[see][for details]{Ivezic_2014book}. $D_{i,j}(\mathrm{E})$ is the metric that measures the maximum distance between two cumulative distribution functions (CDF) built for a given element E and drawn from the observations, and specific pair $(i,j)$ of supernova models, defined as

\begin{equation}
    D_{i,j}(\mathrm{E})=\mathrm{max} \left | \mathrm{eCDF}(R(\mathrm{E})) - \mathrm{tCDF}(R_i^{\text{Ia}}(\mathrm{E}), R_j^{\text{cc}}(\mathrm{E})) | \right.
	\label{eq:KS}
\end{equation}

These two CDFs are built as follows. Firstly, we build the empirical cumulative distribution function (eCDF) from the measured abundance ratios of the ICM of a given element E ($\mathrm{eCDF}(R(\mathrm{E}))$, see Section \ref{subsec:eCDF}). Secondly, the theoretical cumulative distribution function (tCDF) is associated with the theoretical abundance ratios of a pair of SN models, $R_i^{\text{Ia}}(\mathrm{E})$ and $R_j^{\text{cc}}(\mathrm{E})$, composed by a certain SNIa model $i$ and a SNcc model $j$ ($\mathrm{tCDF}(R_i^{\text{Ia}}(\mathrm{E}), R_j^{\text{cc}}(\mathrm{E}))$, see Section \ref{subsec:tCDF}). This tCDF is built using those ratios as the limits of the probability distribution function (PDF) and the assumption that, in between these limits, each value has the same probability, i.e., a flat PDF.

We choose a pre-established significance level $\alpha=0.05$ as the threshold to reject the null hypothesis. Hence, when a \textit{p}-value is lower than the threshold, we can conclude that the two distributions are too dissimilar to be drawn from the same distribution at a 95\% significance level and, thus, we reject the null hypothesis. We implement the two-sample Kolmogorov-Smirnov test adapting the \texttt{scipy.stats.kstest} routine of the \texttt{SciPy-Python} library \citep{SciPy2020} to compute the KS statistics $D_{i,j}(\mathrm{E})$ and the respective \textit{p}-values.

\subsection{Empirical Cumulative Distribution Function (eCDF)} \label{subsec:eCDF}

\begin{figure}
    \centering
    \includegraphics[width=\columnwidth, angle=0]{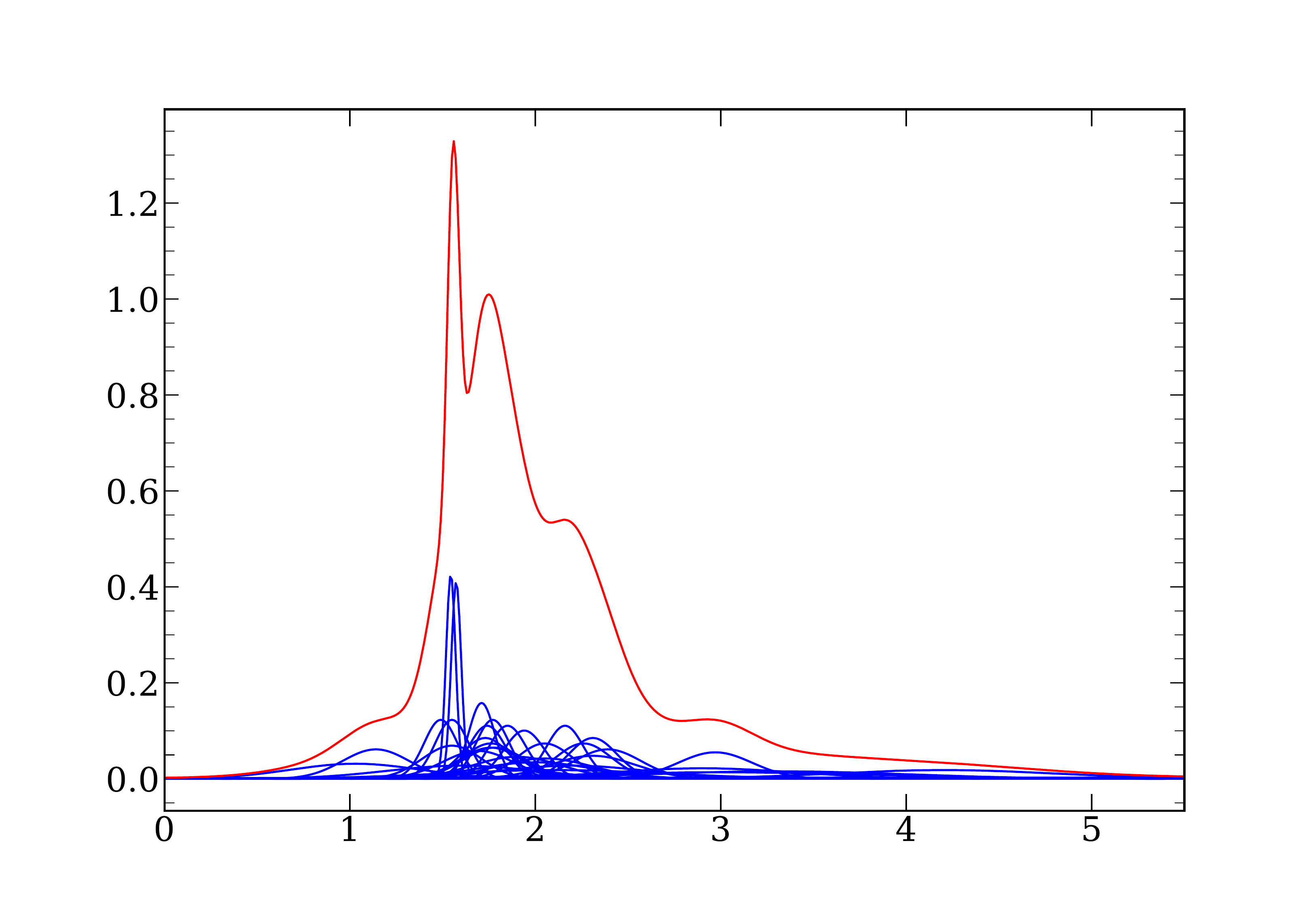}
    \caption{Probability Distribution Function of Si/Fe abundance ratio measurements of inner and outer regions. Blue lines indicate the Si/Fe measurement for each object in our sample, represented by normal distributions whose mean is the nominal Si/Fe value and the standard deviation is the respective error. The red line is the final sum of each normal distribution in our sample. We also present the PDFs for others abundance ratios in the Appendix \ref{sec:pdfs} in Figure \ref{fig:pdfs}.}
   \label{fig:pdf_Si_Fe}
\end{figure}

We built the empirical CDF considering the nominal values of the observed abundance ratios and their associated errors. For the sake of illustration, we describe the procedure using the Si/Fe abundance ratio to evaluate a pair of SN models as an example. We assumed that each Si/Fe measurement could be modeled as a normal distribution centered in the nominal value of the abundance ratio with a standard deviation given by its error. Figure \ref{fig:pdf_Si_Fe} shows the normal distributions of Si/Fe ratios, $R(\mathrm{E})$, for each galaxy cluster and group (blue lines) in our sample. The total probability distribution function of Si/Fe (red line) is obtained by summing and normalizing the individual normal distributions. We can easily calculate the corresponding eCDF of Si/Fe (green line, Figure \ref{fig:cdf_Si_Fe}) using the total probability distribution. This procedure is repeated for each one of the eight abundance ratios considered in this work. The PDFs for others abundance ratios are presented in the Appendix \ref{sec:pdfs} in Figure \ref{fig:pdfs}.

\subsection{Theoretical Cumulative Distribution Function (tCDF)} \label{subsec:tCDF}

The observed abundance ratios can be expressed as a (linear) combination of yields produced by both SN types (see Section \ref{sec:SN_models}). For this reason, we build a single theoretical CDF per pair of SNe models, i.e., composed of one SNIa and one SNcc model.

To illustrate the procedure, as an example, we chose the yields of the SNIa WDD1 delayed-detonation model of \citet{Iwamoto1999} (Iw99\_WDD1 model) and the Type II supernova solar metallicity model of \citep{Nomoto2013} (No13\_SNII\_Z2E-2 model), respectively. We then employ the Si/Fe ratio predictions from the Iw99\_WDD1 model and No13\_SNII\_Z2E-2 model, which are
$R(\texttt{Si})_\text{\tiny WDD1}^{\text{Ia}}=1.067$ and $R(\texttt{Si})_\text{\tiny SNIIZ2E-2}^{\text{cc}}=4.43$, respectively (for definition, see Eq.~\eqref{eq:R_E}).

We built the tCDF as a piecewise function defined as follows.
For a 100\% SNIa enrichment of the ICM (and therefore no contribution by SNcc), the abundance ratio would correspond to $R(\texttt{Si})\sim1.067$, as predicted by the Iw99\_WDD1 model solely. In the opposite extreme case, where all the enrichment comes from SNcc, $R(\texttt{Si})\sim4.43$, as estimated by the No13\_SNII\_Z2E-2 model. This implies that: (i) any intermediate measurement $1.067<R(\texttt{Si})<4.43$ must come from a linear combination of both SN types, and (ii) this pair of SN models can only explain the observed abundance ratio between $1.067$ and $4.43$, so that the probability of measuring a $R(\texttt{Si})$ lower than $1.067$ or higher than $4.43$ is 0. In terms of cumulative distribution function ($\text{tCDF(}x\text{)=P[Si/Fe}\le x$]), any measurement of $R(\texttt{Si}) < 1.067$ corresponds to a value of 0 (i.e., this pair of SN models could not explain finding a measurement of $R(\texttt{Si}) < 1.067$), while the tCDF is equal to 1 for any  measurement of $R(\texttt{Si})>4.43$, i.e., any value of Si/Fe lower or equal than $4.43$, is predicted by this pair of SN models. We assume that any intermediate combination of both SN types produces a linear relation in probability, which is derived from the following tCDF:

\begin{equation}\label{eq:tCDF}
    \text{tCDF}(x) = \left\{
    \begin{array}{@{}c@{\hspace{0.7cm}}l}
          \text{0}                   & \text{if $x < 1.067$,} \\[1mm]
          \dfrac{x-1.067}{4.43-1.067}  & \text{if $1.067 \le x < 4.43$,} \\[1mm]
          \text{1}                   & \text{if $x \ge 4.43$.}
    \end{array}
      \right.
\end{equation}

\begin{figure}
    \centering
    \includegraphics[width=\columnwidth, angle=0]{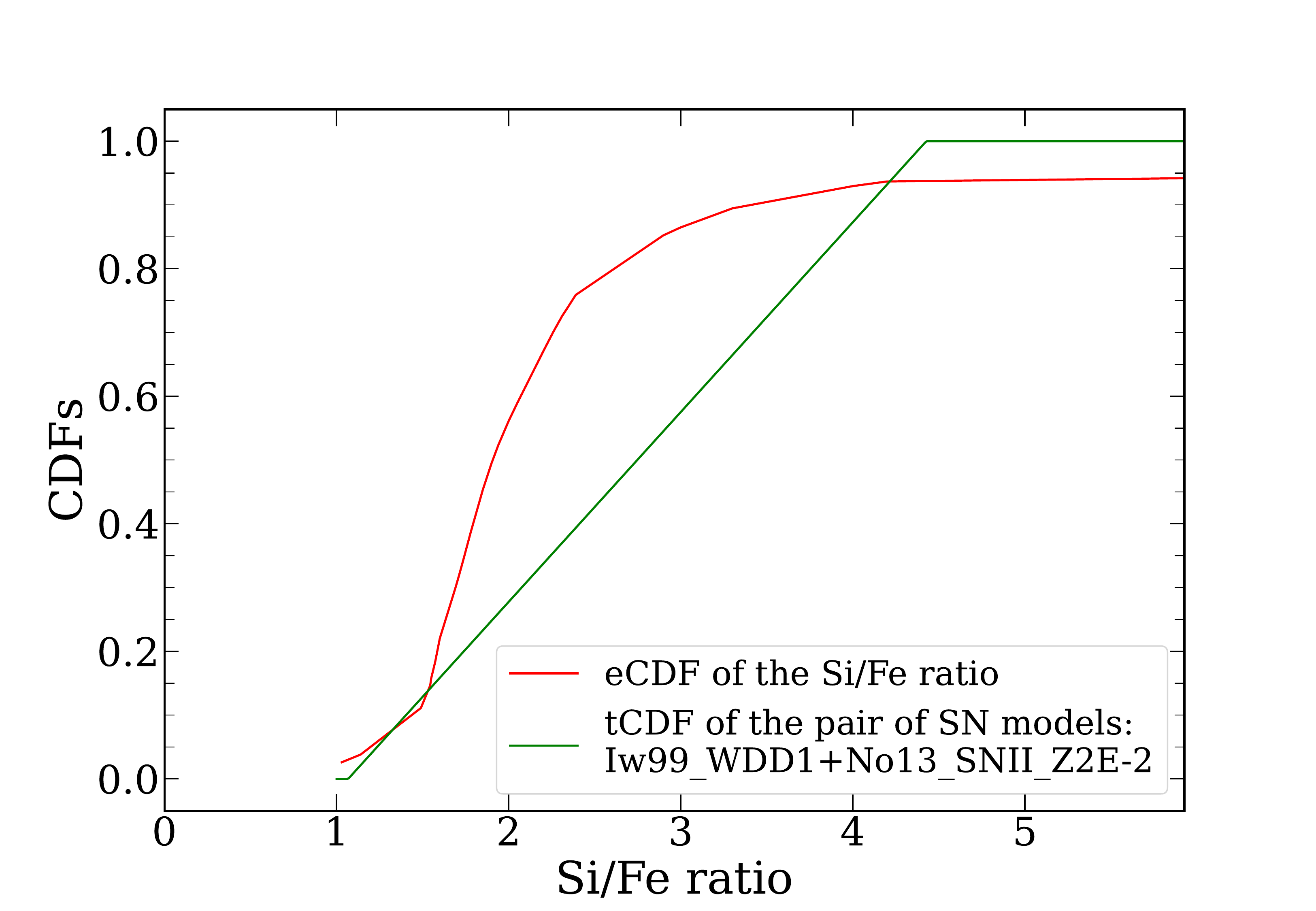}
    \caption{Comparison between the Empirical cumulative distribution function (eCDF) and Theoretical cumulative distribution function (tCDF) for the SN pair composed by Iw99\_WDD1 (SNIa) and No13\_SNII\_Z2E-2 (SNcc). The solid red line represents the eCDF of Si/Fe abundance ratio of the groups and clusters of galaxies of our sample. The solid green line indicates the tCDF built from the pair of SNe models Iw99\_WDD1 and No13\_SNII\_Z2E-2 (see text).}
   \label{fig:cdf_Si_Fe}
\end{figure}

The tCDF of the pair of SN models Iw99\_WDD1 (SNIa) and No13\_SNII\_Z2E-2 (SNcc) is presented in Figure \ref{fig:cdf_Si_Fe} for illustration. We repeat this procedure for each one of the eight abundance ratios considered in this work and for each pair of supernova models.

\subsection{Applying the KS test to the sample} \label{subsec:}

Figure \ref{fig:cdf_Si_Fe} reveals the similarity or differences between the eCDF for Si/Fe and the tCDF for Si/Fe and the pair of SN models Iw99\_WDD1 (SNIa) and No13\_SNII\_Z2E-2 (SNcc). We quantify this (dis)similarity by Equation \eqref{eq:KS} and its \textit{p}-value. Using only Si/Fe ratio, we repeated the above procedure for each of 7192 different pair combinations of SNe models, as shown in Figure \ref{fig:pvalues} for some of the tested SN models. Moreover, this analysis is conducted for each one of the eight abundance ratios listed in Section \ref{subsec:spec_model}, which is also represented in Figure \ref{fig:pvalues}. In the end, we have implemented a total of 57536 KS tests.

As mentioned previously, to reject the null hypothesis associated with the KS test, we use a threshold of 0.05. As a result, a rejected pair of SN models cannot explain our measurements of that particular abundance ratio at a 95\% significance level. As we repeat this process for each of our eight abundance ratios, we can count how many times a particular pair of SN models is rejected according to the KS test. The closer this number is to 8 (total number of abundance ratios), the less representative of the data this pair of models is.

Note that in this work, we assume that there is an equal probability for an abundance ratio to have a value at any point within the predicted range of $R_i^{\text{Ia}}(\mathrm{E})$ and $R_j^{\text{cc}}(\mathrm{E})$ for a given SNIa model $i$ and a SNcc model $j$ when building its $\mathrm{tCDF}(R_i^{\text{Ia}}(\mathrm{E}), R_j^{\text{cc}}(\mathrm{E}))$. There is no theoretical reason (so far) to assume otherwise. However, this assumption may induce some perhaps unfair rejections. Among all the SN model pairs analyzed here, some outlier SN model pairs predict very wide limits in abundance ratios, and the method will tend to reject them, given that ePDFs have a more limited range. Ideally, one would have different tPDFs provided by the modelers, perhaps based on the likelihood of physical parameters used in the models. This is something to encourage the modelers to provide. Another extreme scenario that could cause undesired rejections would be if we measured the same abundance ratio for every cluster region of every cluster with extreme precision. In this case, the ePDF would be a very narrow Gaussian within the range predicted by the tPDFs, increasing the eCDF and tCDF distances. Such an unusual set of observational measurements, however, would imply that the assumption of equal probabilities along with all the theoretically ``predicted" ranges would be likely incorrect. In any case, this particular anomalous case would be easily seen in the ePDF curves.

\section{Published Theoretical Supernova yield models} \label{sec:summary_SNmodels}

Based on the measured ICM abundances of eighteen nearby galaxy clusters/groups, we checked the capacity of theoretical SN models to recover the observed abundance ratio distributions of our sample successfully. For that, we first built linear combinations of the fractional contribution of yields from each pair of SN models. We then verified its consistency with the distribution of observed abundance ratio measurements.  The theoretical SNIa and SNcc yield models used are summarized in Tables \ref{tab:list_SNIa_models} and \ref{tab:list_SNcc_models}, respectively. We briefly describe them here. In the model description below, all masses are in $M_\odot$ and $\rho_{c,9}$ is the central density in units of $10^9 \mathrm{g~ cm^{-3}}$).

We tested the following theoretical SNIa models:

\begin{enumerate}
    \item The classical 1D Chandrasekhar mass ($M_\text{Ch}$) models \citep{Iwamoto1999}, including the ``fast'' deflagration models W7 and W70, as well as the delayed-detonation series of WDD and CDD models. Here, W and C are models based on a white dwarf and a carbon-oxygen core of an asymptotic giant branch star, respectively \citep{Nomoto_1984}. They are artificially transformed from a subsonic to a supersonic regime based on different transition densities ($\rho_{T,7}$). In denoting the models in the tables, we follow their original notation but starting with ``Iw99''.
    
    \item 2D spherical pure deflagration (DF) model and two 2D delayed-detonation (or deflagration-detonation transition, DD) models, one of which assumes a central (denoted as ``ctr'') deflagration and the other an initial off-center (denoted as ``off'') deflagration \citep{Maeda2010}. We denote them as Ma10\_(ctr$\mid$off)\_(DF$\mid$DD), whereby (x$\mid$y$\mid$z) indicates that we choose either x or y or z to be in the displayed name.
    
    \item 3D off-center deflagration model of \citet{Kromer2015} in a near-$M_\text{Ch}$ hybrid white dwarf, i.e., composed by carbon-oxygen (C+O) core and a surrounding layer of oxygen-neon (O+Ne) with five ignition spots placed randomly around the center of the WD as in \citet{Kromer2013.2015} and \citet{Fink2014} (denoted Kr15\_hybrid).

    \item A set of 3D pure deflagration models of $M_\text{Ch}$ C+O WD progenitor of \citet{Fink2014}, where the number of ignition spots, N, varies from 1 to 1600 and with a range of central densities ($\rho_{c,9}=1-5.5$). The authors used the same setup for deflagration explosions as \citet{Seitenzahl2013} without assuming delayed-detonation. Two models initially have a very dense arrangement of the ignition kernels and are referred to as ``compact'' models and denoted with an additional ``c'' in the model name. The notation then follows Fi14\_$\rho_{c,9}\_c$.

    \item A set of 3D delayed-detonation models of \citet{Seitenzahl2013} assuming a range of the number of ignition spots N, which varies from 1 to 1600, central densities ($\rho_{c,9}=1-5.5$) and initial metallicities ($Z_{\rm init}=0.01-1.0 Z_\odot$). We name the models Se13\_$\rho_{c,9}$\_Z$Z_{\rm init}$\_c. The additional ``$Z_{\rm init}$'' or ``c'' are in the name when the initial metallicity is not a unity or if there are ``compact'' ignition kernels, respectively. We also included the more recent 3D gravitationally confined detonation model with one off-centered ignition spot Se16\_GCD \citep{Seitenzahl2016}.
 
    \item A set of 3D delayed-detonation $M_\text{Ch}$ models of \cite{Ohlmann2014}, which investigates the impact of different initial carbon mass fractions ($X_{\rm C}$) on the spontaneous deflagration to detonation transition. They have deflagration ignition conditions identical to the N100 model of \citet{Seitenzahl2013}. The WD progenitor has a homogeneous core or a homogeneous carbon depleted core with different $X_{\rm C}$, and we denote these models following Oh14\_DD\_$X_{\rm C}$.

    \item A set of 2D pure turbulent deflagration models with and without transition to detonation for different central densities ($\rho_{c,9}=0.5-5.0$) and initial metallicities ($Z_{\rm init}=0-5 Z_\odot$) of the progenitor \citep{Leung2018}. Four models assume only pure turbulent deflagration and are denoted with an additional ``1P'' in the model name. We then denote the models as Le18\_$\rho_{c,9}\_{\rm Z}Z_{\rm init}$\_1P.
    
    \item We also consider, the sub-MCh double-detonation (DDet) models of \citet{Leung2020a}, for a wide range of progenitor masses ($M_\text{WD}=0.9-1.2$), Helium envelope masses ($M_\text{He}=0.05-0.1$), initial metallicities ($Z_{\rm init}=0-0.1$) and also probed carbon detonations triggered by different He detonation configurations: spherical (S), bubble (B) and ring (R) shaped. We denote their models as Le20a\_$M_\text{WD}$\_$M_\text{He}$\_$Z_{\rm init}$\_(S$\mid$B$\mid$R). 
    
   \item 2D pure turbulent deflagration models of \citet{Leung2020b} for a range of initial central densities ($\rho_{c,9}=1.0-9.0$) and WD masses ($M_\text{WD}=1.33-1.40$) of the progenitor with solar metallicity. These models are named as Le20b\_$\rho_{c,9}$\_$M_\text{WD}$.
   We also include the set of hybrid C+O core WDs with an O+Ne+Mg surrounding layer (hybrid) for the same range of total masses composed by different C+O masses ($M_{\rm C+O}=0.43-0.5$) with constant oxygen and neon mass of $0.9 M_\odot$. We name these hybrid models as Le20b\_Hybrid\_$M_{\rm C+O}$.
   
   \item The 3D violent merger models of \citet{Kromer2013,Kromer2016} including a merger between two sub-$M_\text{Ch}$ WDs (with masses $0.9 M_\odot$ and $0.76 M_\odot$) for two different metallicities ($Z_{\rm init}=1$ and $Z_{\rm init}=0.01$). These models are respectively, named Kr13\_0.9\_0.76 and Kr16\_0.9\_0.76\_Z1E-2.
    
    \item The 3D violent merger model between two sub-$M_\text{Ch}$ C+O WDs of $\approx 0.9 M_\odot$ \citep{Pakmor2010}, as well as the 3D violent merger model between two sub-$M_\text{Ch}$ C+O WDs of $0.9 M_\odot$ and $1.1 M_\odot$ \citep{Pakmor2012}, named
    Pr10\_0.9\_0.9 and Pr12\_1.1\_0.9, respectively.
    
    \item The 3D models of \citet{Papish2016} with published yields, simulating a direct collision of two sub-$M_\text{Ch}$ C+O WDs of $0.6 M_\odot$, with and without the presence of a low-mass He shell of $0.01M_\odot$) in the WD progenitors. These models are named Pa16\_1A and Pa16\_1C\_He, respectively.
    
    \item 3D Dynamically-driven double-degenerate double-detonation models of \citet{Shen2018} for different progenitor masses ($M_\text{WD}=0.8-1.10 $), carbon to oxygen ratios (${\rm C/O}=30/70$ or $50/50$), initial metallicities ($Z_{\rm init}=0-0.02$) and ${\rm ^{12}C+^{16}O}$ reaction rates ($\text{f}_{\rm ^{12}C+^{16}O}=0.1$ or $1.0$). The model parameters are inserted in their names following the notation  Sh18\_$M_\text{WD}$\_${\rm C/O}$\_$Z_{\rm init}$\_$\text{f}_{\rm ^{12}C+^{16}O}$.
    
    \item 1D pure detonation (Det) of sub-$M_\text{Ch}$ C+O WD models for a range of $M_\text{WD}=(0.81-1.15)$, initial central densities ($\rho_{c,7}=1.0-7.9$) of the progenitor of \citet{Sim2010}. We follow the notation Si10\_Det\_$M_\text{WD}$\_$\rho_{c,7}$. One of the cases, Si10\_Det\_1.06\_4.15 is also simulated for a non-zero Ne mass fraction and detonated by Si10\_Det\_1.06\_4.15\_Ne. 
    
    \item Both sets of 2D converging-shock (CS) and edge-lit (EL) double-detonation models and He only (He) detonation models of \citet{Sim2012}, for two $M_\text{WD}$ masses of 0.66 and 0.79 $M_\odot$ (with a fixed $M_{\rm He}=0.21$) and central densities of $\rho_{c,7}$ of 0.38 and 0.85, for each case. We named them Si12\_DDet\_$M_\text{WD}$\_$\rho_{c,7}$\_(CS$\mid$EL$\mid$He). 
    
    \item The 3D delayed-detonation models of O+Ne WD for a range of masses $M_\text{WD}=1.18-1.25$ and central densities ($\rho_{c,9}=0.1-0.2$) of \citet{Marquardt2015}. We also include their comparison model of a 1.23 M$_\odot$ C+O WD. We denote them as Mk15\_DD\_$M_\text{WD}$\_$\rho_{c,9}$\_(ONe$\mid$CO).

    \item 2D and 3D centrally ignited deflagration models, as well as 3D multi-point ignition models with 5 and 30 ignition bubbles (IB) of \citet{Travaglio_2004}. Their WD central density, $\rho_{c,9}$, is 2.9 using a variety of linear grid sizes (GS) of 256, 512, and 768. We name the models Tr04\_3D\_IB\_GS if the ignition bubbles are off-center and Tr04\_(2D$\mid$3D)\_GS if they are centered. In one of the cases, we also considered the Tr04\_3D\_256 for the nucleosynthesis calculations starting only at 90\% of the temperature peak ($\sim8.5\ \mathrm{x}\ 10^9$ K) for those tracer particles that reach nuclear statistical equilibrium (NSE) conditions. We add a ``T'' to its notation.
    \end{enumerate}

In the context of SNcc models, we test the following SNcc models assuming a Salpeter IMF (see Equation \eqref{eq:IMF_integration}):

\begin{enumerate}
    \item Type II supernova model of \citet{Nomoto1997}, which provides the IMF-weighted yields calculated for a range of $10-50M_{\odot}$ with $Z_{\rm init}=1 Z_\odot$ and we denote by No97\_SNII\_Z1.

    \item Three older versions of the SNII and Hypernovae (HN) models from \citet{Nomoto2006} as a function of the initial metallicity and mass of the progenitor star; The models are named No06\_(SNII$\mid$HN)\_$Z_{\rm init}$. Their updated models are in Table \ref{tab:list_SNcc_models} referenced by \citet{Kobayashi2011}.

    \item A set of SNcc models including normal SNII, HN, and pair-instability SN (PISN) taken from \citet{Nomoto2013} as a function of the mass of the progenitor star (up to $40 \text{M}_\odot$) for a range of initial metallicity ($Z_{\rm init}=0-0.05$). The PISN model has only zero initial metallicity. The models are named No13\_(SNII$\mid$HN$\mid$PISN)\_Z$Z_{\rm init}$. We also consider the extended version for yields normal SNII and HN models (up to $140 \text{M}_\odot$) at $Z_{\rm init}=0$, which we denoted by No13\_(SNII$\mid$HN)\_Z0\_ext. We also include a PISN \& SNII model with a progenitor mass range from 11 to 300 M$_{\odot}$ and zero initial metallicity called No13\_SNII\&PISN\_Z0.
    
    \item \citet{Heger2002} SNII yields for $10-100 \text{M}_\odot$ mass range and \citet{Heger2010} PISN yields for $140-260 \text{M}_\odot$ denoted by He02\_SNII\_$Z_{\rm init}$ and He10\_PISN\_$Z_{\rm init}$ model, respectively. We also consider the combination of the two. They are named He10\_SNII\&PISN\_0, for $10-260 \text{M}_\odot$ mass range. All of the above models use $Z_{\rm init}=0$.
    
    \item SNII yields for a $13-35 \text{M}_\odot$ mass range of the progenitor with different metallicities ($Z_{\rm init}=0-0.02$), which correspond to models named Ch04\_SNII\_Z$Z_{\rm init}$ \citep{Chieffi2004}.

    \item A recent alternative set of SNcc models of \citet{Sukhbold2016} that provides yields for a progenitor mass range of $12.25-120 \text{M}_\odot$ with solar metallicity. These models include neutrino transport and use the well-studied SN1987A as a calibrated point, denoted Su16\_W18 and Su16\_N20.
\end{enumerate}

\section{Results} \label{sec:results}

\subsection{Abundance Distributions} \label{subsec:abund}

\begin{figure}
    \centering
    \includegraphics[width=\columnwidth, angle=0]{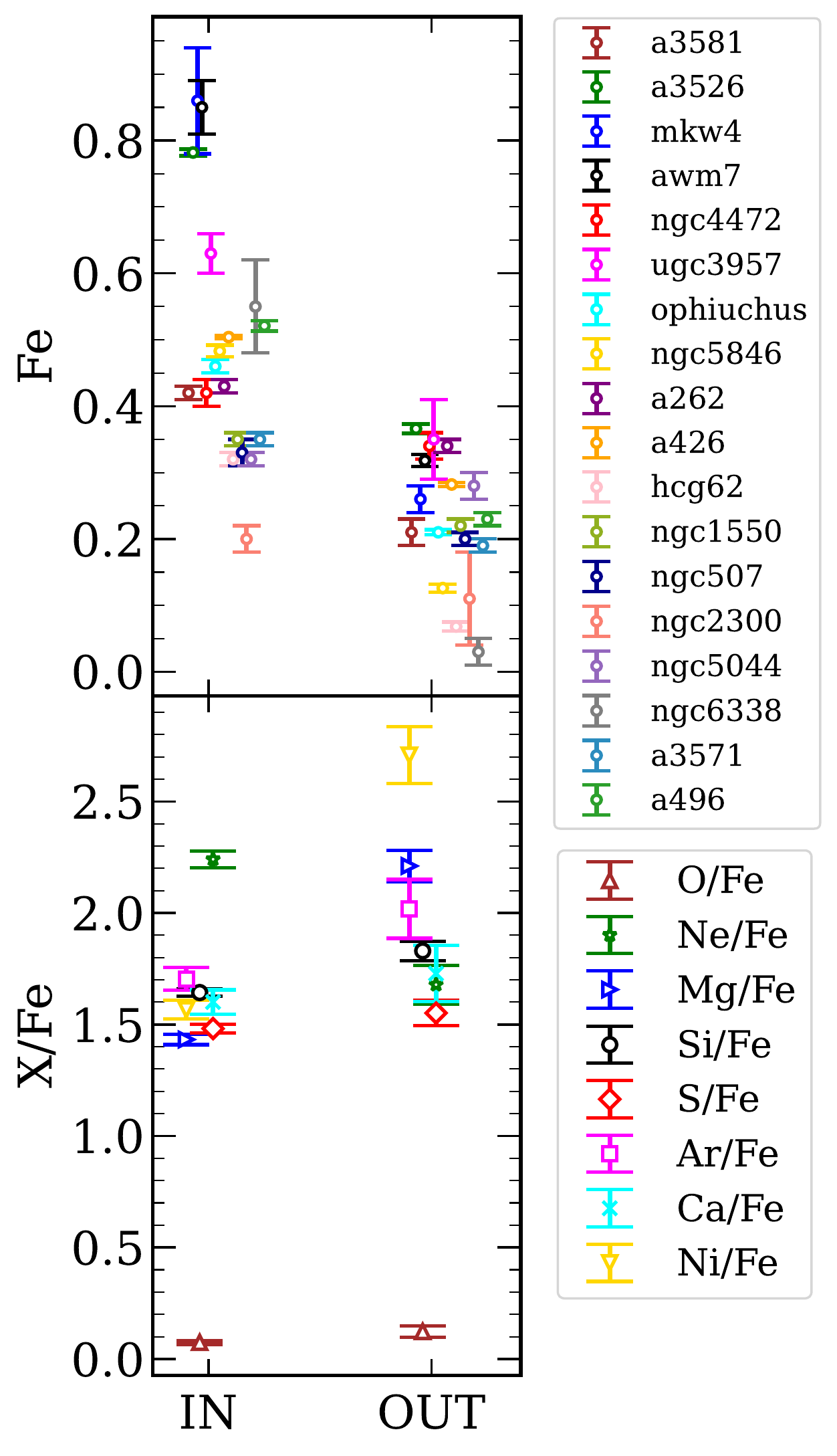}
    \caption{\textit{Top}: \textit{Suzaku} Fe abundance from the inner and outer regions (see text) for each group/cluster in our sample. \textit{Bottom}: The error-weighted average over our sample of each abundance ratio (X/Fe) for the inner and outer regions.}
   \label{fig:IN_OUT_average}
\end{figure}

Table \ref{tab:Inner_Outer_Abunds} presents our results of the X-ray spectral fitting for inner, outer, and entire FoV regions. We measured the elemental abundances (relative to Solar) of O, Ne, Mg, Si, S, Ar, Ca, Ni, and Fe for the eighteen groups and clusters of galaxies up to redshift 0.0391. The error-weighted average over our sample of each abundance ratio is presented in Figure \ref{fig:IN_OUT_average}.

By comparing Fe abundance measurements of inner and outer regions, we have confirmed the central Fe abundance enhancement using \textit{Suzaku} observations for the clusters and groups of our sample (see Figure \ref{fig:IN_OUT_average}). The high Fe abundance in central regions compared to outer regions is consistent with our sample being composed of cool-core clusters/groups (see Section \ref{subsec:regions}). This gradient is possibly due to a combination of effects, including ram-pressure stripping of galaxy flybys near the center of clusters, winds from the BCG, uplifting gas by AGN produced bubbles \citep[e.g.][]{Nulsen2002, Simionescu2009, Kirkpatrick2015} or even intracluster type Ia supernovae \citep{Omar2019}. As mentioned previously, if the Fe gradient is coupled to an abundance ratio gradient, we can use it to improve the discriminating power of our technique. We can do this by separating observations of internal and external regions instead of using only the entire FoV measurements. Despite the obvious gain in number counts (statistics), the use of the whole FoV averages out the abundance inhomogeneities (potentially due to different enrichment types) inherent to the ICM and makes it insensitive to the real span of elemental abundance ratios. By sampling, instead, abundance measurements in internal and external regions, we maximize the observed range of abundance ratios and improve the discriminating power between the SN explosion models, which is the goal of this work.

Figure \ref{fig:IN_OUT_average} shows the elemental abundance ratio of element ``X" to iron (X/Fe) radial inner and outer region measurements. While some ratios are consistent with flat profiles, within the errors, others show indications of radial gradients within $\sim$0.2\,R$_{200}$.

It is worth noticing that aside from the expected abundance ratio radial gradients mentioned above, the distribution of abundance ratios often shows significant differences not just radially but also azimuthally \citep[e.g.][]{Dupke2007ApJ...668..781D}.
These inhomogeneities also can be found at small spatial scales near the cluster core \citep[e.g.][]{Dupke2007ApJ...671..181D,Dupke2007ApJ...668..781D} and can be due to different mechanisms, e.g., core sloshing, AGN induced bubble uplifting \citep[e.g.][]{Churazov2001,Guo2010}, residual substructures from a merger, and will vary from cluster to cluster. The precise level of inhomogeneity and its causes for each cluster are scientifically interesting, but not relevant for the purposes of this work. However, even if we had the necessary spatial resolution to map the inhomogeneities in detail, it could only result in minor variations of the \textit{range} of abundance ratios, which would only slightly improve the discriminating power of the method. Given that the ICM is optically thin throughout, we apply the simplest models that account for the collisionally ionized equilibrium (CIE) plasma, Galactic Hydrogen absorption, and the AGN emission, the latter being applied only to the inner and total regions. Nevertheless, as a sanity check, we compare some of our results to other works that aim to determine the phenomenological idiosyncrasies of the ICM, especially near the core. For that, we chose the Perseus cluster, overwhelmingly studied with multiple independent abundance ratio measurements, including RGS/\textit{XMM-Newton} and \textit{Hitomi} \citep[][hereafter S19]{Simionescu2019} aside from \textit{Suzaku} \citep[][hereafter T09]{Tamura2009}. The detailed discussion of that comparison is presented in Appendix \ref{sec:comparison_perseus}, where one can see that despite the differences in extracted spatial regions, spectral modeling, systematics related to the effective area, and line de-blending capabilities, our results are consistent overall with previous measurements.

Since the systems studied here exhibit positive radial temperature gradients within the regions analyzed, and they are overall ``cold'' it is pertinent to check the effects of the so-called Fe bias \citep[][]{Buote1998,Buote2000,Gastaldello2021}. In principle, one would expect this effect to be small\citep[e.g.][]{rasia2008} especially using our coarse regions and extending the energy range down to 0.5\,keV, which allows us to obtain a robust determination of the continuum. Nevertheless, we investigated whether neglecting the multi-temperature structure in cool groups significantly impacts the X/Fe ratio. We re-fitted the cool clusters (below kT=2\,keV) with a vapec+vapec (2T) model, where the individual abundance parameters were tied between the vapec models. We found that the single vapec (1T) model measures Fe abundance was lower than the 2T model for 55\% of the cool groups/clusters. In fact, the Fe error-weighted average increased from $0.553 \pm 0.005$ (1T) to $0.779 \pm 0.007$ (2T) for our sample of groups/clusters below 2\,keV. However, the abundance ratios derived from the 2T and 1T models are generally within the errors, except for very few specific and non systematic cases, which do not affect the assessment of SN pair rankings, given the large set of abundance ratio measures to build the eCDF.

We also compare our X/Fe measurements with previous results in the literature. A brief discussion of that comparison is presented in the Appendix \ref{sec:comparison_ratios}, where one can see that despite using different samples, X-ray satellites, detectors, extraction regions, atomic database versions, and spectral models, the results presented here are, in general, consistent for most X/Fe measurements.

\subsection{Comparison between ICM abundance ratios with Supernovae models yields}

\begin{figure*}
    \centering
    \includegraphics[width=0.497\textwidth, clip]{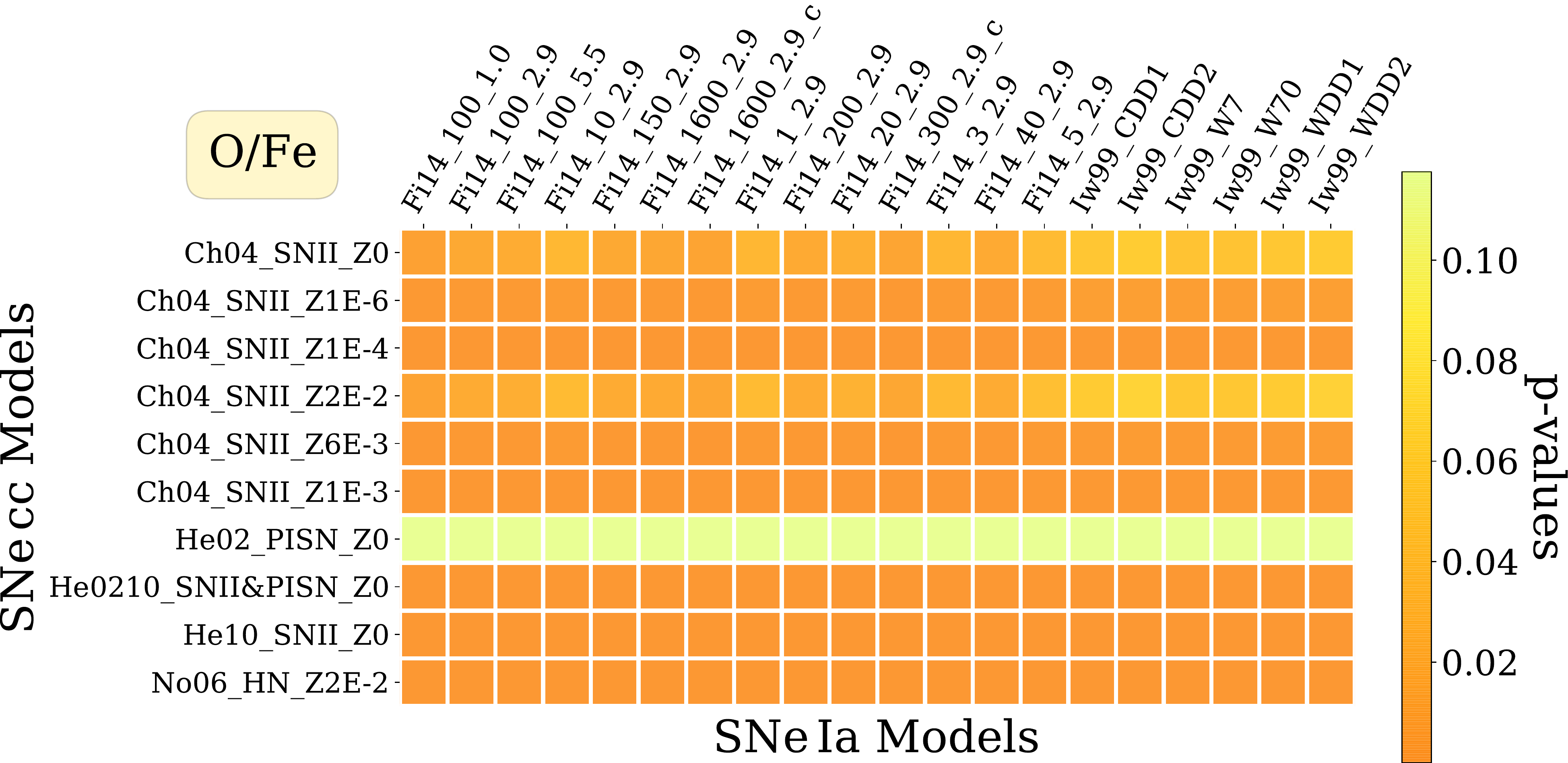}
    \includegraphics[width=0.497\textwidth, clip]{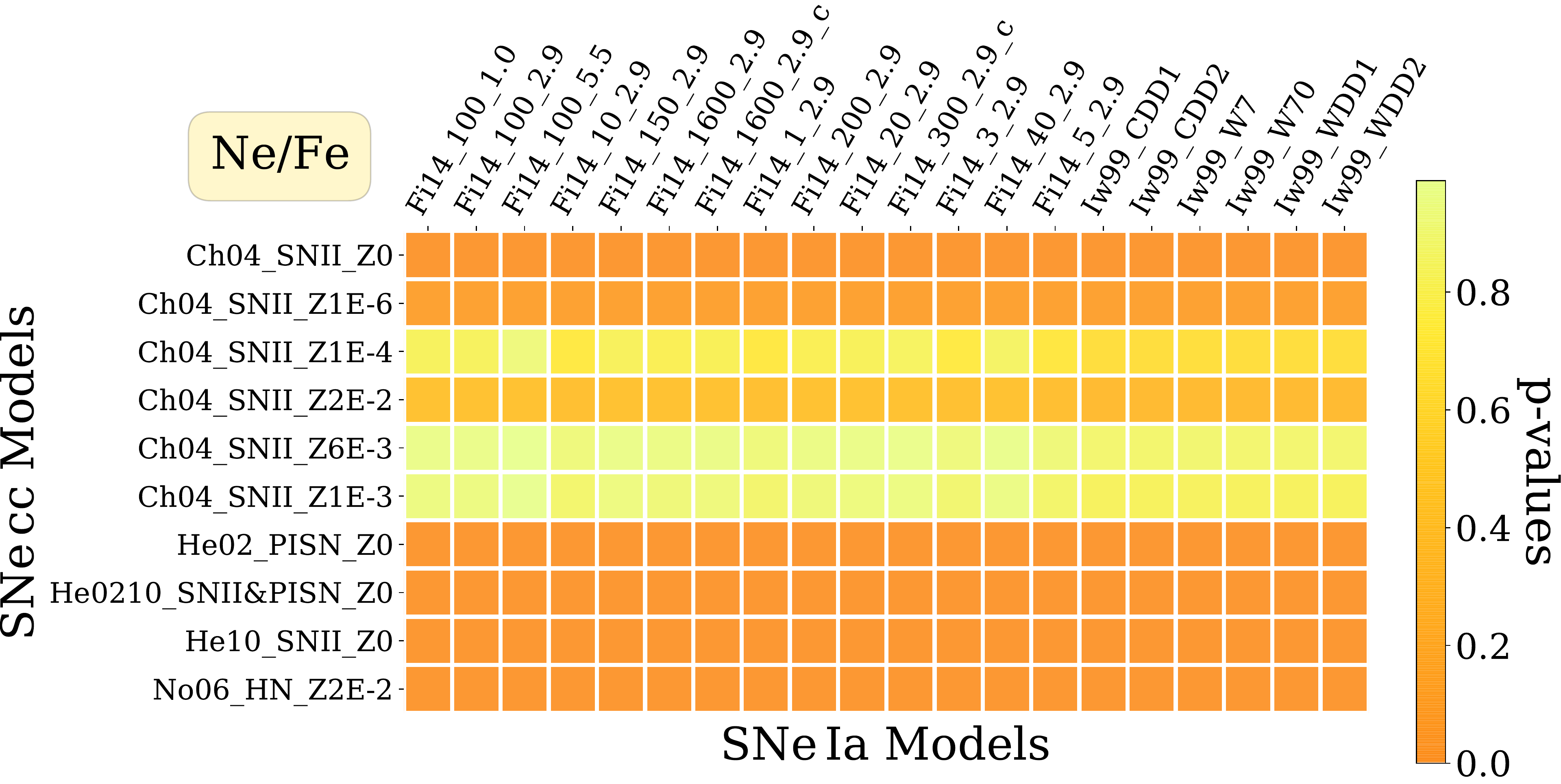}
    \includegraphics[width=0.497\textwidth, clip]{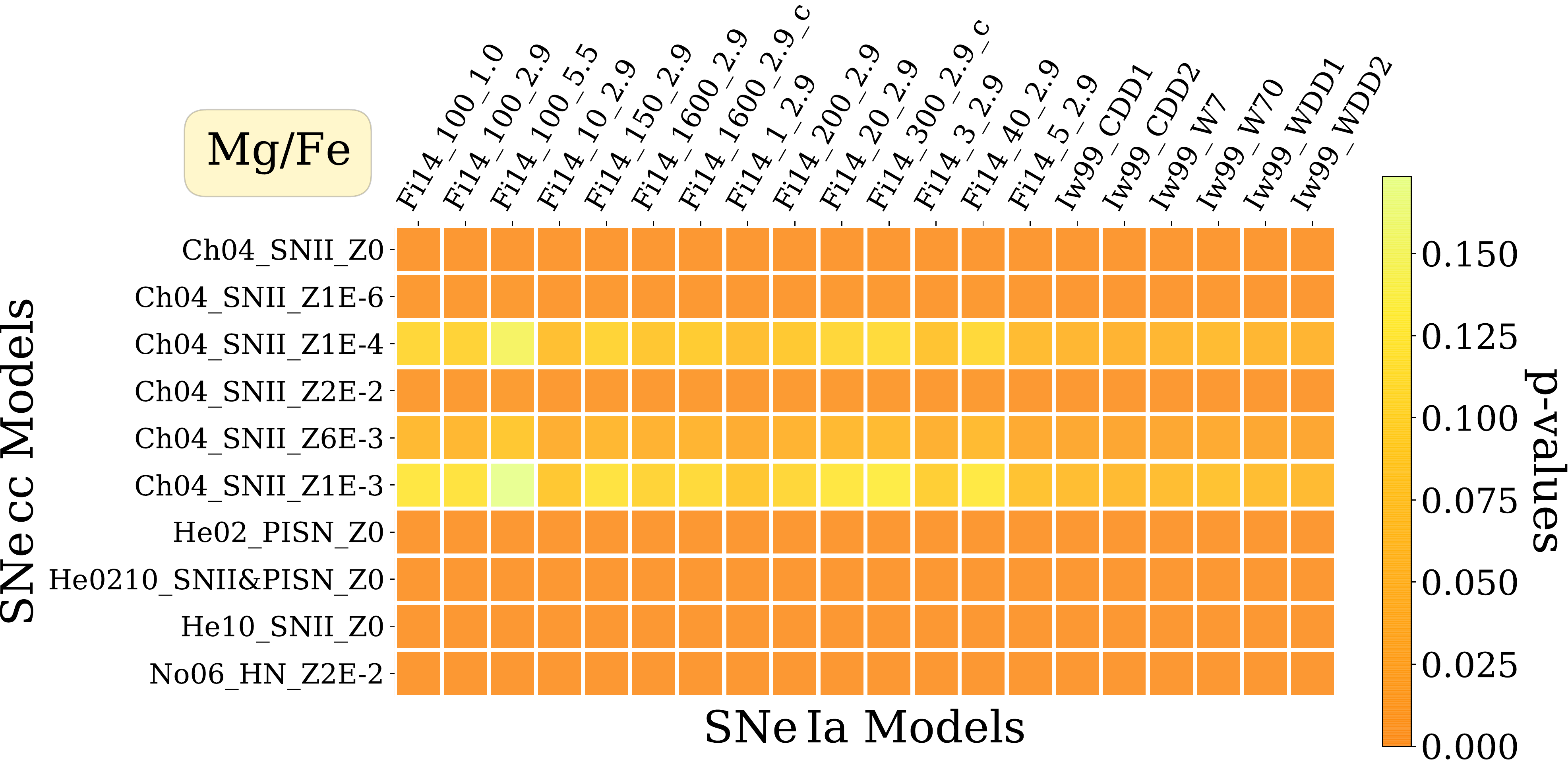}
    \includegraphics[width=0.497\textwidth, clip]{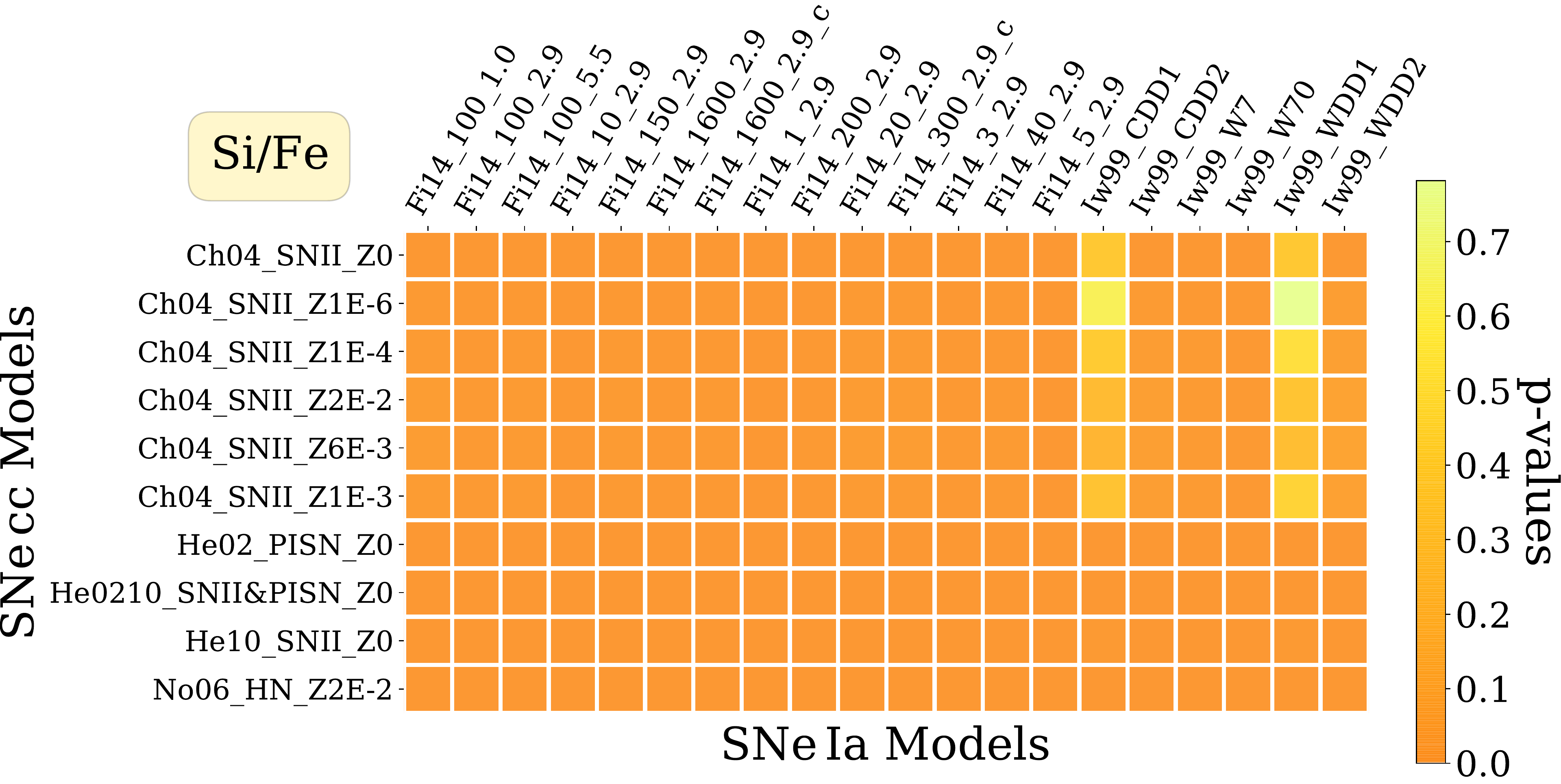}
    \includegraphics[width=0.497\textwidth, clip]{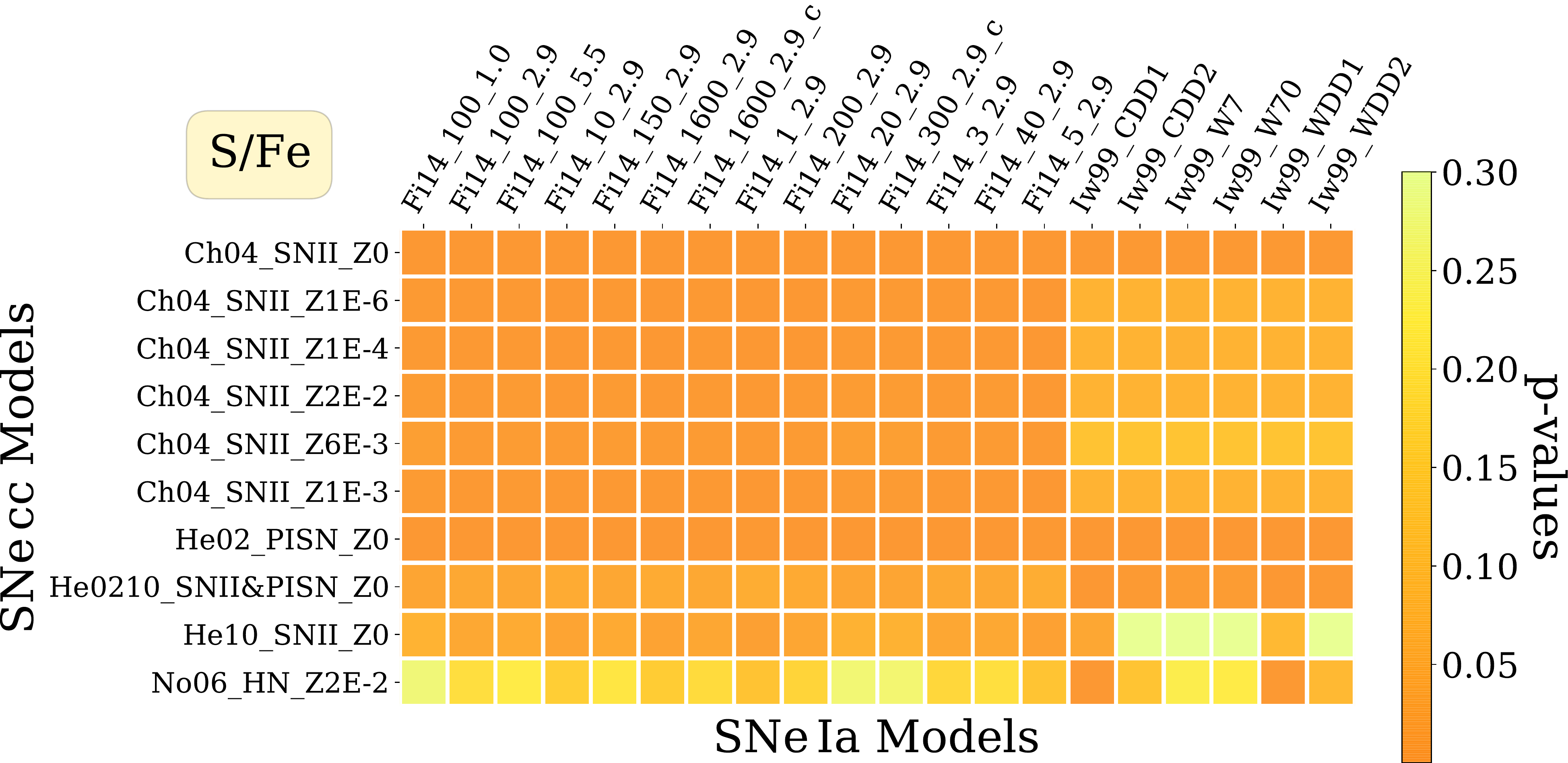}
    \includegraphics[width=0.497\textwidth, clip]{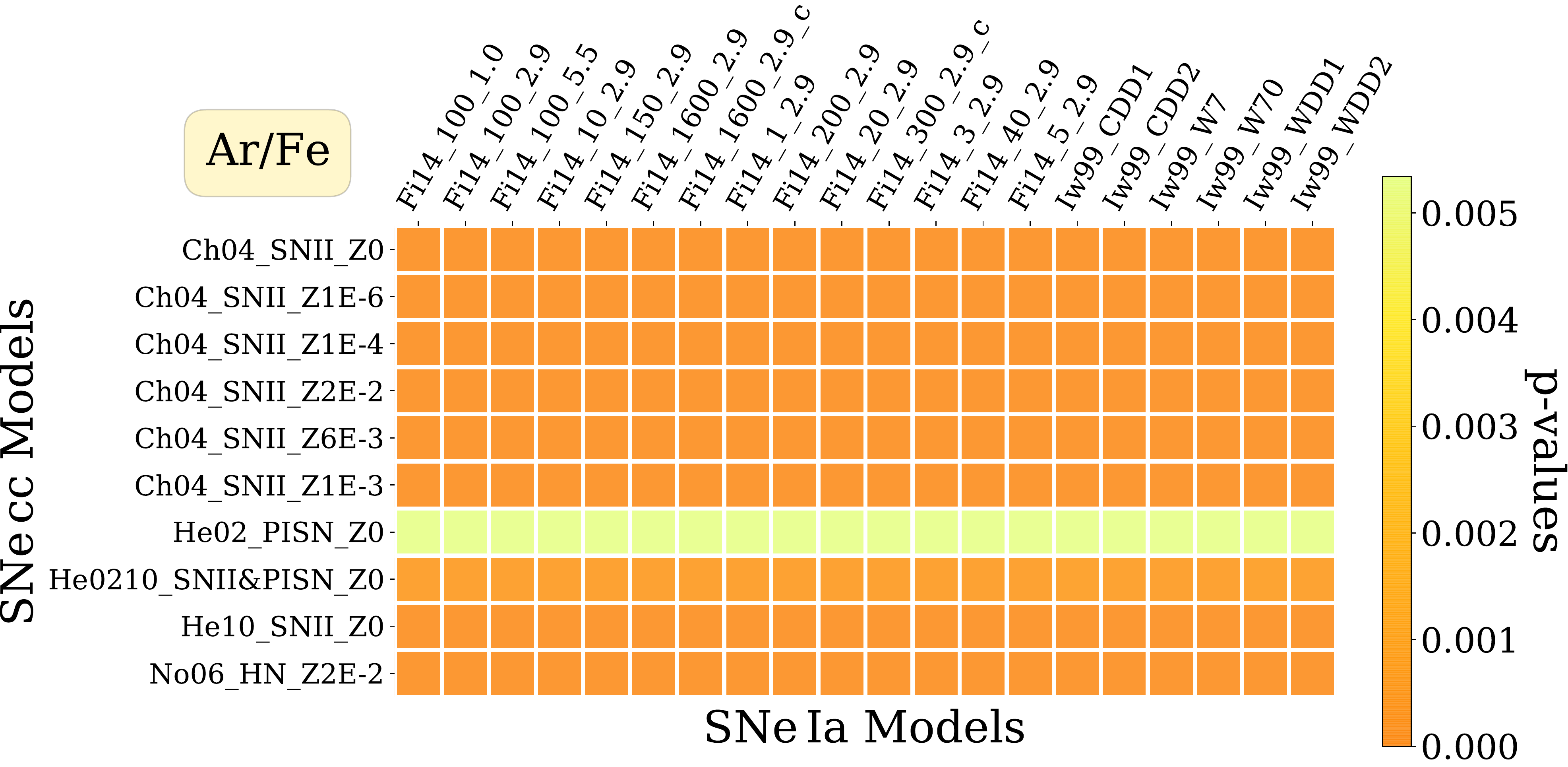}
    \includegraphics[width=0.497\textwidth, clip]{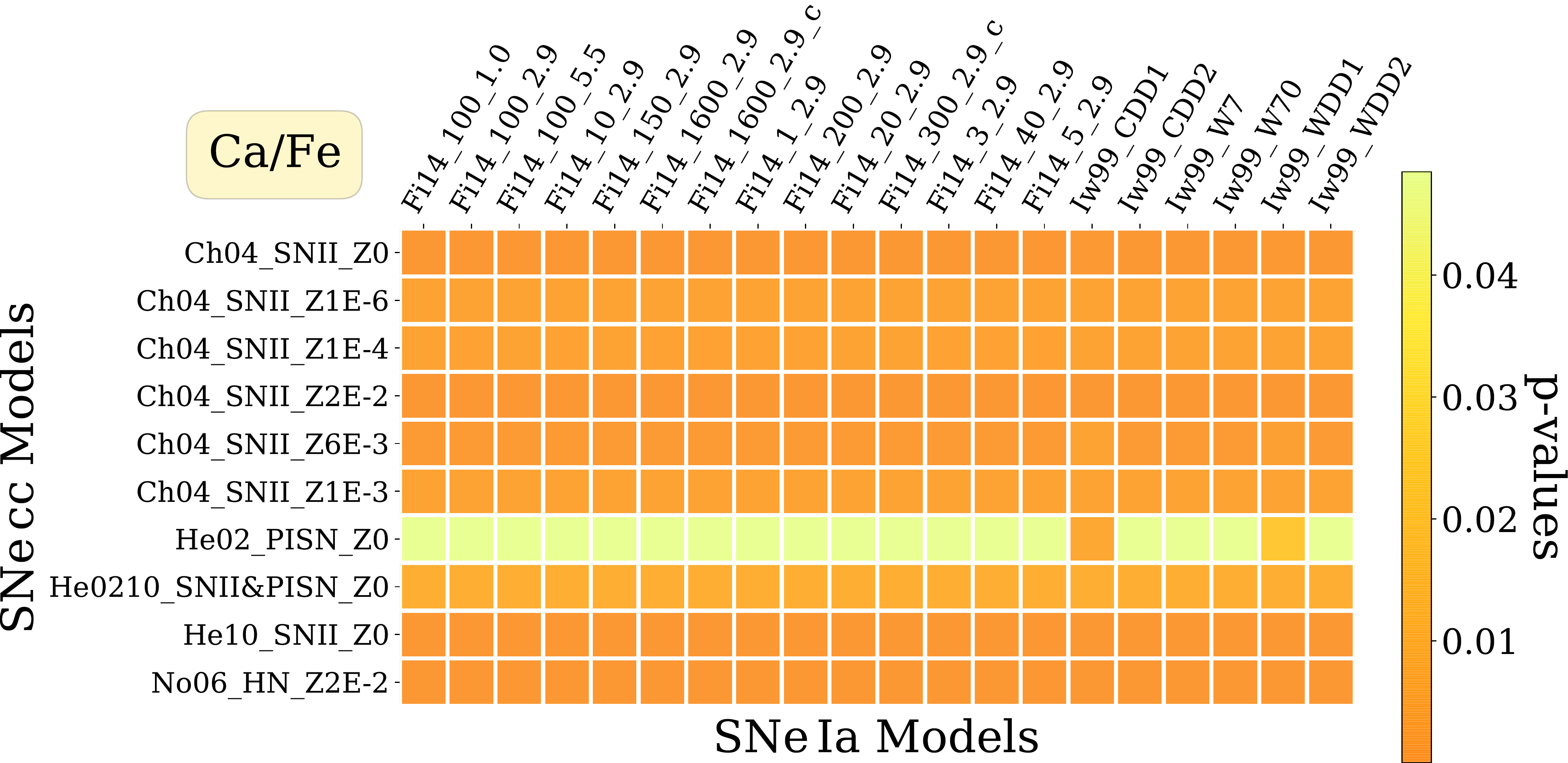}
    \includegraphics[width=0.497\textwidth, clip]{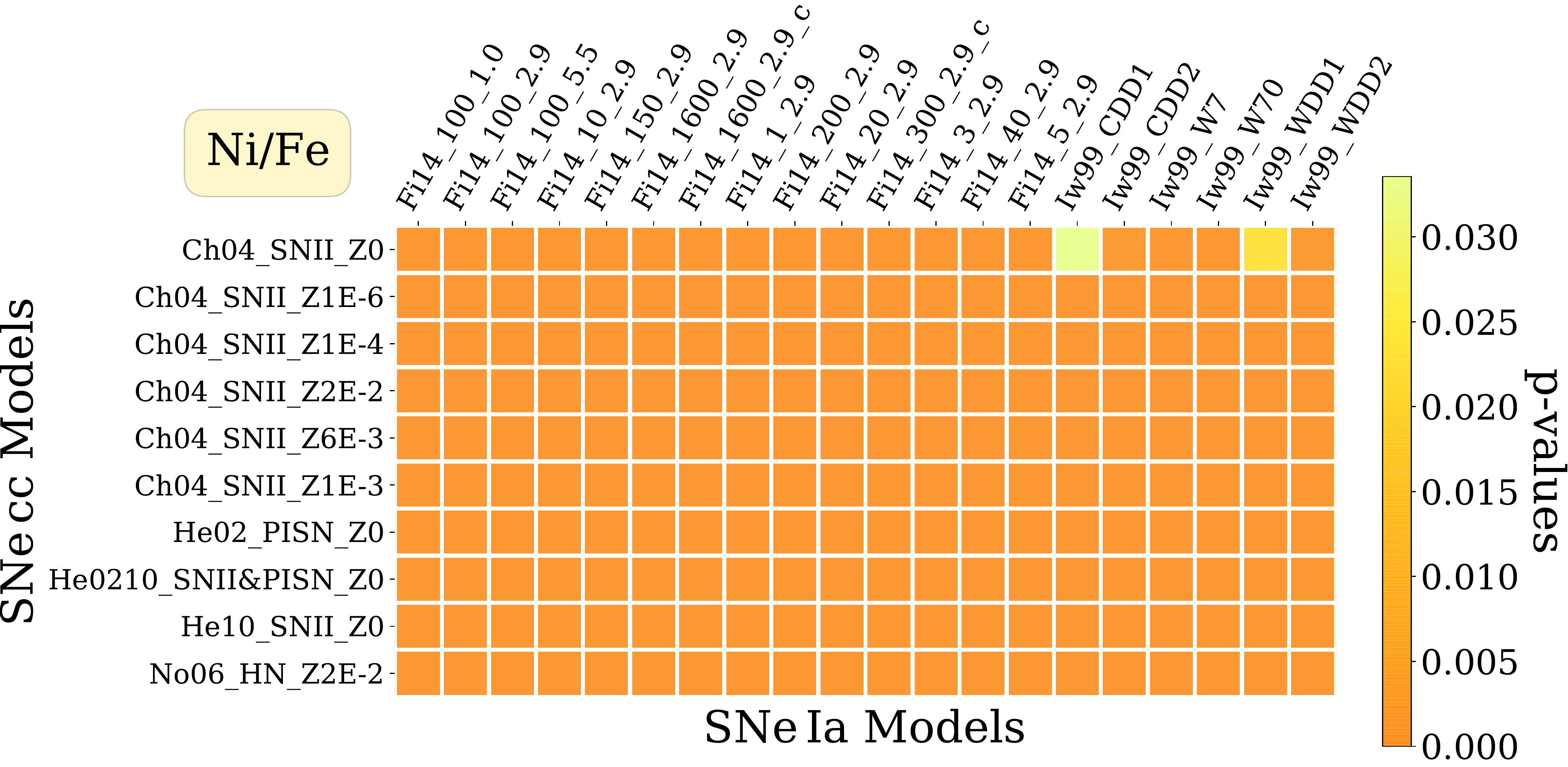}
    \caption{KS test \textit{p}-values of a subset of SN model pairs (SNIa+SNcc) tested in this work using inner/outer regions for each of the eight abundance ratios (see Section \ref{sec:KStest}). The color intensity map represents the values in the color bar, where more yellow cells have greater \textit{p}-values. Part of SN model pairs are shown for illustration.}
    \label{fig:pvalues}
\end{figure*}

\begin{figure}
    \centering
    \includegraphics[width=\columnwidth, angle=0]{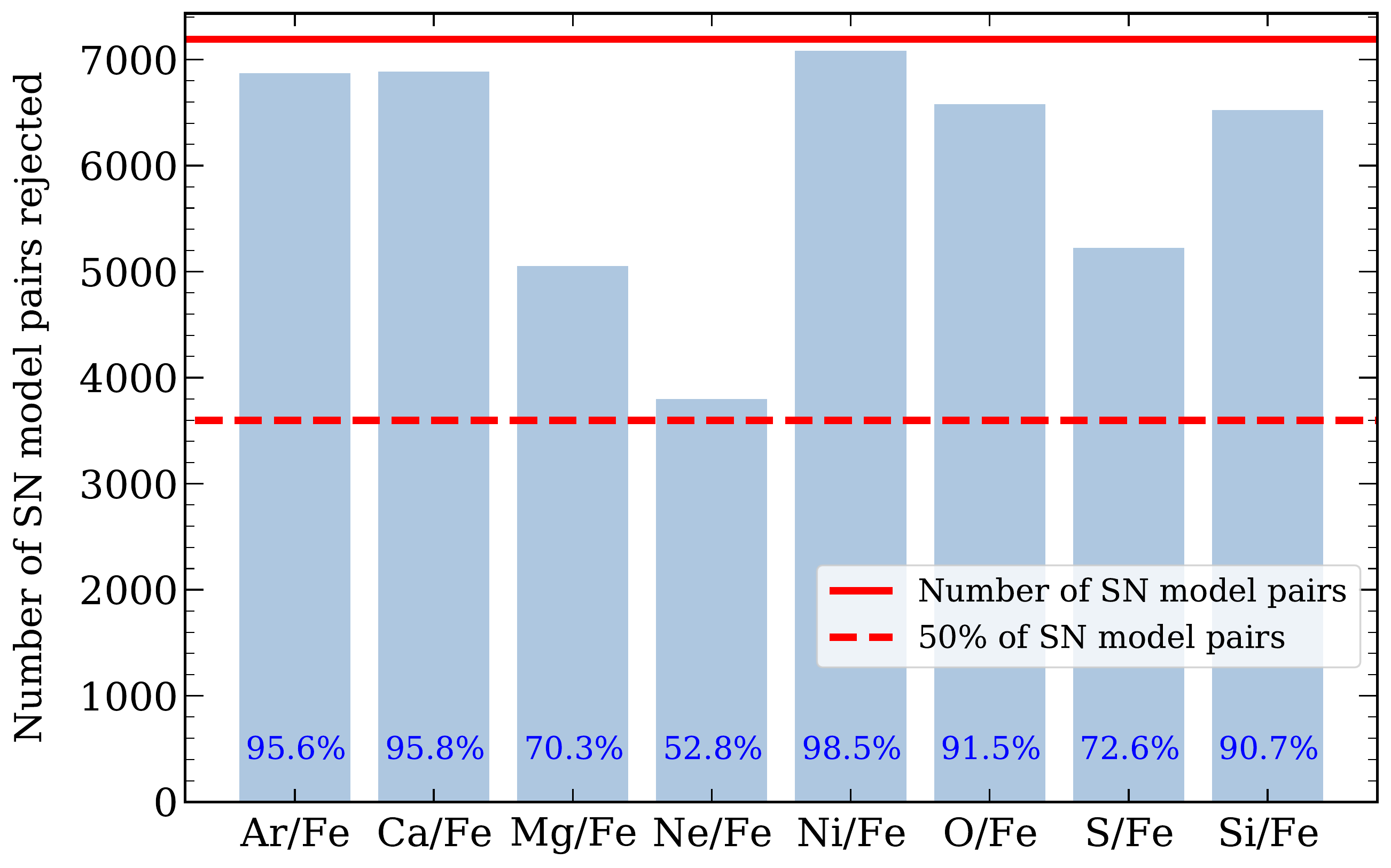}
    \caption{The number (light blue bars) and the fraction (dark blue on the bottom of the bars) of SN model pairs rejected for each abundance ratio at a 95\% significance level. Solid and dashed red lines represent the total number of SN model pairs tested (7192 pairs) and its half (3,596 pairs), respectively.}
   \label{fig:rejection_times_per_ratio}
\end{figure}

\begin{figure}
    \centering
    \includegraphics[width=\columnwidth, angle=0]{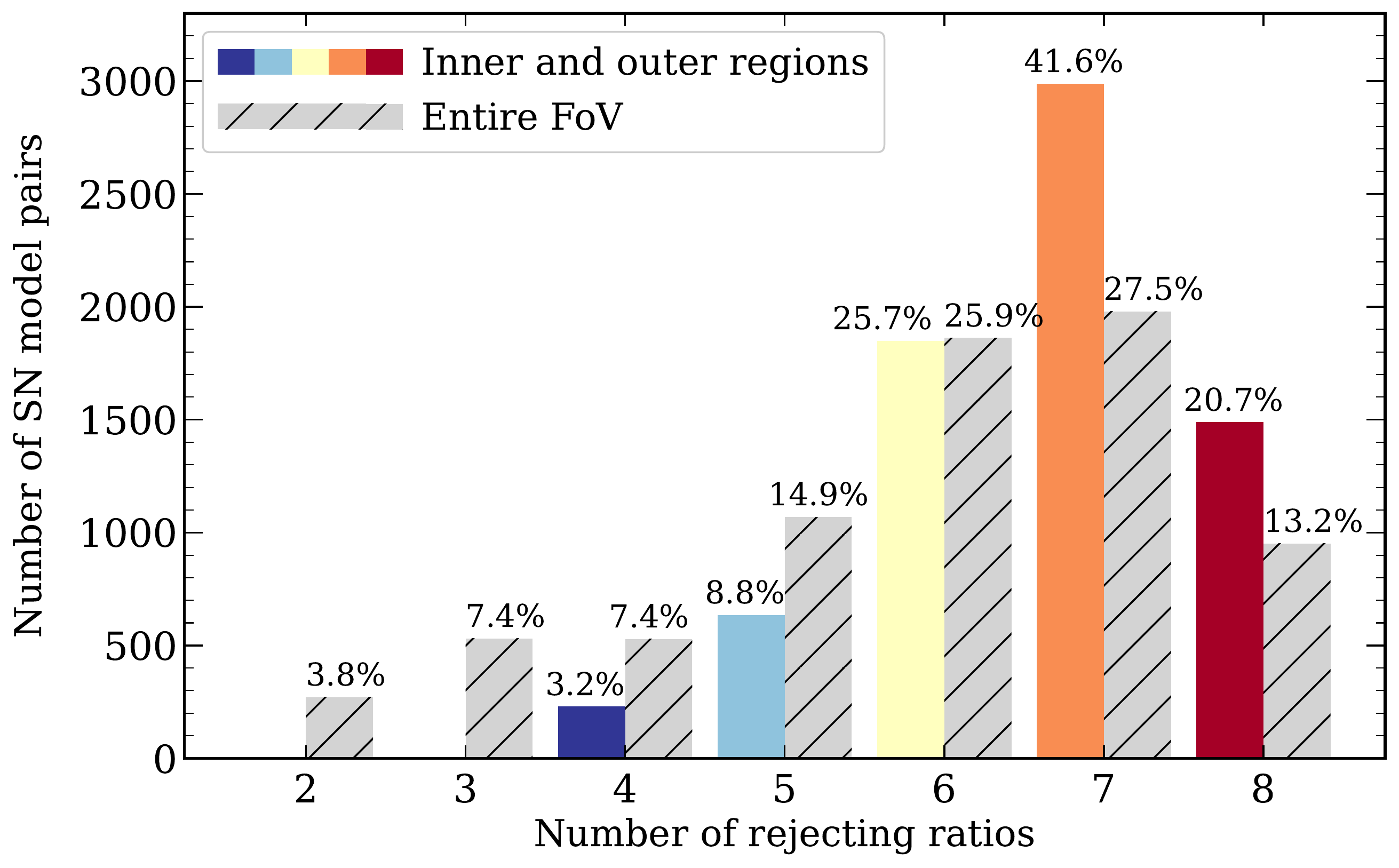}
    \caption{The amount of SN model pairs rejected by the number of rejecting ratios when using the inner/outer regions individually (colored bars) or instead, using the full FoV, denoted by ``entire FoV'' (gray hatched bars). The fraction of SN model pairs rejected per number of rejecting ratios (out of the 7192 SN model pairs tested) is also indicated on the top of the bars.}
  \label{fig:hist_rejection_pair_per_rejected_times}
\end{figure}

\begin{figure}
    \centering
    \includegraphics[width=1\columnwidth, angle=0]{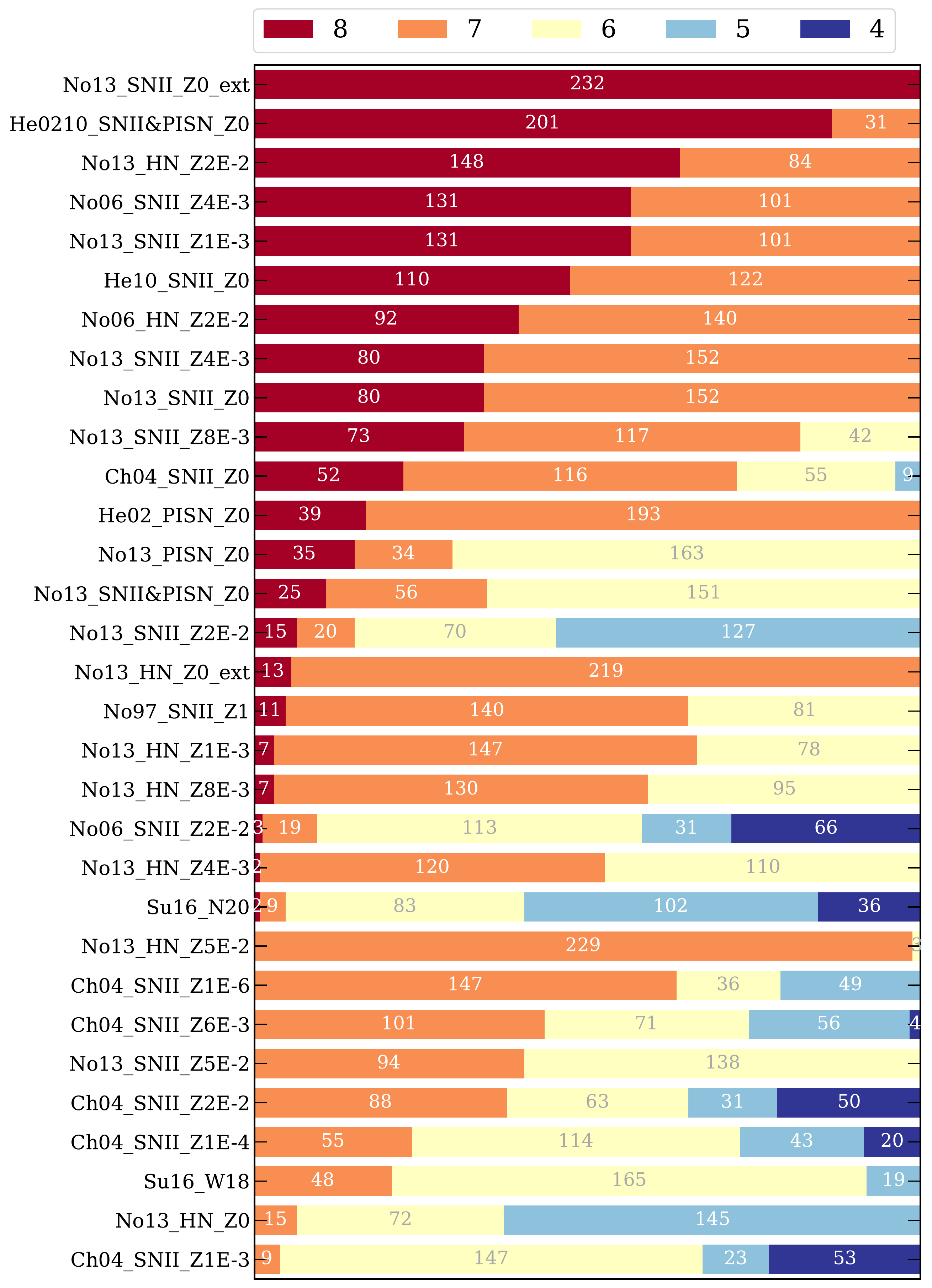}
    \caption{The number of SNIa models rejected when combined with each of the SNcc models. The number of times a SN model pair can be rejected is 8 (red), 7 (orange), 6 (yellow), 5 (light blue), and 4 (dark blue) times, as indicated by the numbers in the stacked bars. The redder the bar, the greater is the number of the most rejected pairs of SN models (i.e., the highest the rate of rejection).}
   \label{fig:hist_numberSNIa_sncc}
\end{figure}

\begin{figure}
    \centering
    \includegraphics[width=0.80\columnwidth, angle=0]{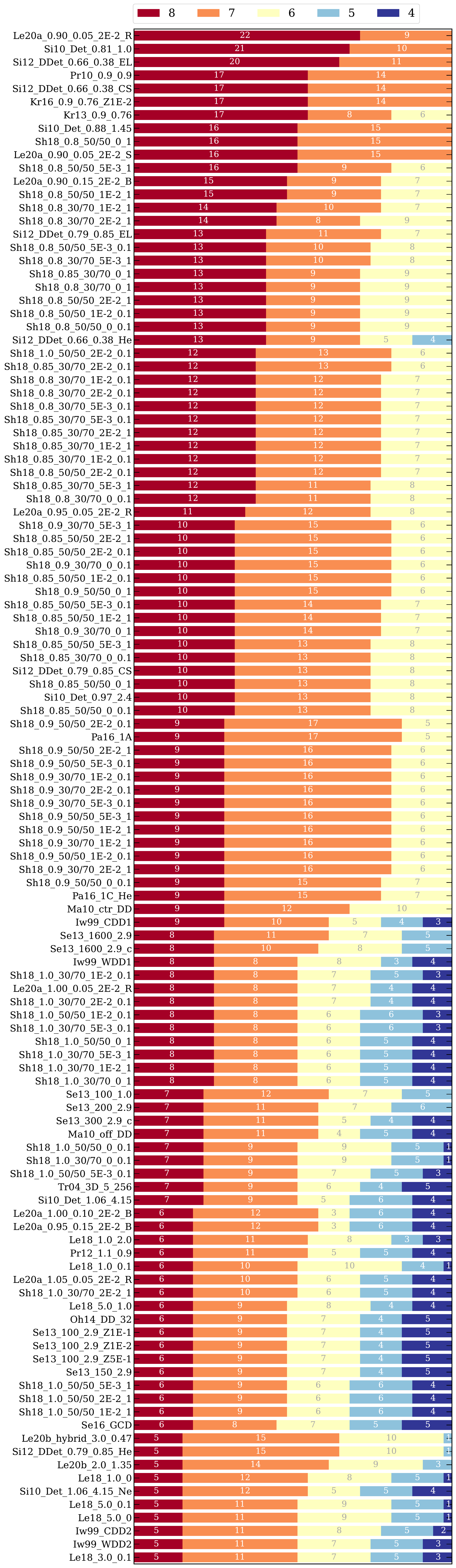}
    \caption{Same as Figure \ref{fig:hist_numberSNIa_sncc}, but for the rejected SNcc models.}
   \label{fig:hist_numberSNcc_SNIa_vertical_withnumbers_part1}
\end{figure}

\begin{figure}
    \centering
    \includegraphics[width=0.80\columnwidth, angle=0]{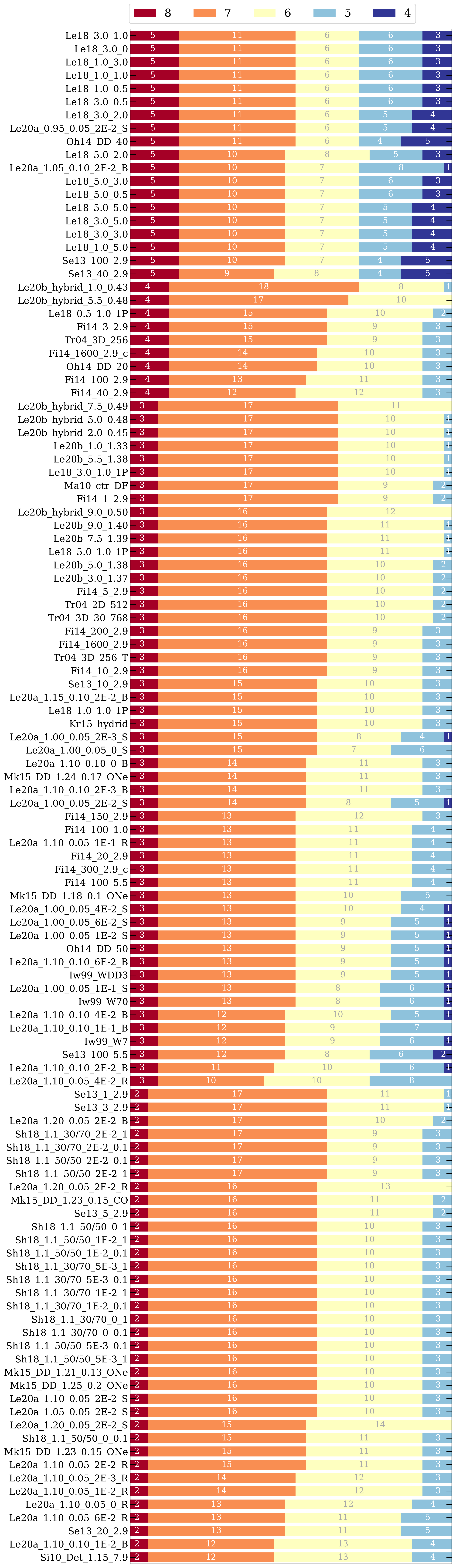}
    \caption{Continued from Figure \ref{fig:hist_numberSNcc_SNIa_vertical_withnumbers_part1}}
   \label{fig:hist_numberSNcc_SNIa_vertical_withnumbers_part2}
\end{figure}

We assess the (dis)similarity of two sets of abundance ratio distributions: the ones predicted by a particular pair of SN models to those observed in the ICM/IGrM. Figure \ref{fig:pvalues} shows the compilation of the \textit{p}-values for each abundance ratio for some pair of SN models. In these figures, we only show a snippet of the entire figure for illustration here. The color bar represents the \textit{p}-value range, increasing from yellow to orange. The color bars throughout this paper indicate the number of rejecting ratios.

We found a significant number of rejected SN model pairs by each abundance ratio individually; every abundance ratio rejected at least 50\% of the total pairs (see Figure \ref{fig:rejection_times_per_ratio}). In principle, one could choose a single abundance ratio as the ``best'' one and carry out the comparison with the SN model pairs, but this is prone to inaccuracies, given the different systematics of different detectors, such as the capacity of de-blending specific lines and different frequency range covered. Therefore, any robust comparative evaluation of SN model pairs should use as many abundance ratios as possible. Here, we used eight abundance ratios to perform this comparison without attributing reliability weights to any individual one (see Section \ref{sec:KStest}).

It is important to note that the discriminative power of the method increases with the number of rejections as well as with the number of rejecting ratios. This combination is maximized when we treat inner and outer regions separately, as shown in Figure \ref{fig:hist_rejection_pair_per_rejected_times}. Using the abundance ratios measured from the entire FoV instead of the inner and outer regions reduces the amount of rejected SN pairs at a higher number of rejecting ratios. We found that 2,989 SN pairs have been rejected by 7 (out of 8 ratios) using the inner/outer region, while only about 1,979 have been rejected by seven ratios using the entire FoV. 
Since this paper aims to compare and rank different pairs of SN models, the more extensive the observationally sampled range in the SNIa/SNcc contribution to ICM metal mass fraction, the higher discriminative power we have. The fact that separating the inner/outer regions increases the discriminative power of the method 
is consistent with the presence of radial gradients in abundance ratios. Moreover, the fact that no pair was rejected by only one ratio (or none) is consistent with the assumption that all ratios should tend to converge to a single combination of SNIa model and SNcc model mix. If the individual ratios did not converge, they would reject the pairs randomly and one could expect many ratios to reject all SN model pairs.
This is not the case as shown in Figure \ref{fig:rejection_times_per_ratio}, where it can be seen that the rejection number per ratio varies from 52.8\% (Ne/Fe) to 98.5\%(Ni/Fe). The reasons for particular individual ratios to reject pairs may be due not only to the models being inadequate but also to different intrinsic systematics related to abundance measurements, such as the conditions of the organic deposition layer and periods where Spaced-Row Charge Injection was turned on and off among the four detectors\footnote{\url{heasarc.gsfc.nasa.gov/docs/astroe/prop\_tools/suzaku\_td/node10.html\#SECTION001031230000000000000}}, or the ATOMDB used. The latter is especially important for ratios depending on abundances of elements measured within the L-shell region. However, here, we found no convincing indication that would lead us to systematically exclude data from any of the detectors, time periods or individual abundance ratios. In the next section, we discuss the results for the most and the least rejected \textit{pairs} of SN models.

\begin{figure*}
    \centering
    \includegraphics[width=1\linewidth, angle=0]{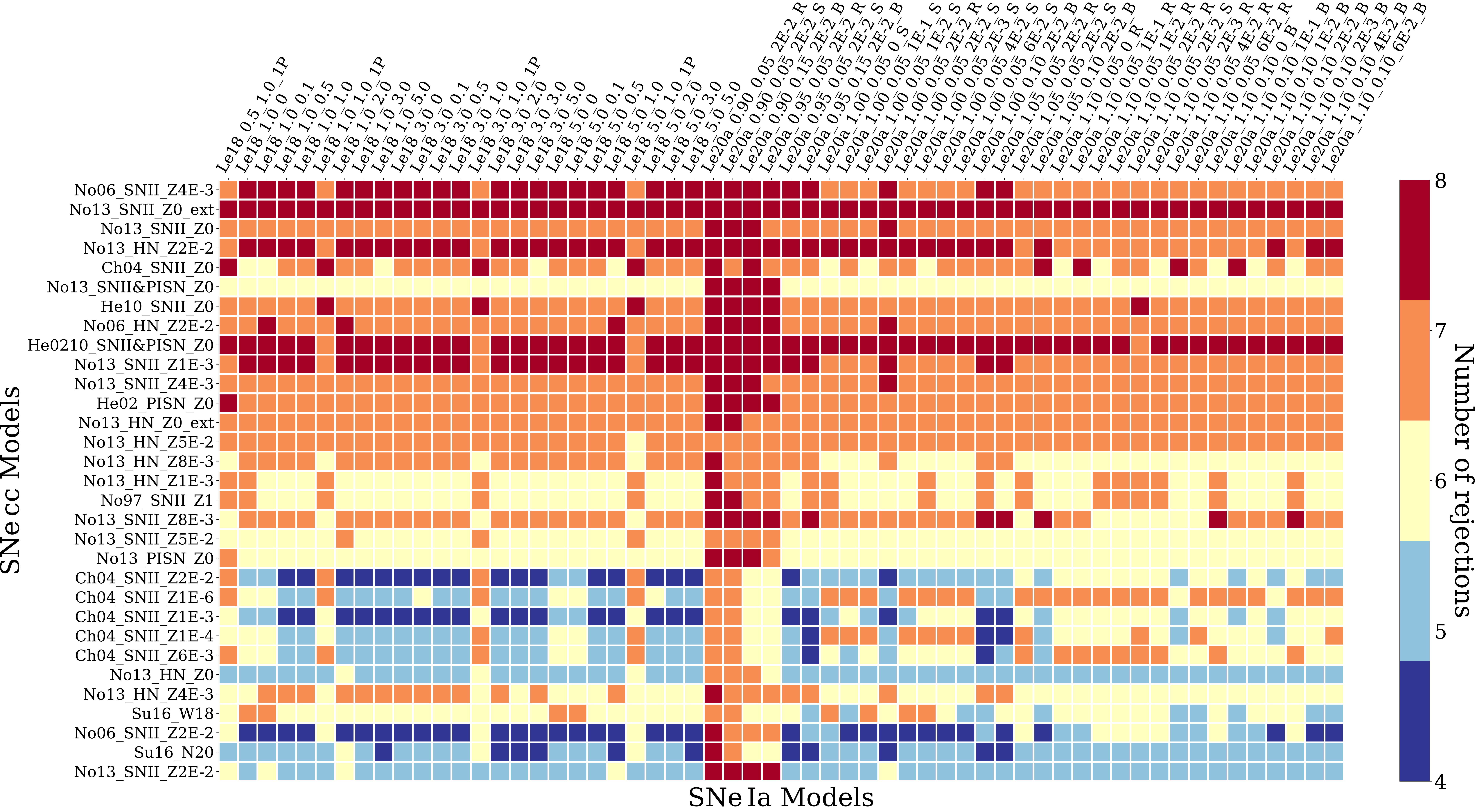}\caption{The number of times each SN model pair has been rejected at the 95\% significance level via KS test using inner/outer regions. A pair of SN models can be rejected at a maximum of 8 times due to the 8 abundance ratios considered in this work (see Figure \ref{fig:pvalues}). The red and dark blue squares represent the most and least rejected pairs of SN models found in this work. The remaining results can be found in the Appendix \ref{sec:rejecting_times} in Figures \ref{fig:rejected_times_2of4}, \ref{fig:rejected_times_3of4} and \ref{fig:rejected_times_4of4}.}
    \label{fig:rejected_times_1of4}
\end{figure*}

\subsubsection{The most rejected SN model pairs} \label{sec:most_rejected}

The most rejected pairs of SN models have complete incompatibility with all abundance ratio distributions. They correspond to about $21\%$ of the total pairs of SN models tested (i.e., 1,489 pairs) and, we indicate them in red color throughout this paper (e.g., Figure \ref{fig:hist_numberSNIa_sncc}). In the following, we highlight some general common trends noticed. We show the main results in Figures \ref{fig:hist_numberSNIa_sncc}--\ref{fig:hist_numberSNcc_SNIa_vertical_withnumbers_part2}.

Figure \ref{fig:hist_numberSNIa_sncc} indicates the number of SNIa models that are rejected when combined with a particular SNcc model; the color bar represents the number of rejecting ratios that the corresponding SN pair is rejected at a 95\% significance level. The No13\_SNII\_Z0\_ext (SNcc) model paired with each one of the 232 SNIa models has the maximum rejection number that our test can obtain.
It represents a Type II Supernova model with zero initial metallicity progenitor stars ($Z_{\rm init}=0$), where the average yield spans from $11$ to $140$ $\text{M}_\odot$. Even within the wide range of SNIa models tested, all SN pairs that include the No13\_SNII\_Z0\_ext model provide a poor correspondence of the observed abundance ratio pattern of the ICM/IGrM analyzed here. We note that in this case, the only stellar yields available above 40M$_\odot$ are 100M$_\odot$ and 140M$_\odot$. As a result of this limitation, the final yield integration (Eq. \eqref{eq:IMF_integration}) may be affected by the IMF weighting for the broad-spaced mass binning applied within the 40-140M$_\odot$ interval mass \citep{Mernier2017}. Therefore, one could compare it with the No13\_SNII\_Z0 model instead, in which the progenitor's mass upper limit is 40M$_\odot$. Despite the slight performance improvement for the No13\_SNII\_Z0 SNcc model, where the number of the most rejected combinations drops from 232 to 80, the 152 remaining pairs are still notably incompatible for all ratios except Ca/Fe. Even though these family models' performance changes with the initial metallicity, no clear trend could be found.

The performance of SN pairs that include pair-instability SN (PISN) yields is significantly poorer (in the bottom 50\%) than other types of SNcc models, none of them being rejected by less than six abundance ratios. In particular, the theoretical Ni/Fe, Ca/Fe, Si/Fe, and Ne/Fe ratios (and also S/Fe most of the time) of these pairs are entirely incompatible with the observed ratio distributions of our sample. For instance, SN pairs including the SNII \& PISN yields of \citep{Heger2002,Heger2010} (He0210\_SNII\&PISN\_Z0) are rejected 201 times for all abundance ratios tested (red bars in Figure \ref{fig:hist_numberSNIa_sncc}) whereas, the remaining 31 pairs are still rejected significantly by 7 ratios, except for S/Fe (orange bars in Figure \ref{fig:hist_numberSNIa_sncc}). This result suggests that PISNe, as characterized by their predicted elemental yields, are less likely to contribute to the chemical enrichment of the ICM as other models, based on the ICM measurements taken from our sample and on the PISN models considered.

Analogously, it may be insightful to look also at the most rejected SNIa models.
Interestingly, no individual SNIa model was rejected by all abundance ratios when paired with each one of all SNcc models (see Figure \ref{fig:hist_numberSNcc_SNIa_vertical_withnumbers_part1}). 
However, one can see from Figures \ref{fig:hist_numberSNcc_SNIa_vertical_withnumbers_part1} and \ref{fig:hist_numberSNcc_SNIa_vertical_withnumbers_part2} that nine SNIa models were rejected by at least 7 ratios for all combinations tested (red and orange bars). 
Two 3D violent C+O WDs merger models (Pr10\_0.9\_0.9 \& Kr16\_0.9\_0.76\_Z1E-2), two 2D double-detonation models both with $M_\text{WD}=0.90$ but with different He detonation configurations (Le20a\_0.90\_0.05\_2E-2\_R \& Le20a\_0.90\_0.05\_2E-2\_S), two pure detonation 1D sub-Chandrasekhar mass C+O WDs models (Si10\_Det\_0.81\_1.0 \& Si10\_Det\_0.88\_1.45), two 2D double-detonation models with different initial detonation configurations (Si12\_DDet\_0.66\_0.38\_EL \& Si12\_DDet\_0.66\_0.38\_CS) and a 1D dynamically-driven double-degenerate double-detonation model with high ${\rm ^{12}C+^{16}O}$ reaction rate (Sh18\_0.8\_50/50\_0\_1). 

By comparing how the variants (parameter values) of the mentioned SN models have performed, we can draw some conclusions about the possible commonalities that could explain the high rejection level. 
Out of the four violent merger models, the one that fared well was Pr12\_1.1\_0.9 \citep{Pakmor2012}, a merger of two WDs of $0.9 M_\odot$ and $1.1 M_\odot$. All other violent merger models where both WD masses were $\le 0.9 M_\odot$ were highly rejected. There was a tendency to be less rejected as WD masses got larger: $\geq 0.9 M_{\odot}$ for the double-detonation models of \citet{Leung2020a},  $\geq 1.06 M_\odot$ for the pure detonation models of \citet{Sim2010} and $\geq 0.79 M_{\odot}$ for all cases of the double-detonation models of \citet{Sim2012}. Although a similar trend can be seen in the double-degenerate double-detonation models of \citet{Shen2018} for low masses, the 1 M$_{\odot}$ models outperformed the 1.1 M$_{\odot}$ ones.
 
The complete rejecting maps of all pairs of SN models tested in this work are presented in Figure \ref{fig:rejected_times_1of4} and in the Appendix \ref{sec:rejecting_times} in Figures \ref{fig:rejected_times_2of4}, \ref{fig:rejected_times_3of4} and \ref{fig:rejected_times_4of4}.

\subsubsection{The least rejected SN model pairs} \label{subsec:least_rejected}

\begin{figure*}
    \centering
    \includegraphics[width=1\linewidth, angle=0]{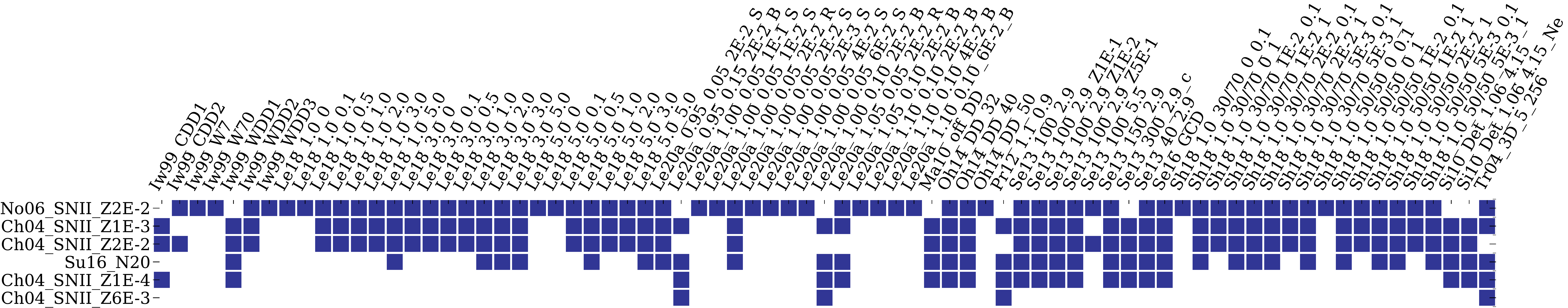}
    \caption{Subset of 229 SN pairs (SNIa+SNcc) ranked as the least rejected pairs of SN models in this work (dark blue). The blank squares refer to the SN pairs rejected by more than four abundance ratios.}
    \label{fig:least_rejected}
\end{figure*}

In this section, we analyze the subset of pairs of SN models rejected only 4 out of 8 times tested, which are the least rejected pairs found (see Figure \ref{fig:least_rejected}). It is worth stressing again that when a pair of models is rejected repeatedly, we can conclude that this pair does not represent the observations satisfactorily. On the other hand, even if a pair of models is not systematically rejected, one can only conclude that the observed data set provides insufficient evidence against the null hypothesis. As a result, one cannot formally state that the pair is the best fit for the observed data. As in the previous section, we can highlight some common trends among the SN models that fared best for ICM enrichment studies. 

Figure \ref{fig:hist_rejection_pair_per_rejected_times} shows that only 229 (out of 7192) SN model pairs are rejected only by 4 abundance  ratios, when paired within the set of SNIa models (see dark blue bars in Figure \ref{fig:hist_numberSNIa_sncc}), corresponding to a small fraction of $\sim3.2\%$ of the total number of pairs tested in this study. Among them, we identified only 6 (out of 31) different SNcc models: the \citet{Nomoto2006} Type II model with $Z_{\rm init}=2$E-2, four Type II models from \citet{Chieffi2004} with 1E-4$\le Z_{\rm init}\le$2E-2 and a nonrotating model calibrated for SN1987A with $Z_{\rm init}=1 Z_\odot$ \citep{Sukhbold2016}.

All models above are SNcc models with nonzero initial metallicity progenitor stars exploding as Type II Supernova, which outperforms PISN and Hypernova models with any initial metallicity and also Type II Supernovae with $Z_{\rm init}=0$. The total number of SNIa models combined with each SNcc model mentioned above with only 4 rejection ratios are 66, 53, 50, 36, 20, and 4, respectively (see Figure  \ref{fig:hist_numberSNIa_sncc}). 
Since our work ranks SN model \textit{pairs} and not individual models, it is interesting to look at the ``flexibility'' of the SNIa models individually to maximize the number of SN pairs that rank best in the least-rejected-pair space. Doing so, we notice that the 3D delayed-detonation models of \citet{Seitenzahl2013} with 40-150 ignition spots for central densities ($\rho_{c,9}=2.9$) and initial metallicities ($Z_{\rm init}\neq0$), the delayed-detonation models of \citet{Ohlmann2014} with 100 near-center ignition spots for carbon depleted core WD (Oh14\_DD\_32/40), and also the gravitationally confined detonation model with one off-centered ignition spot of \citet{Seitenzahl2016} outperform all other matches when paired with the SNcc models listed above. In addition, the Ch04\_SNII\_Z6E-3 model performed equally well when paired with 3D multi-point ignition, five bubbles, and $\rho_{c,9}=2.9$ model of \citet{Travaglio_2004}.

\section{Conclusions} \label{sec:conclusion}

We have determined O/Fe, Ne/Fe, Mg/Fe, Si/Fe, S/Fe, Ar/Fe, Ca/Fe, and Ni/Fe elemental abundance ratios in the inner, outer, and entire FoV regions ($\sim$0.2\,R$_{200}$) of 18 nearby clusters and groups of galaxies ($z\leq0.0391$) with \textit{XIS/Suzaku} observations, which were subsequently used to compare how well the predicted yields of different combinations of SNe yield models (i.e., an SNcc and an SNIa model) recover the observed abundance ratios in the intracluster medium. By implementing the developed non-parametric analysis based on the KS test, we ranked 7192 SNe explosion yields models from the literature using these eight different abundance ratios.

Our methodology reinforces the potential that X-ray spectroscopy of the intracluster gas has to rank among the variety of SNe yield models in the literature. Despite the purely statistical nature of the method, it provides a helpful general guidance to those interested in both the chemical enrichment mechanisms in galaxy clusters and those interested in the nature of supernova progenitor systems and explosion mechanisms. Our main findings can be summarized as follows. 

\begin{enumerate}
    \item We confirm the Fe central abundance enhancement for our sample, even for those systems where the cool-core is not easily resolved, suggesting that the Fe abundance (negative) gradient has a shallower profile than that of the temperature (positive) gradient (see Table \ref{tab:Inner_Outer_Abunds}). 

    \item By building probability distributions for each elemental abundance ratio measurement, we quantified the degree of incompatibility between the prediction by each pair of SN models and the observations. More than half of the total number of SN pairs are rejected by each abundance ratio individually (see Figure \ref{fig:rejection_times_per_ratio}) and $\approx 21\%$ of the total pairs of SN models tested (i.e., 1489 SN pairs) have complete incompatibility with all abundance ratio distributions (see Figure \ref{fig:hist_rejection_pair_per_rejected_times}). 
    
    \item The separate measurements of clusters' inner and outer regions enhance the discriminative power of our technique among the 7192 SN model pairs tested (see Figure \ref{fig:hist_rejection_pair_per_rejected_times}). This is consistent with the premise of abundance ratio gradients in the central regions of cool-core clusters and groups, as raised by different authors previously.
    
    \item Among the 7192 different SN model pairs tested, all 232 SN model combinations that include No13\_SNII\_Z0\_ext are rejected by all abundance ratios (see Figure \ref{fig:hist_numberSNIa_sncc}). This model simulates a Type II Supernova with zero initial metallicity progenitor stars ($Z_{\rm init}=0$) for yield mass range from $11$ to $140 \text{M}_\odot$. On the other hand, no SNIa model paired with all 31 SNcc models, achieves the highest rejection number (see Figures \ref{fig:hist_numberSNcc_SNIa_vertical_withnumbers_part1} and \ref{fig:hist_numberSNcc_SNIa_vertical_withnumbers_part2}). 

    \item Overall, PISN and Hypernova models with any initial metallicity and Type II Supernova models with $Z_{\rm init}=0$ do not predict the observed abundances as well as Type II Supernova with nonzero initial metallicity progenitors.

    \item Narrowing down to the least rejected pairs of SN models, we see that 3D models where the WD progenitor central density of $2.9 \times 10^9\,{\rm gcm^{-3}}$ seem to dominate. Among them, \citet{Ohlmann2014} delayed-detonation models with 100 near-center ignition spots, for carbon depleted core (Oh14\_DD\_32/40), \citet{Seitenzahl2013} delayed-detonation models with 40-150 near-center ignition spots and their one-ignition, and \citet{Seitenzahl2016} gravitationally confined detonation model (Se16\_GCD) are more observationally robust, in the sense of predicting the range of the observed abundance ratios and so is the off-center 5-bubble ignition of \citet{Travaglio_2004} (Tr04\_3D\_5\_256)(see Table \ref{tab:list_SNIa_models}).
    
    On the SNcc side, among the least rejected pairs, we find \citet{Nomoto2006} Type II model with $Z_{\rm init}=2$E-2, four Type II models from \citet{Chieffi2004} with $1{\rm E}-4\le Z_{\rm init}\le 2{\rm E}-2$ and a nonrotating model calibrated for SN1987A with $Z_{\rm init}=1 Z_\odot$ \citep{Sukhbold2016} (see Table \ref{tab:list_SNcc_models} and Section \ref{sec:summary_SNmodels} for main characteristics about SN models tested in this work).

    \item The results favor the scenario where the massive progenitor stars of core-collapse Supernova that contributed significantly to the ICM enrichment must have been initially enriched. The zero metallicity SNcc models ($Z_{\rm init}=0$) do not reasonably reproduce the ICM abundance ratios distribution pattern, independent of the SNIa pairing choice. Also, the results suggest the origin of the majority of the intracluster gas metal enriching SNIa as near-Chandrasekhar mass progenitors with near-center, multiple ignition spots.
    
\end{enumerate}

The PDF non-parametric analysis carried out in this work is a potentially powerful tool to constrain competing SNe explosion models. This technique will become more and more refined as the number of ICM abundance measurements grows, as the spectral resolution improves, and as the number of predicted yields from different types of SN are made available. Hence, the algorithm is suitable for new X-ray measurements and SN models,  as they appear in the literature. The analysis is robust because we are not ``picking and choosing" the preferred ratios, even though the method is open to the implementation of particular sets of weights to individual ratios based on information on the instrument efficiency, background contamination susceptibility, and energy range used, etc. This analysis can be helpful to the cluster and the SNe communities, serving as a guide to improving SN models based on observations and helping to choose the most adequate set of SN models to analyze many critical problems related to metal enrichment in clusters, groups, and even galaxies. In this work, no combination of SN models is fully compatible with all observed abundance ratios, even testing an extensive set of 7192 pairs of SN models. 
The upcoming micro-calorimeters onboard X-ray observatories will achieve more precise measurements of ICM abundances of (rare) elements, allowing better discrimination among SNe models enriching the ICM \citep[e.g.][]{Mernier2020}. The power of XRISM Resolve high-resolution spectroscopy \citep{XRISM2020} will enable accurate measurements of elements such as Cr and Mn in the ICM for multiple nearby systems within the next few years. Incorporating these different abundance ratios in this analysis can better constrain SN models by having accurate measurement abundances of these and other elements and different abundance ratios. By combining low and stable detector background, high effective area, and large FoV, the Advanced X-ray Imaging Satellite (AXIS) camera \citep{Mushotzky2019} will provide measurements of abundance with low particle background and an overwhelmingly better spatial resolution than Suzaku, offering invaluable information to map the enrichment in different cluster regions. Complementary, the X-IFU Athena \citep[][]{Pointecouteau2013} will trace the detailed metal distribution in these systems at high redshifts (z$\sim$2) and allow us to explore the nearby systems at unprecedented distances \citep[][]{Cucchetti2018}. It will be crucial to investigate the role of pre-enrichment in the history of metal enrichment mechanisms. Compiling these achievements with this analysis should provide significantly better constraints on the variety of SN models and serve as a testbed for the different physical ingredients input in the SN simulations.

\section*{Acknowledgements}

We thank the referee for the extremely helpful comments, which allowed us to improve this paper.
We thank Drs. Eric Miller and Jimmy Irwin for the support and helpful discussions. This work uses archived \textit{Suzaku} satellite observations. This study was financed in part by the Coordenação de Aperfeiçoamento de Pessoal de Nível Superior – Brasil (CAPES) – Finance Code 001. R.A.D. acknowledges partial support from NASA grants 80NSSC20P0540 and 80NSSC20P0597 and the CNPq grant 308105/2018-4. Y. J-T has received funding from the
European Union’s Horizon 2020 research and innovation programme under the Marie Skłodowska-Curie grant agreement No 898633. Y. J-T. also acknowledges financial support from the State Agency for Research of the Spanish MCIU through the ‘‘Center of Excellence Severo Ochoa’’ award to the Instituto de Astrofísica de Andalucía (SEV-2017-0709). We thank Dr. Francois Mernier for kindly suggesting the Heidelberg Supernova Model Archive (HESMA), \url{ https://hesma.h-its.org}, which this work made use of.

\startlongtable
\begin{longrotatetable}
\begin{deluxetable*}{lcccccccccc}
\tablecaption{Error-weighted average of all XIS instruments for temperature and chemical abundance from inner, outer, and total (entire) Field of View spectra regions (see Section \ref{sec:data}). \label{tab:Inner_Outer_Abunds}}
\tablehead{\colhead{Name} & \colhead{Temperature}  & \colhead{Ar} & \colhead{Ca} & \colhead{Fe} & \colhead{Mg} & \colhead{Ne} & \colhead{Ni} & \colhead{O} & \colhead{S} & \colhead{Si}\\
\colhead{$\mathrm{}$} & \colhead{$\mathrm{[keV]}$} & \colhead{[$Z_\odot$]} & \colhead{[$Z_\odot$]} & \colhead{[$Z_\odot$]} & \colhead{[$Z_\odot$]} & \colhead{[$Z_\odot$]} & \colhead{[$Z_\odot$]} & \colhead{[$Z_\odot$]} & \colhead{[$Z_\odot$]} & \colhead{[$Z_\odot$]}}
\startdata
\multicolumn{11}{c}{\textsc{Inner Regions}}\\
\hline
NGC 5846 Group    & $0.750\pm0.003$ & $2.6\pm0.5$ & $2.5\pm0.9$ & $0.44\pm0.01$ & $0.75\pm0.03$ & $1.16\pm0.07$ & $0.32\pm0.09$ & $0.34\pm0.06$ & $0.92\pm0.08$ & $0.96\pm0.04$ \\
NGC 4472 Group    & $0.873\pm0.005$ & $0.7\pm0.4$ & $0.3\pm0.2$ & $0.42\pm0.02$ & $0.81\pm0.05$ & $1.3\pm0.1$ & $1.6\pm0.2$ & $0.35\pm0.06$ & $0.85\pm0.09$ & $0.86\pm0.05$ \\
HCG62 Group Group Group             & $0.893\pm0.005$ & $0.5\pm0.3$ & $0.4\pm0.2$ & $0.32\pm0.01$ & $0.60\pm0.04$ & $0.6\pm0.1$ & $2.0\pm0.2$ & $0.07\pm0.03$ & $0.80\pm0.07$ & $0.72\pm0.04$ \\
Ophiuchus Cluster & $7.09\pm0.08$ & $1.4\pm0.3$ & $0.7\pm0.3$ & $0.46\pm0.01$ & $0.7\pm0.2$ & $1.4\pm0.3$ & $0.6\pm0.2$ & $0.3\pm0.2$ & $1.1\pm0.1$ & $0.9\pm0.1$ \\
NGC 1550 Group    & $1.272\pm0.004$ & $0.8\pm0.2$ & $1.0\pm0.3$ & $0.35\pm0.01$ & $0.48\pm0.05$ & $0.4\pm0.1$ & $0.7\pm0.1$ & $0.17\pm0.06$ & $0.56\pm0.04$ & $0.55\pm0.03$ \\
Abell 3581        & $1.573\pm0.009$ & $0.9\pm0.1$ & $0.9\pm0.1$ & $0.42\pm0.01$ & $0.57\pm0.03$ & $0.54\pm0.09$ & $1.01\pm0.08$ & $0.64\pm0.05$ & $0.59\pm0.03$ & $0.72\pm0.02$ \\
NGC 507 Group     & $1.213\pm0.007$ & $0.8\pm0.1$ & $1.4\pm0.3$ & $0.33\pm0.02$ & $0.57\pm0.04$ & $0.3\pm0.1$ & $1.7\pm0.1$ & $0.21\pm0.07$ & $0.65\pm0.04$ & $0.75\pm0.02$ \\
Perseus Cluster	  & $3.835\pm0.005$ & $0.88\pm0.04$ & $0.76\pm0.04$ & $0.504\pm0.002$ & $0.79\pm0.02$ & $1.33\pm0.03$ & $0.84\pm0.04$ & $0.47\pm0.03$ & $0.68\pm0.02$ & $0.78\pm0.01$ \\
Abell 496         & $3.17\pm0.02$ & $0.9\pm0.1$ & $0.8\pm0.1$ & $0.521\pm0.008$ & $0.93\pm0.08$ & $1.6\pm0.1$ & $0.7\pm0.1$ & $0.4\pm0.1$ & $0.79\pm0.05$ & $0.92\pm0.04$ \\
Abell 3571	      & $6.53\pm0.05$ & $0.9\pm0.4$ & $0.5\pm0.2$ & $0.35\pm0.01$ & $0.5\pm0.2$ & $0.09\pm0.06$ & $0.4\pm0.2$ & $0.3\pm0.2$ & $0.5\pm0.1$ & $0.4\pm0.1$ \\
Abell 262         & $1.658\pm0.007$ & $0.9\pm0.1$ & $1.0\pm0.2$ & $0.43\pm0.01$ & $0.60\pm0.06$ & $0.5\pm0.1$ & $0.9\pm0.1$ & $0.07\pm0.04$ & $0.77\pm0.05$ & $0.74\pm0.04$ \\
NGC 2300 Group    & $0.78\pm0.01$ & $0.8\pm0.5$ & $0.8\pm0.2$ & $0.20\pm0.02$ & $0.46\pm0.09$ & $0.7\pm0.2$ & $0.2\pm0.1$ & $0.2\pm0.1$ & $0.9\pm0.2$ & $0.60\pm0.08$ \\
Centaurus Cluster & $1.820\pm0.002$ & $1.11\pm0.07$ & $1.37\pm0.08$ & $0.782\pm0.005$ & $0.76\pm0.03$ & $1.54\pm0.06$ & $1.62\pm0.06$ & $0.011\pm0.007$ & $1.16\pm0.03$ & $1.23\pm0.02$ \\
MKW4 Cluster      & $1.66\pm0.02$ & $1.6\pm0.5$ & $1.1\pm0.6$ & $0.86\pm0.08$ & $0.7\pm0.2$ & $1.5\pm0.6$ & $1.9\pm0.5$ & $0.7\pm0.2$ & $1.3\pm0.2$ & $1.6\pm0.2$ \\
NGC 5044 Group    & $0.878\pm0.004$ & $0.6\pm0.2$ & $0.4\pm0.2$ & $0.32\pm0.01$ & $0.75\pm0.03$ & $0.82\pm0.08$ & $2.5\pm0.1$ & $0.20\pm0.04$ & $0.75\pm0.05$ & $0.63\pm0.03$ \\
AWM7 Cluster      & $3.03\pm0.06$ & $1.3\pm0.3$ & $1.5\pm0.3$ & $0.85\pm0.04$ & $1.4\pm0.2$ & $1.2\pm0.3$ & $2.3\pm0.3$ & $0.3\pm0.1$ & $1.0\pm0.1$ & $1.5\pm0.1$ \\
NGC 6338 Group    & $1.87\pm0.04$ & $0.4\pm0.3$ & $0.4\pm0.2$ & $0.55\pm0.07$ & $0.3\pm0.2$ & $1.9\pm0.7$ & $0.4\pm0.2$ & $0.3\pm0.2$ & $1.2\pm0.3$ & $0.9\pm0.2$ \\
UGC 3957 Group    & $2.50\pm0.04$ & $1.6\pm0.3$ & $1.2\pm0.3$ & $0.63\pm0.03$ & $1.3\pm0.2$ & $0.7\pm0.2$ & $1.4\pm0.4$ & $0.5\pm0.2$ & $0.9\pm0.1$ & $1.1\pm0.1$ \\
\hline
\multicolumn{11}{c}{\textsc{Outer Regions}}\\
\hline
NGC 5846 Group    & $0.95\pm0.01$ &   $0.2\pm0.2$ &   $0.3\pm0.2$ &   $0.13\pm0.02$ & $0.15\pm0.08$ &   $0.5\pm0.2$ &   $1.4\pm0.3$ &   $0.2\pm0.1$ &   $0.7\pm0.2$ & $0.44\pm0.07$ \\
NGC 4472 Group    &  $1.296\pm0.006$ &   $1.4\pm0.3$ &   $0.5\pm0.2$ &   $0.34\pm0.02$ & $0.80\pm0.06$ &   $0.7\pm0.1$ &   $1.7\pm0.2$ &   $0.2\pm0.1$ & $0.97\pm0.07$ & $1.01\pm0.05$ \\
HCG62 Group Group             & $1.12\pm0.01$ &   $0.2\pm0.1$ &   $0.4\pm0.2$ & $0.068\pm0.007$ & $0.26\pm0.06$ & $0.20\pm0.06$ & $0.74\pm0.09$ &   $0.3\pm0.1$ & $0.27\pm0.06$ & $0.29\pm0.03$ \\
Ophiuchus Cluster & $8.25\pm0.03$ &   $1.8\pm0.1$ &   $1.5\pm0.1$ & $0.210\pm0.004$ & $1.53\pm0.09$ &   $0.2\pm0.1$ &   $0.2\pm0.1$ & $0.13\pm0.08$ & $0.63\pm0.06$ & $0.49\pm0.05$ \\
NGC 1550 Group    & $1.344\pm0.008$ &   $0.3\pm0.2$ &   $0.3\pm0.2$ &   $0.22\pm0.01$ & $0.50\pm0.06$ &   $0.5\pm0.1$ &   $0.9\pm0.1$ & $0.06\pm0.05$ & $0.41\pm0.05$ & $0.39\pm0.03$ \\
Abell 3581        & $1.65\pm0.02$ &   $0.2\pm0.1$ &   $0.7\pm0.3$ &   $0.21\pm0.02$ & $0.14\pm0.06$ &   $0.3\pm0.1$ & $0.13\pm0.08$ & $0.13\pm0.07$ & $0.24\pm0.06$ & $0.24\pm0.03$ \\
NGC 507 Group     & $1.327\pm0.008$ &   $0.3\pm0.1$ &   $0.3\pm0.1$ &   $0.20\pm0.01$ & $0.51\pm0.05$ &   $0.6\pm0.1$ &   $1.5\pm0.1$ &   $0.2\pm0.1$ & $0.42\pm0.05$ & $0.47\pm0.02$ \\
Perseus Cluster	  & $5.13\pm0.01$ & $0.58\pm0.07$ & $0.52\pm0.07$ & $0.282\pm0.003$ & $0.55\pm0.04$ & $0.62\pm0.05$ & $0.50\pm0.06$ & $0.24\pm0.05$ & $0.33\pm0.03$ & $0.42\pm0.02$ \\
Abell 496         & $4.00\pm0.03$ &   $0.9\pm0.2$ &   $0.5\pm0.2$ &   $0.23\pm0.01$ &   $1.0\pm0.1$ &   $0.9\pm0.2$ &   $1.3\pm0.2$ & $0.02\pm0.01$ & $0.34\pm0.08$ & $0.47\pm0.07$ \\
Abell 3571	      & $5.20\pm0.05$ &   $1.1\pm0.4$ & $0.08\pm0.06$ &   $0.19\pm0.01$ &   $0.7\pm0.2$ & $0.05\pm0.03$ &   $0.4\pm0.2$ &   $0.8\pm0.3$ &   $0.3\pm0.1$ &   $0.4\pm0.1$ \\
Abell 262         & $2.13\pm0.01$ &   $0.4\pm0.1$ &   $0.8\pm0.2$ &   $0.34\pm0.01$ & $0.80\pm0.08$ &   $0.7\pm0.1$ &   $1.3\pm0.1$ & $0.04\pm0.03$ & $0.46\pm0.05$ & $0.58\pm0.04$ \\
NGC 2300 Group    & $1.05\pm0.06$ &       $1\pm1$ &   $2.5\pm0.6$ &   $0.11\pm0.07$ &   $0.4\pm0.3$ &   $2.0\pm0.9$ &   $1.1\pm0.9$ &   $0.2\pm0.2$ &       $3\pm1$ &   $0.4\pm0.3$ \\
Centaurus Cluster & $3.35\pm0.01$ &   $0.5\pm0.1$ &   $0.6\pm0.1$ & $0.366\pm0.007$ & $0.69\pm0.06$ & $0.81\pm0.08$ &   $0.8\pm0.1$ & $0.02\pm0.01$ & $0.50\pm0.04$ & $0.68\pm0.04$ \\
MKW4 Cluster      &   $1.72\pm0.02$ &   $1.3\pm0.4$ &   $0.4\pm0.2$ &   $0.26\pm0.02$ &   $0.4\pm0.2$ &   $0.3\pm0.2$ &   $1.1\pm0.3$ & $0.14\pm0.09$ &   $0.5\pm0.1$ & $0.57\pm0.09$ \\
NGC 5044 Group    & $1.22\pm0.01$ &   $0.5\pm0.2$ &   $0.4\pm0.2$ &   $0.28\pm0.02$ & $0.53\pm0.05$ &   $0.8\pm0.1$ &   $1.1\pm0.2$ & $0.21\pm0.07$ & $0.47\pm0.06$ & $0.43\pm0.03$ \\
AWM7 Cluster      & $3.44\pm0.02$ &   $0.7\pm0.2$ &   $0.5\pm0.1$ & $0.318\pm0.009$ &   $0.9\pm0.1$ &   $0.4\pm0.1$ &   $1.5\pm0.2$ & $0.03\pm0.02$ & $0.41\pm0.06$ & $0.56\pm0.05$ \\
NGC 6338 Group    & $1.67\pm0.07$ &   $0.4\pm0.2$ &   $0.8\pm0.1$ &   $0.03\pm0.02$ &   $0.3\pm0.2$ &   $0.4\pm0.2$ &   $0.5\pm0.2$ &   $0.3\pm0.2$ &   $0.5\pm0.2$ & $0.54\pm0.09$ \\
UGC 3957 Group    & $2.30\pm0.08$ &   $0.3\pm0.1$ &   $1.3\pm0.6$ &   $0.35\pm0.06$ &   $1.4\pm0.4$ &   $0.5\pm0.2$ &   $3.0\pm0.8$ &   $0.2\pm0.1$ &   $1.1\pm0.3$ &   $1.0\pm0.2$ \\
\hline
\multicolumn{11}{c}{\textsc{Full Field of View Regions}}\\
\hline
NGC 5846 Group    & $0.773\pm0.003$ & $0.5\pm0.3$ & $0.2\pm0.2$ & $0.242\pm0.007$ & $0.47\pm0.02$ & $0.79\pm0.04$ & $1.2\pm0.1$ & $0.04\pm0.03$ & $0.85\pm0.05$ & $0.66\pm0.02$\\
NGC 4472 Group    & $1.107\pm0.002$ & $1.00\pm0.09$ & $0.03\pm0.02$ & $0.325\pm0.004$ & $0.43\pm0.02$ & $0.22\pm0.04$ & $0.48\pm0.03$ & $0.02\pm0.02$ & $0.74\pm0.02$ & $0.66\pm0.01$\\
HCG62 Group Group             & $0.996\pm0.002$ & $0.49\pm0.09$ & $0.5\pm0.2$ & $0.139\pm0.003$ & $0.28\pm0.01$ & $0.29\pm0.03$ & $0.97\pm0.03$ & $0.009\pm0.006$ & $0.46\pm0.02$ & $0.41\pm0.01$\\
Ophiuchus Cluster & $8.62\pm0.02$ & $1.26\pm0.08$ & $0.78\pm0.08$ & $0.317\pm0.003$ & $0.82\pm0.06$ & $0.34\pm0.08$ & $0.05\pm0.03$ & $1.6\pm0.1$ & $0.51\pm0.04$ & $0.31\pm0.03$\\
NGC 1550 Group    & $1.322\pm0.002$ & $0.50\pm0.05$ & $0.6\pm0.1$ & $0.221\pm0.004$ & $0.42\pm0.02$ & $0.48\pm0.04$ & $0.90\pm0.04$ & $0.17\pm0.04$ & $0.44\pm0.02$ & $0.45\pm0.01$\\
Abell 3581        & $1.657\pm0.003$ & $0.68\pm0.06$ & $0.81\pm0.09$ & $0.331\pm0.004$ & $0.35\pm0.02$ & $0.56\pm0.05$ & $0.50\pm0.04$ & $0.50\pm0.04$ & $0.45\pm0.02$ & $0.49\pm0.01$\\
NGC 507 Group     & $1.275\pm0.003$ & $0.63\pm0.08$ & $1.0\pm0.1$ & $0.235\pm0.005$ & $0.47\pm0.02$ & $0.54\pm0.05$ & $1.38\pm0.05$ & $0.2\pm0.1$ & $0.50\pm0.02$ & $0.59\pm0.01$\\
Perseus Cluster	  & $4.297\pm0.004$ & $0.48\pm0.03$ & $0.46\pm0.03$ & $0.403\pm0.001$ & $0.56\pm0.02$ & $1.35\pm0.02$ & $0.72\pm0.03$ & $0.91\pm0.02$ & $0.39\pm0.01$ & $0.543\pm0.009$\\
Abell 496         & $3.76\pm0.01$ & $0.77\pm0.09$ & $0.75\pm0.09$ & $0.381\pm0.005$ & $0.75\pm0.05$ & $1.26\pm0.07$ & $0.7\pm0.1$ & $0.16\pm0.06$ & $0.57\pm0.04$ & $0.63\pm0.03$\\
Abell 3571	      & $6.22\pm0.03$ & $0.8\pm0.2$ & $0.7\pm0.2$ & $0.301\pm0.007$ & $0.17\pm0.08$ & $0.2\pm0.1$ & $0.2\pm0.1$ & $0.9\pm0.1$ & $0.46\pm0.08$ & $0.29\pm0.06$\\
Abell 262         & $2.057\pm0.007$ & $0.56\pm0.07$ & $0.79\pm0.09$ & $0.360\pm0.006$ & $0.56\pm0.04$ & $0.94\pm0.06$ & $0.61\pm0.07$ & $0.05\pm0.03$ & $0.51\pm0.03$ & $0.57\pm0.02$\\
NGC 2300 Group    & $0.94\pm0.02$ & $1.0\pm0.6$ & $0.3\pm0.2$ & $0.15\pm0.01$ & $0.24\pm0.07$ & $0.6\pm0.2$ & $0.1\pm0.1$ & $0.3\pm0.2$ & $0.5\pm0.1$ & $0.40\pm0.07$\\
Centaurus Cluster & $2.814\pm0.006$ & $0.73\pm0.05$ & $0.97\pm0.05$ & $0.668\pm0.004$ & $0.57\pm0.03$ & $2.44\pm0.04$ & $0.69\pm0.05$ & $0.10\pm0.04$ & $0.75\pm0.02$ & $0.90\pm0.02$\\
MKW4 Cluster      & $1.723\pm0.008$ & $0.9\pm0.1$ & $1.1\pm0.2$ & $0.43\pm0.01$ & $0.65\pm0.06$ & $0.7\pm0.1$ & $1.3\pm0.1$ & $0.11\pm0.06$ & $0.61\pm0.04$ & $0.84\pm0.03$\\
NGC 5044 Group    & $0.987\pm0.002$ & $0.53\pm0.09$ & $0.21\pm0.08$ & $0.254\pm0.005$ & $0.58\pm0.02$ & $0.67\pm0.04$ & $1.75\pm0.06$ & $0.12\pm0.02$ & $0.55\pm0.02$ & $0.50\pm0.01$\\
AWM7 Cluster      & $3.50\pm0.01$ & $0.42\pm0.09$ & $0.67\pm0.09$ & $0.435\pm0.006$ & $0.74\pm0.06$ & $0.83\pm0.08$ & $1.0\pm0.1$ & $0.09\pm0.04$ & $0.54\pm0.04$ & $0.66\pm0.03$\\
NGC 6338 Group    & $2.00\pm0.03$ & $0.7\pm0.2$ & $1.3\pm0.3$ & $0.32\pm0.02$ & $0.11\pm0.06$ & $1.0\pm0.2$ & $0.2\pm0.1$ & $0.2\pm0.1$ & $0.48\pm0.07$ & $0.49\pm0.05$\\
UGC 3957 Group    & $2.54\pm0.02$ & $1.1\pm0.2$ & $0.8\pm0.2$ & $0.49\pm0.02$ & $1.0\pm0.1$ & $1.0\pm0.2$ & $1.6\pm0.2$ & $0.3\pm0.1$ & $0.74\pm0.08$ & $0.86\pm0.06$\\
\enddata
\tablecomments{Chemical abundances are in units ANGR solar values.}
\end{deluxetable*}
\end{longrotatetable}

\appendix

\section{Probability distribution functions for abundance ratios used in this work}\label{sec:pdfs}

\begin{figure*}[!h]
    \centering
    \includegraphics[width=0.36\textwidth, clip]{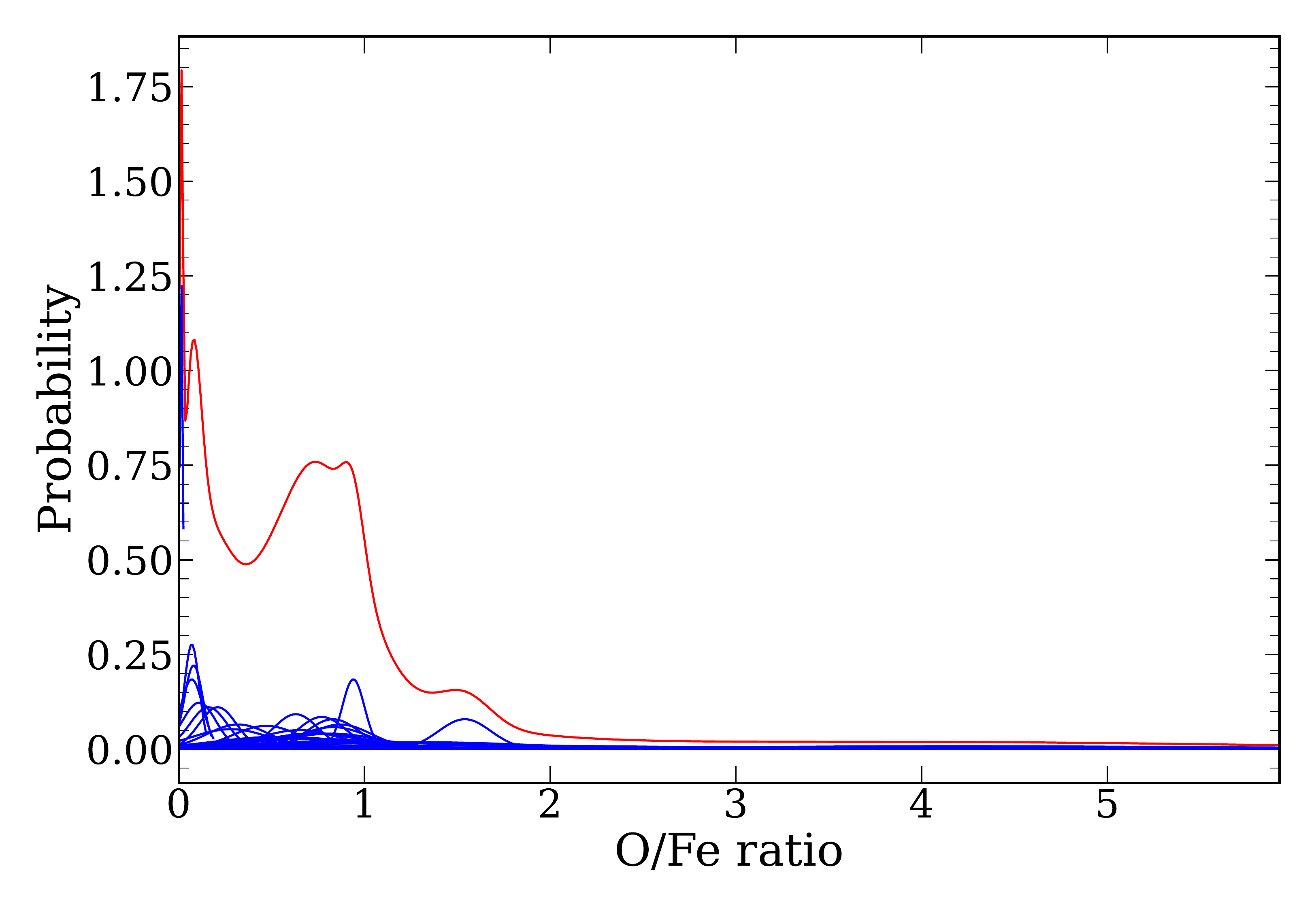}
    \includegraphics[width=0.36\textwidth, clip]{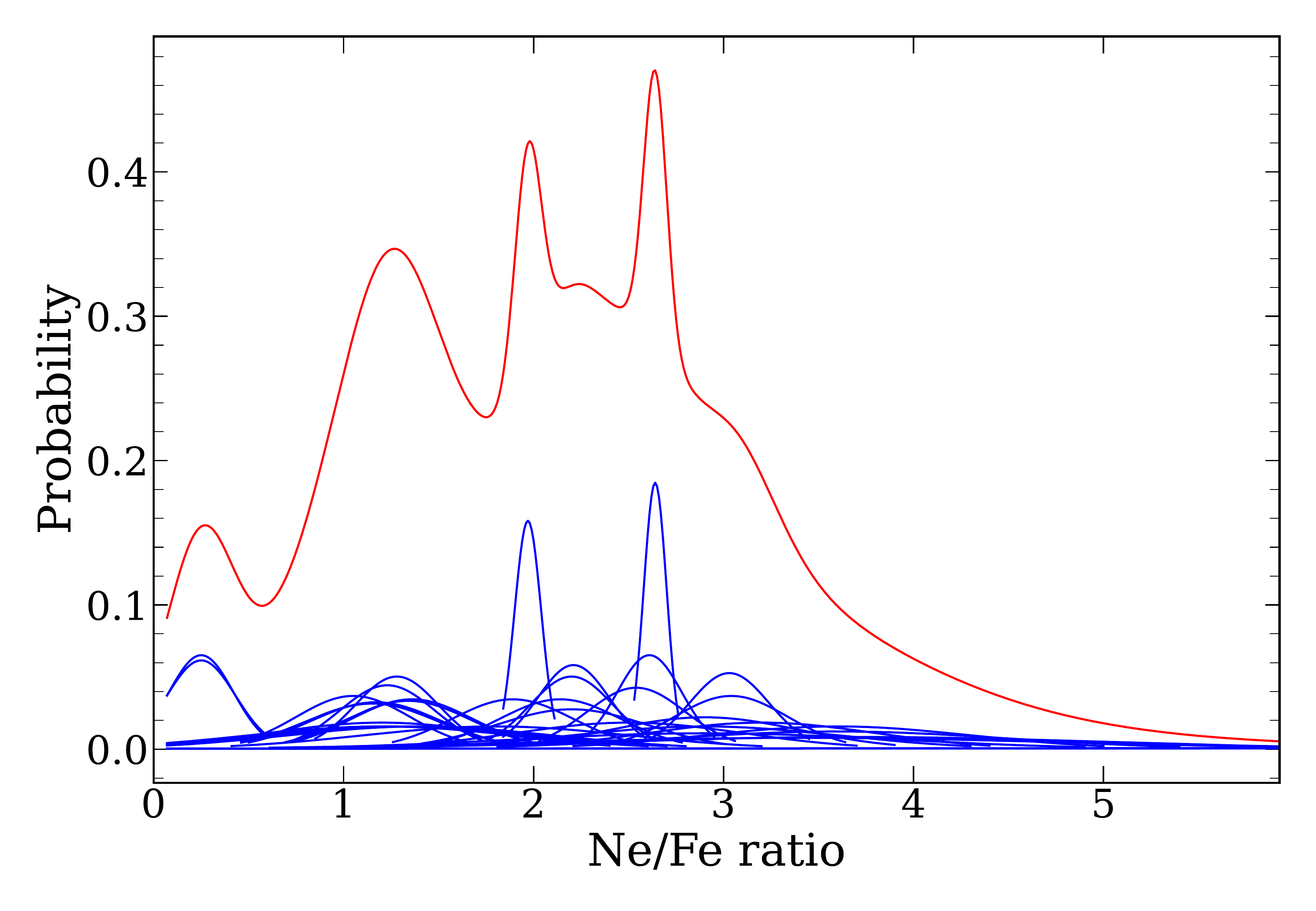}
    \includegraphics[width=0.36\textwidth, clip]{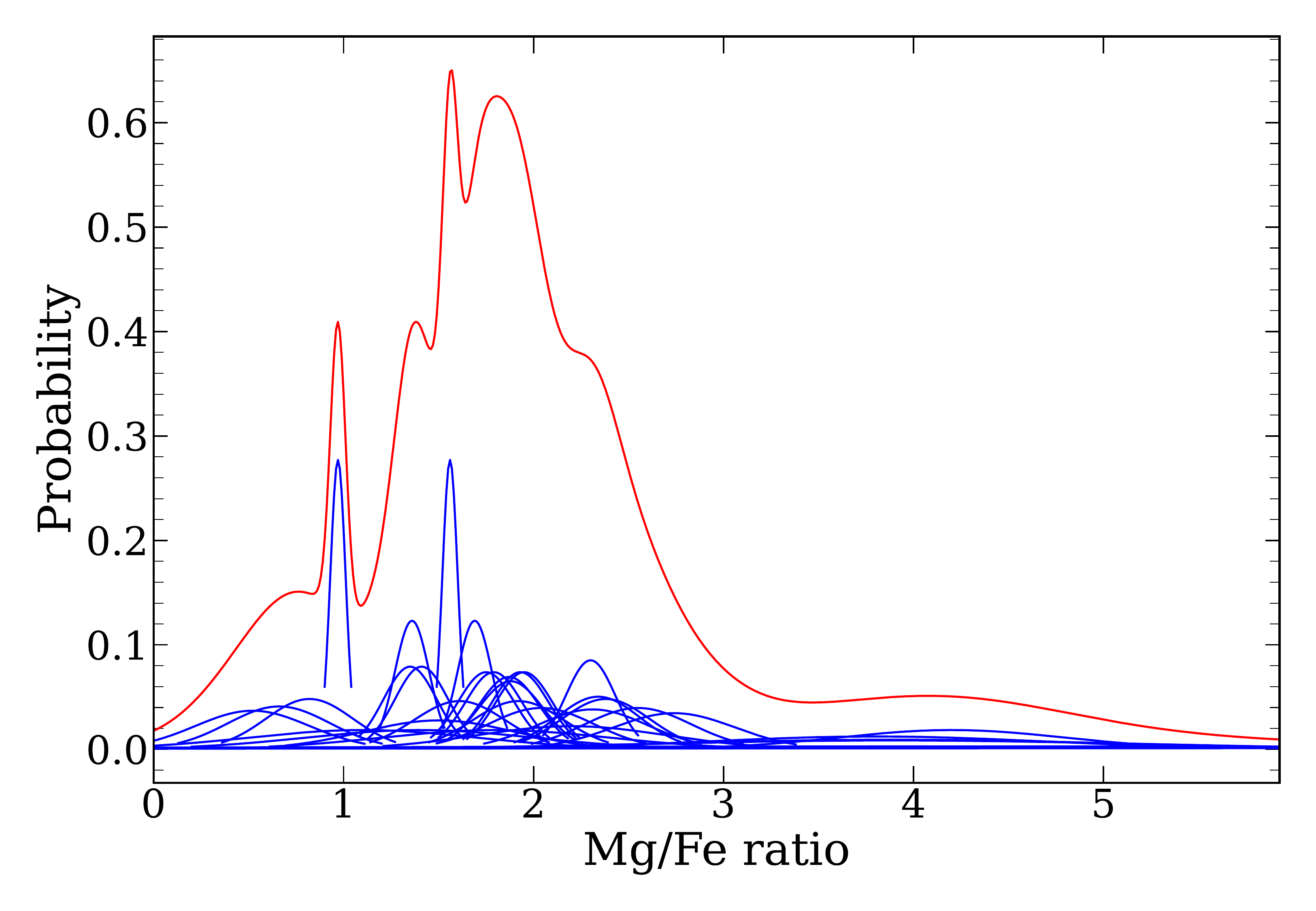}
    \includegraphics[width=0.36\textwidth, clip]{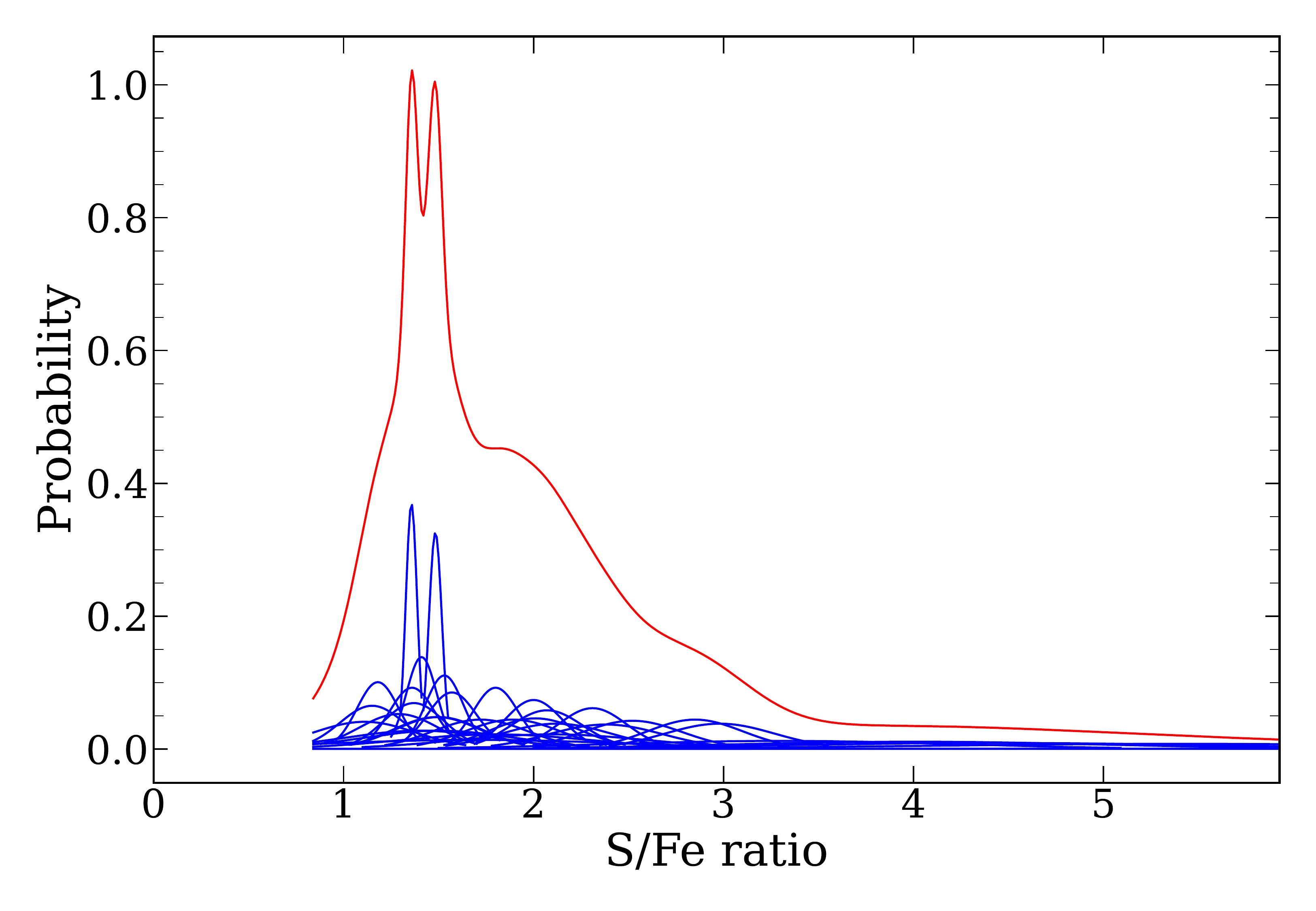}
    \includegraphics[width=0.36\textwidth, clip]{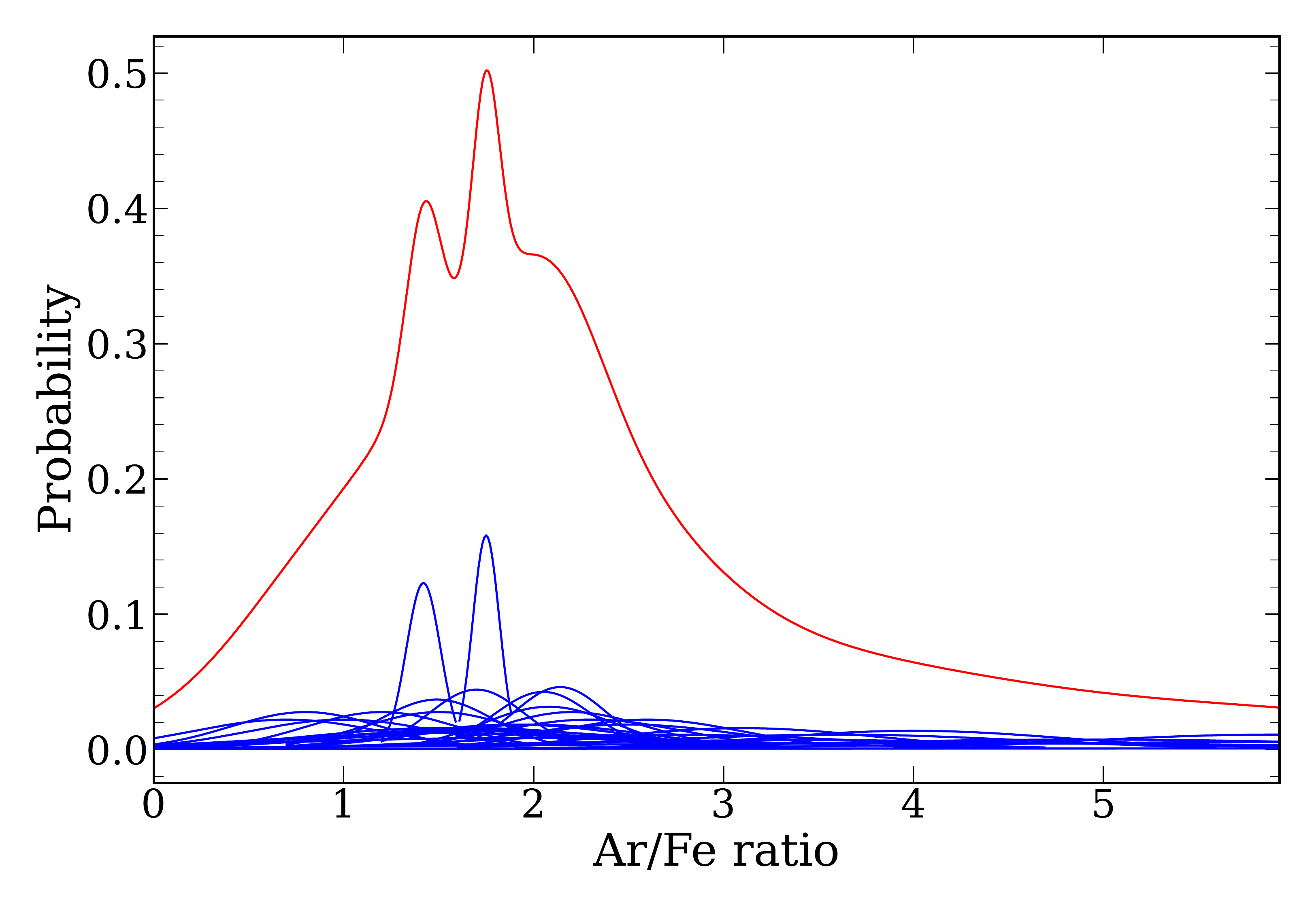}
    \includegraphics[width=0.36\textwidth, clip]{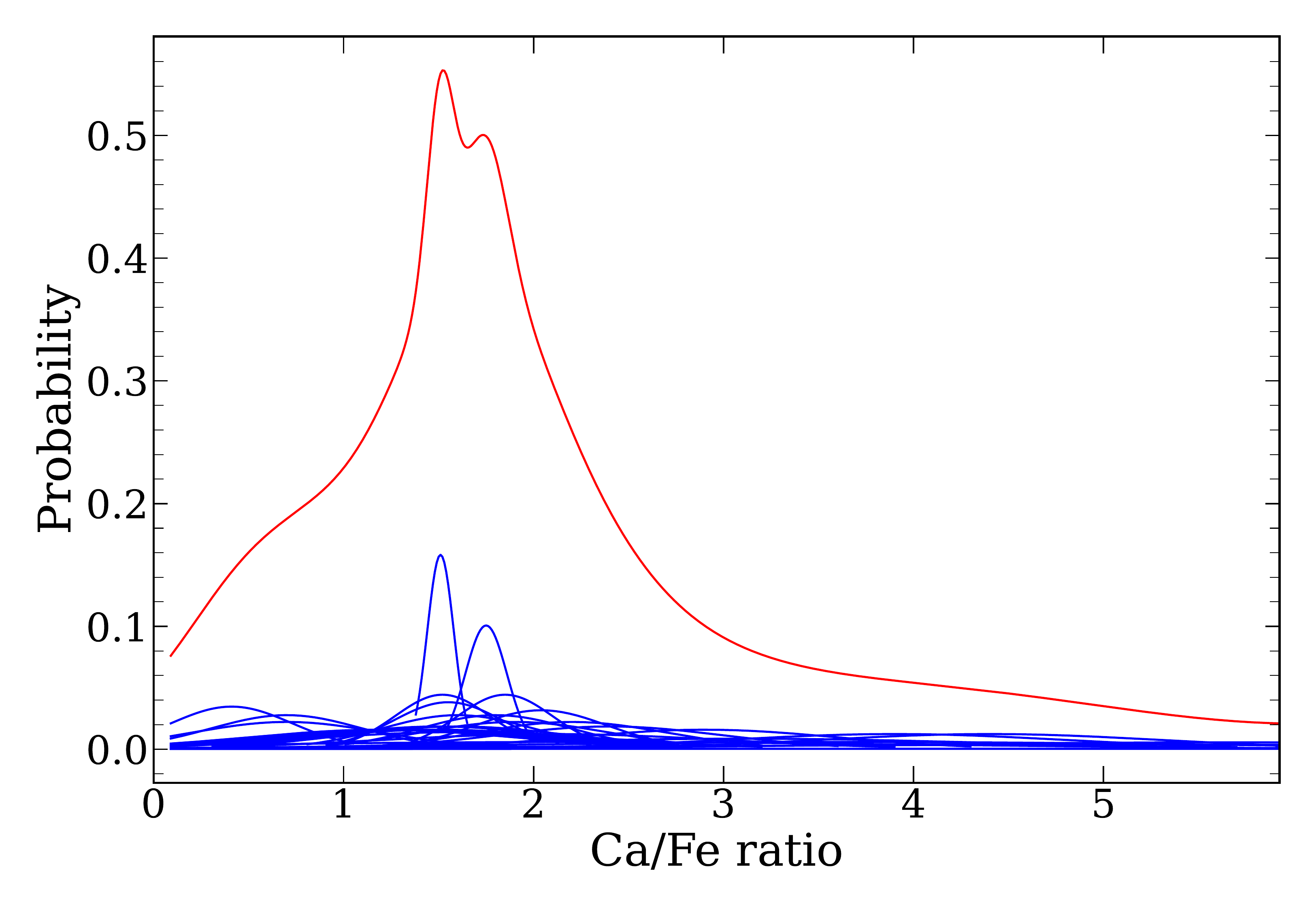}
    \includegraphics[width=0.36\textwidth, clip]{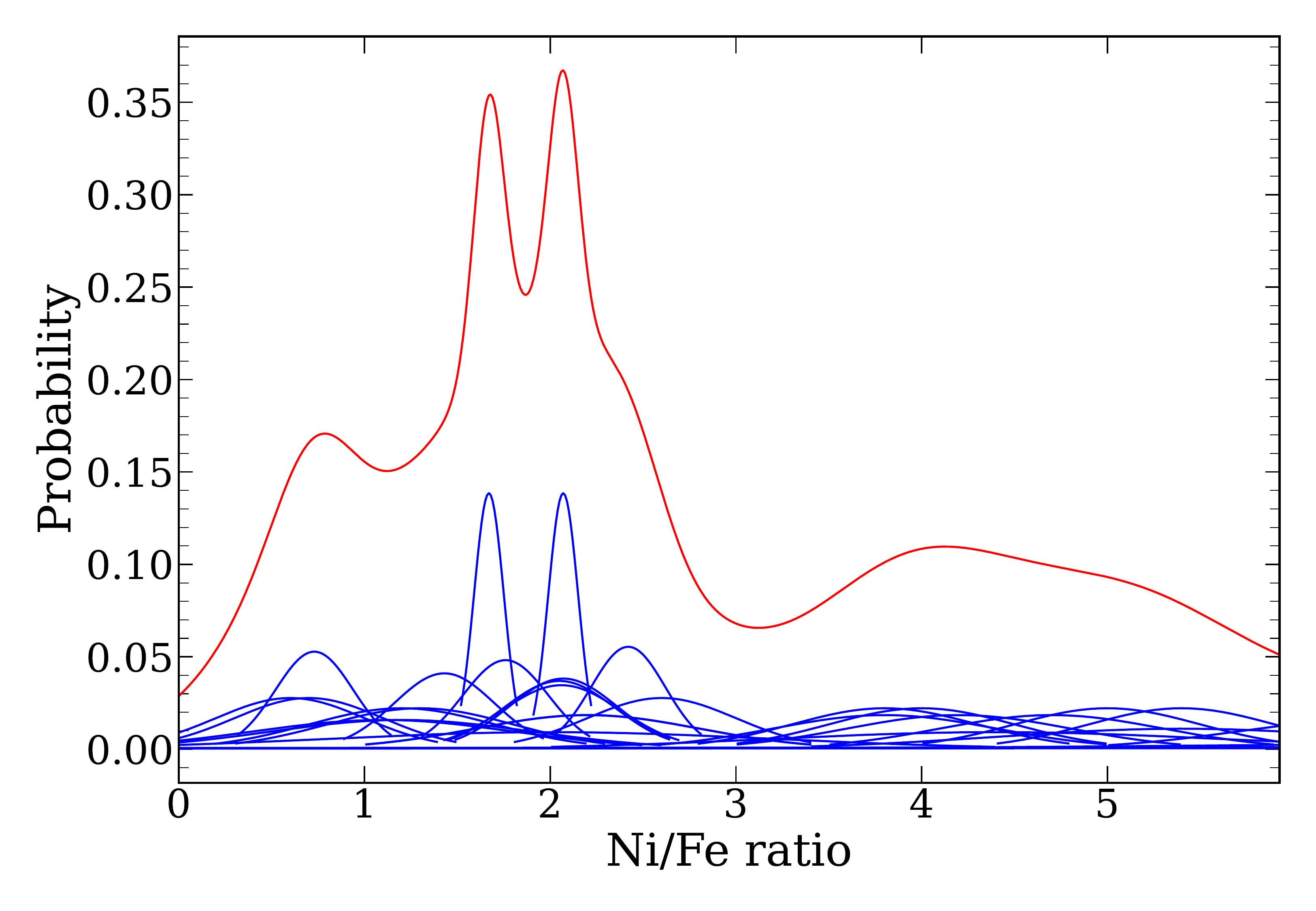}
    \caption{Same as Figure \ref{fig:pdf_Si_Fe} for each X/Fe abundance ratio measurements.}
    \label{fig:pdfs}
\end{figure*}

This section presents the probability distribution functions for other abundance ratios used throughout this work. Figure \ref{fig:pdfs}  illustrates the PDFs from both inner and outer regions (solid blue lines) for each object and the combined PDF distribution (solid red lines).

\clearpage

\section{Comparison of Perseus measurements with previous results}\label{sec:comparison_perseus}

Here we estimate the impact of averaging out the possible idiosyncratic phenomenologies of the ICM near the core on the abundance measurements, due to \text{Suzaku}'s limited spatial resolution. We chose the Perseus cluster as the golden case for comparison because it has been overwhelmingly studied with multiple independent abundance measurements, including the \textit{XMM-Newton} RGS,  \textit{Hitomi} \citep[][hereafter S19]{Simionescu2019} and also \textit{Suzaku} \citep[][hereafter T09]{Tamura2009}.

 T09 have measured several abundance ratios in the central regions of the Perseus cluster with \textit{Suzaku}, using the observation taken in Feb 2006 (that we used in this work) and seven more pointings taken for calibration purposes between Aug. 2006 and Feb 2009. Their results for the two central bins within the central $4\arcmin$ are very similar to each other, which is not surprising given that they are effectively within \textit{Suzaku}'s PSF. Their observed abundance ratios generally agree with what we found for a single absorbed CIE model, except for the Ne/Fe ratio. We found a higher Ne/Fe value even considering the $\sim25\%$ systematic errors suggested by T09 to account for the simultaneous fittings of all instruments and observations in their work. Our measured Ne/Fe ratio becomes compatible with the values obtained by T09 when we use a two-temperature model (with the same spectral fitting setup as that of T09). However, as we mentioned previously, a more complex spectral fitting model is not justified in our work since a detailed mapping of the ICM thermal inhomogeneities is not our goal. In the worst-case scenario, this would increase the span covered by the Ne/Fe ratio values, making that index less discriminative than the others. Indeed, the Ne/Fe ratio has the lowest impact on our analyses since it is the most irrelevant ratio to indicate the discrepancies between the SN model pairs (see Ne/Fe ratio in Figure \ref{fig:rejection_times_per_ratio}). The evaluation of the most rejected pair of models relies not only on Ne/Fe but also on seven other ratios. Moreover, excluding Ne/Fe from the analysis increases the absolute number of the least rejected pairs of SN models (i.e., incompatible with only four ratios -- see section \ref{subsec:least_rejected}) from 229 to 238. It corresponds to only a slight change from $\approx3.2\%$ to $3.3\%$ of the total pairs of SN models tested, respectively.

We also chose to use only the earliest Perseus observation to avoid the systematics due to the contamination layer of the filters, which is stronger in lower frequencies and may artificially influence the measurements of the abundances of O and Ne. The amplitude and scatter of the degradation of the soft-energy effective area due to the organic contamination layer in the optical path of each sensor of \textit{Suzaku} is well documented \citep{Petre2008} also in \textit{The Suzaku Data Reduction Guide} v5.0\footnote{\url{https://heasarc.gsfc.nasa.gov/docs/suzaku/analysis/abc/}}, and affects more strongly the late (post 2006) observations, which were included in both \citet{Tamura2009} and \citet{Simionescu2019}.

The \textit{Hitomi} satellite carried on-board a high-resolution spectrometer, which unfortunately stopped working during the commissioning phase in 2016. However, it partially observed the core of the Perseus cluster with four pointings, three overlapping in the central $\sim2^{\prime}-3^{\prime}$, and an offset pointing about 4$^{\prime}$ SW of the center \citep{Hitomi2018}. S19 performed an analysis of the metallicity of the Perseus core using data from \textit{Hitomi} (above 2\,keV), \textit{Suzaku} (all available pointings to date), and \textit{XMM-Newton},\textit{ including} the RGS for abundances measured below 2\,keV.

When comparing the results from S19 to our analysis of the main \textit{Suzaku} pointing of Perseus, one should keep in mind that, even with the large PSF of \textit{Suzaku} (and Hitomi), S19 so-called \textit{Hitomi}'s ``entire core" region is asymmetrically configured and reaches at most $\sim3^{\prime}$ distance from the center. In area, it is $\sim$ 50\% smaller than our central region and significantly larger (by 4 to 5 times) than that analyzed with the RGS. One notable discrepancy likely due to the differences in extraction regions is the Fe abundance, which is measured to be $0.46\pm0.05$ (RGS/\textit{XMM-Newton}), $0.71\pm0.003$ (this work, but with the same spectral models as that of S19), $0.81\pm0.01$ (\textit{Hitomi}), and $0.821\pm0.001$ (\textit{Suzaku}) in solar units converted from \citet{Lodders2009}, used in their paper, for proper comparison.

The Fe abundance distribution in Perseus is known to have an overall negative radial gradient first noted by \citet{Ulmer1987} and later confirmed by many different works \citep[e.g.][]{Ponman1990, Dupke2001, Simionescu2011, Ueda2013}. Also, it has significant anisotropic inhomogeneities measured with \textit{Chandra} \citep{Sanders2004}. The slight differences between the Fe abundances measured by us with \textit{Suzaku} and with \textit{Hitomi} are not surprising, given the presence of a radial Fe abundance gradient. However, the discrepant lower Fe abundance measured by the RGS is puzzling and possibly related to the degeneracy between the Fe absolute abundance and the normalization of the CIE component due to the RGS limited wavelength coverage (S19).

The best-fit values of Ne/Fe, Mg/Fe, and Ni/Fe ratios obtained by us agree with those from the RGS, while the O/Fe ratio is somewhat smaller. \citet{dePlaa2017} extensively discusses the systematics affecting the O/Fe ratio with the RGS, and we refer the reader to that work. However, given the very small area of analysis of the RGS (basically the central $10\,{\rm kpc}$), it cannot be ruled out that this discrepancy could be caused by a large amount of dips and peaks of the individual elemental abundances in that region near the core \citep{Sanders2004}. In addition, if we fix the value of the Fe abundance to that of \textit{Hitomi}, which is measured in the Fe-K complex and very reliable, the O/Fe ratio measured by RGS becomes entirely consistent with our measured values. Analogously, if we fix the Fe abundance of \textit{Hitomi} as the ``golden" value and use it to compensate for the possible impact of the central Fe abundance gradient in our measurements with \textit{Suzaku}, the abundance ratios of Si, S, and Ni to Fe become entirely consistent, while Ca and Ar to Fe become marginally consistent with those derived by our work. Therefore, we consider the overall consistency with previous measurements satisfactory.

\section{Comparison between abundance ratio measurements from the literature with our measurements}\label{sec:comparison_ratios}

We compare the results of T09 and S19 from the Perseus cluster in detail in Appendix \ref{sec:comparison_perseus}, showing a satisfactory agreement with ours. Even though we cannot compare our measurements directly to previous works as closely as we could with Perseus, we can do a ``general" comparison using a common solar abundance table. We converted our abundance ratio measurements from photospheric solar ANGR to proto-solar \citet[][hereafter L09]{Lodders2009}, used by \citet{Mernier2016,Mernier2018} and S19. Similarly, we also converted the measurements of \citet{dePlaa2007} from proto-solar \citet{Lodders2003} to proto-solar L09. After doing the proper abundance table conversion, we show the comparison in Figure \ref{fig:comparison}, where we use the `entire FoV' (all) region. In Figure \ref{fig:comparison}, indeed, Mg/Fe, S/Fe, Ar/Fe, and Ni/Fe measurements are directly consistent with those from \citet{Mernier2016,Mernier2018} and S19. There are some discrepant ratios, which can be due to several reasons, including the sample choice. For example, the error-weighted average of the Ca/Fe ratio becomes about unity (0.997$\pm$0.03 Z$_\odot$) when ratios of NGC 4472 and 5846 groups are not considered. Overall, this illustrates the need for caution when comparing our results to others, which use different samples, satellites, detectors, extraction regions, atomic database versions, and spectral models. Even though a detailed comparison with those other works could be enlightening, it is beyond the scope of the proposed work. In any case, it is important to note that the proposed methodology does not depend on solar abundance table choice.

\begin{figure}[!h]
    \centering
    \includegraphics[width=0.75\textwidth, angle=0]{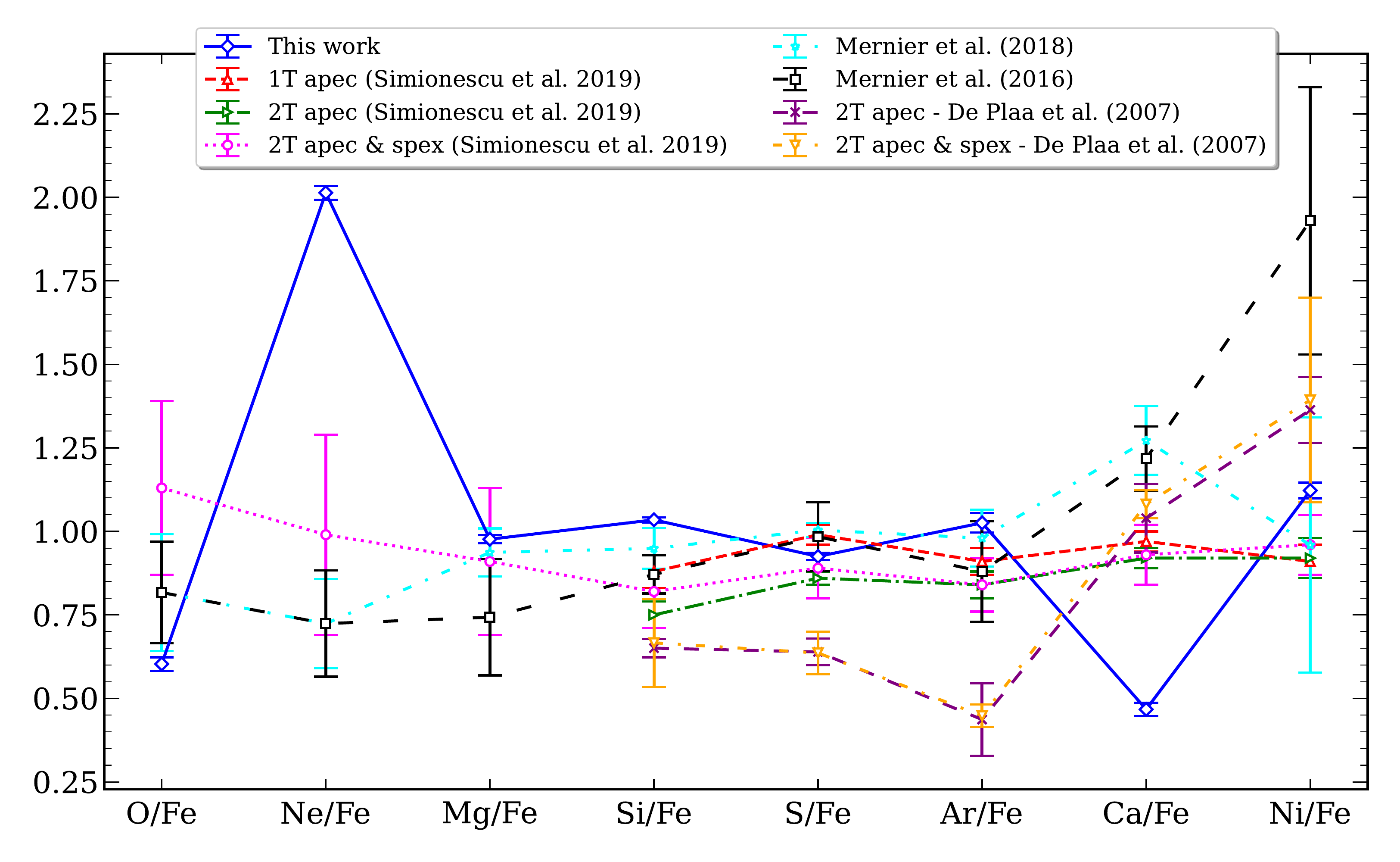}
    \caption{Comparison between abundance ratio measurements from the literature with our measurements from the entire FoV (triangle symbol connected by the solid blue line).}
   \label{fig:comparison}
\end{figure}
\clearpage

\section{The complete evaluation of SN pairs of models in this work.}\label{sec:rejecting_times}

Here, we present the complete result obtained in this work and partially shown in Figure \ref{fig:rejected_times_1of4}. It represents the number of times a pair of SNIa+SNcc models has been rejected at the 95\% level of significance via the KS test using the inner and outer regions. A certain pair of SN models can be rejected at a maximum of 8 times due to the 8 abundance ratios considered in this work (see Figure \ref{fig:pvalues}). The red squares represent the most rejected pairs of SN models in this work. Figures \ref{fig:rejected_times_2of4}, \ref{fig:rejected_times_3of4} and \ref{fig:rejected_times_4of4} together with Figure \ref{fig:rejected_times_1of4} are the fully rejected maps of the 7192 SN pairs tested by the eight abundance ratios.

\begin{figure*}[!h]
    \centering
    \includegraphics[width=1\linewidth, angle=0]{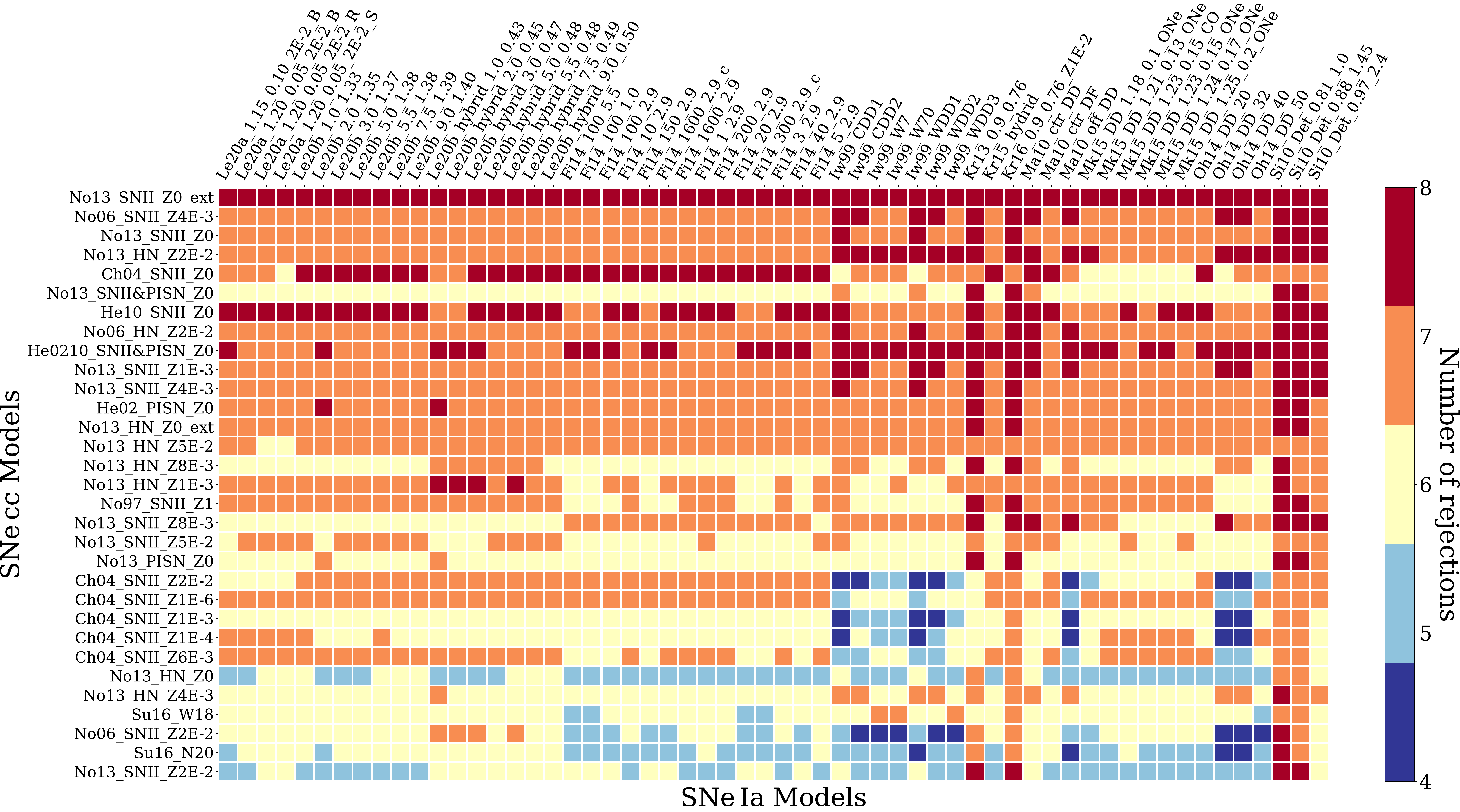}\caption{Continuation of Figure \ref{fig:rejected_times_1of4}.}
   \label{fig:rejected_times_2of4}
\end{figure*}

\begin{figure*}
    \centering
    \includegraphics[width=1\linewidth, angle=0]{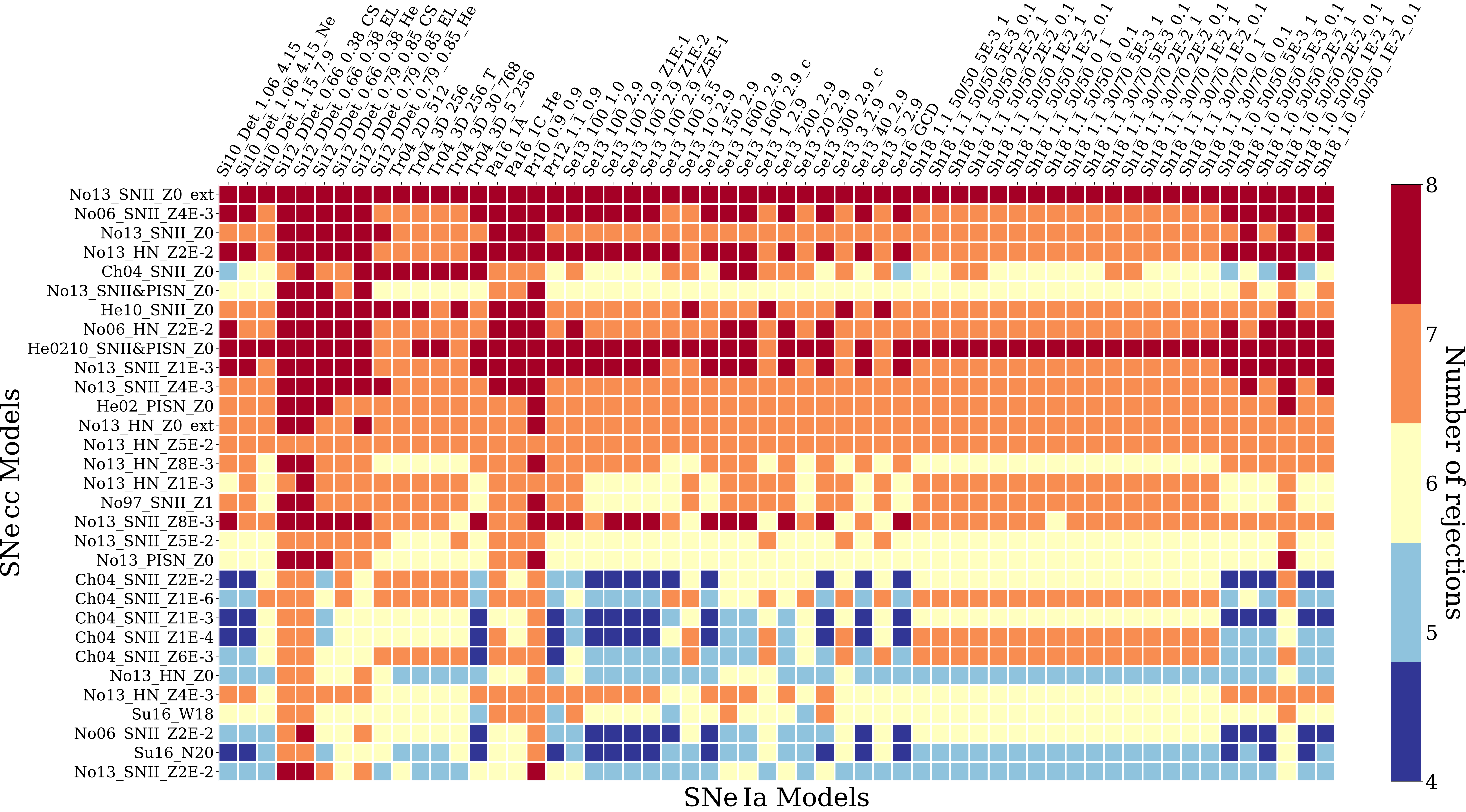}\caption{Continuation of Figure \ref{fig:rejected_times_1of4}.}
   \label{fig:rejected_times_3of4}
\end{figure*}

\begin{figure*}
    \centering
    \includegraphics[width=1\linewidth, angle=0]{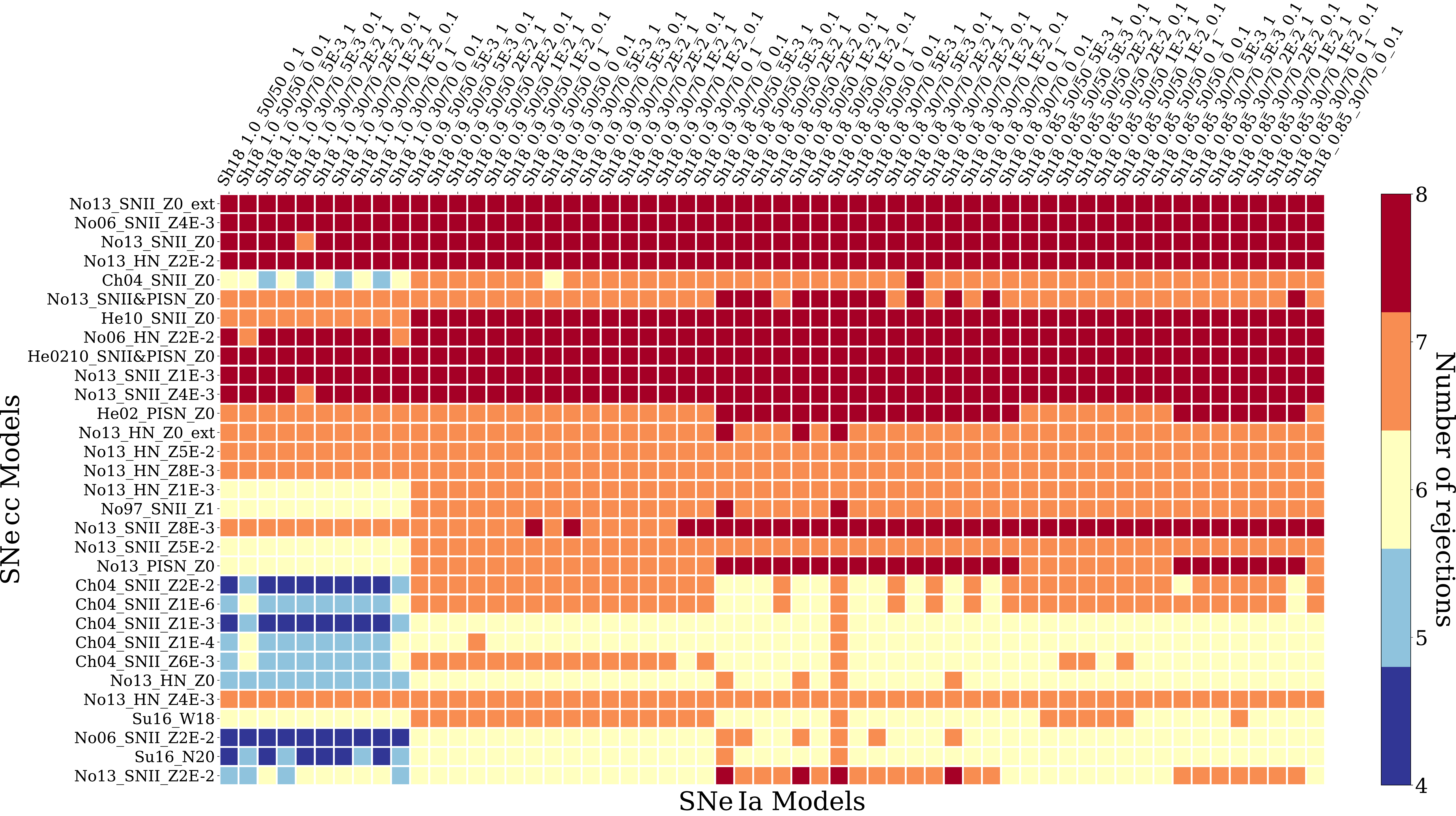}\caption{Continuation of Figure \ref{fig:rejected_times_1of4}.}
   \label{fig:rejected_times_4of4}
\end{figure*}

\clearpage
\section{List of SNe models with yields} \label{sec:list_SNe_models}

Given the substantial number of theoretical SN model combinations tested in this work and summarized in Section \ref{sec:summary_SNmodels}, we present here the main remarks of each SN model collected from the literature. The nomenclature for the 232 SNIa models and 31 SNcc models in Table \ref{tab:list_SNIa_models} and \ref{tab:list_SNcc_models}, respectively.

\startlongtable
\begin{deluxetable*}{lll}
\tablecaption{List of the 232 SNIa explosion models with yields from the literature tested with our method.\label{tab:list_SNIa_models}}
\tablehead{\colhead{Model Name} & \colhead{Ref.} & \colhead{Basic Proprieties}
}
\startdata
Fi14\_1\_2.9 & 1 & 3D deflagration, $\rho_{c,9}=2.9$, 1 slightly off-centre ignition bubble \\
Fi14\_3\_2.9 & 1 & 3D deflagration, $\rho_{c,9}=2.9$, 3 centred ignition bubbles \\
Fi14\_5\_2.9 & 1 & 3D deflagration, $\rho_{c,9}=2.9$, 5 centred ignition bubbles \\
Fi14\_10\_2.9 & 1 & 3D deflagration, $\rho_{c,9}=2.9$, 10 centred ignition bubbles \\
Fi14\_20\_2.9 & 1 & 3D deflagration, $\rho_{c,9}=2.9$, 20 centred ignition bubbles \\
Fi14\_40\_2.9 & 1 & 3D deflagration, $\rho_{c,9}=2.9$, 40 centred ignition bubbles \\
Fi14\_100\_5.5 & 1 & 3D deflagration, $\rho_{c,9}=5.5$, 100 centred ignition bubbles \\
Fi14\_100\_2.9 & 1 & 3D deflagration, $\rho_{c,9}=2.9$, 100 centred ignition bubbles \\
Fi14\_100\_1.0 & 1 & 3D deflagration, $\rho_{c,9}=1.0$, 100 centred ignition bubbles \\
Fi14\_150\_2.9 & 1 & 3D deflagration, $\rho_{c,9}=2.9$, 150 centred ignition bubbles \\
Fi14\_200\_2.9 & 1 & 3D deflagration, $\rho_{c,9}=2.9$, 200 centred ignition bubbles \\
Fi14\_300\_2.9\_c & 1 & 3D deflagration, $\rho_{c,9}=2.9$, 300 compact centred ignition bubbles (highly sph. symmetry) \\
Fi14\_1600\_2.9 & 1 & 3D deflagration, $\rho_{c,9}=2.9$, 1600 centred ignition bubbles \\
Fi14\_1600\_2.9\_c & 1 & 3D deflagration, $\rho_{c,9}=2.9$, 1600 compact centred ignition bubbles (highly sph. symmetry) \\[1mm]
Iw99\_W7 & 2 & Deflagration, $\rho_{c,9}=2.12$, $\mathbf{Z_{\rm \textbf{{init}}}}=1$ \\
Iw99\_W70 & 2 & Deflagration, $\rho_{c,9}=2.12$, $\mathbf{Z_{\rm \textbf{{init}}}}=0$ \\
Iw99\_WDD1 & 2 & Delayed Detonation, $\rho_{c,9}=2.12$, $\rho_{T,7}=1.7$, $\mathbf{Z_{\rm \textbf{{init}}}}=1$ \\
Iw99\_WDD2 & 2 & Delayed Detonation, $\rho_{c,9}=2.12$, $\rho_{T,7}=2.2$, $\mathbf{Z_{\rm \textbf{{init}}}}=1$ \\
Iw99\_WDD3 & 2 & Delayed Detonation, $\rho_{c,9}=2.12$, $\rho_{T,7}=3.0$, $\mathbf{Z_{\rm \textbf{{init}}}}=1$ \\
Iw99\_CDD1 & 2 & Delayed Detonation, $\rho_{c,9}=1.37$, $\rho_{T,7}=1.7$, $\mathbf{Z_{\rm \textbf{{init}}}}=1$ \\
Iw99\_CDD2 & 2 & Delayed Detonation, $\rho_{c,9}=1.37$, $\rho_{T,7}=2.2$, $\mathbf{Z_{\rm \textbf{{init}}}}=1$ \\[1mm]
Kr13\_0.9\_0.76 & 3 & 3D C+O WDs violent merger, $(M_{1,_\text{WD}}, M_{2,_\text{WD}})=(0.9,\,0.76)$, $\mathbf{Z_{\rm \textbf{{init}}}}=1$ \\
Kr15\_hybrid & 4 & 3D Hybrid WD deflagr. (C+O core w/ O+Ne layers), five off-centre spots, $\rho_{c,9}=2.12$ \\
Kr16\_0.9\_0.76\_Z1E-2 & 3,5 & 3D C+O WDs violent merger, $(M_{1,_\text{WD}}, M_{2,_\text{WD}})=(0.9,\,0.76)$, $\mathbf{Z_{\rm \textbf{{init}}}}=0.01$ \\[1mm]
Le18\_0.5\_1.0\_1P & 6 & 2D Pure Turbulent Deflagr., $\rho_{c,9}=0.5$, $\mathbf{Z_{\rm \textbf{{init}}}}=1$, C/O=1.0 \\
Le18\_1.0\_1.0\_1P & 6 & 2D Pure Turbulent Deflagr., $\rho_{c,9}=1.0$, $\mathbf{Z_{\rm \textbf{{init}}}}=1$, C/O=1.0 \\
Le18\_1.0\_0 & 6 & 2D Deflagration-to-Detonation, $\rho_{c,9}=1.0$, $\mathbf{Z_{\rm \textbf{{init}}}}=0$ \\
Le18\_1.0\_0.1 & 6 & 2D Deflagration-to-Detonation, $\rho_{c,9}=1.0$, $\mathbf{Z_{\rm \textbf{{init}}}}=0.1$ \\
Le18\_1.0\_0.5 & 6 & 2D Deflagration-to-Detonation, $\rho_{c,9}=1.0$, $\mathbf{Z_{\rm \textbf{{init}}}}=0.5$ \\
Le18\_1.0\_1.0 & 6 & 2D Deflagration-to-Detonation, $\rho_{c,9}=1.0$, $\mathbf{Z_{\rm \textbf{{init}}}}=1.0$ \\
Le18\_1.0\_2.0 & 6 & 2D Deflagration-to-Detonation, $\rho_{c,9}=1.0$, $\mathbf{Z_{\rm \textbf{{init}}}}=2.0$ \\
Le18\_1.0\_3.0 & 6 & 2D Deflagration-to-Detonation, $\rho_{c,9}=1.0$, $\mathbf{Z_{\rm \textbf{{init}}}}=3.0$ \\
Le18\_1.0\_5.0 & 6 & 2D Deflagration-to-Detonation, $\rho_{c,9}=1.0$, $\mathbf{Z_{\rm \textbf{{init}}}}=5.0$ \\
Le18\_3.0\_1.0\_1P & 6 & 2D Pure Turbulent Deflagr., $\rho_{c,9}=3.0$, $\mathbf{Z_{\rm \textbf{{init}}}}=1.0$, C/O=1.0 \\
Le18\_3.0\_0 & 6 & 2D Deflagration-to-Detonation, $\rho_{c,9}=3.0$, $\mathbf{Z_{\rm \textbf{{init}}}}=0$ \\
Le18\_3.0\_0.1 & 6 & 2D Deflagration-to-Detonation, $\rho_{c,9}=3.0$, $\mathbf{Z_{\rm \textbf{{init}}}}=0.1$ \\
Le18\_3.0\_0.5 & 6 & 2D Deflagration-to-Detonation, $\rho_{c,9}=3.0$, $\mathbf{Z_{\rm \textbf{{init}}}}=0.5$ \\
Le18\_3.0\_1.0 & 6 & 2D Deflagration-to-Detonation, $\rho_{c,9}=3.0$, $\mathbf{Z_{\rm \textbf{{init}}}}=1.0$ \\
Le18\_3.0\_2.0 & 6 & 2D Deflagration-to-Detonation, $\rho_{c,9}=3.0$, $\mathbf{Z_{\rm \textbf{{init}}}}=2.0$ \\
Le18\_3.0\_3.0 & 6 & 2D Deflagration-to-Detonation, $\rho_{c,9}=3.0$, $\mathbf{Z_{\rm \textbf{{init}}}}=3.0$ \\
Le18\_3.0\_5.0 & 6 & 2D Deflagration-to-Detonation, $\rho_{c,9}=3.0$, $\mathbf{Z_{\rm \textbf{{init}}}}=5.0$ \\
Le18\_5.0\_1.0\_1P & 6 & 2D Pure Turbulent Deflagr., $\rho_{c,9}=5.0$, $\mathbf{Z_{\rm \textbf{{init}}}}=1.0$, C/O=1.0 \\
Le18\_5.0\_0 & 6 & 2D Deflagration-to-Detonation, $\rho_{c,9}=5.0$, $\mathbf{Z_{\rm \textbf{{init}}}}=0$ \\
Le18\_5.0\_0.1 & 6 & 2D Deflagration-to-Detonation, $\rho_{c,9}=5.0$, $\mathbf{Z_{\rm \textbf{{init}}}}=0.1$\\
Le18\_5.0\_0.5 & 6 & 2D Deflagration-to-Detonation, $\rho_{c,9}=5.0$, $\mathbf{Z_{\rm \textbf{{init}}}}=0.5$\\
Le18\_5.0\_1.0 & 6 & 2D Deflagration-to-Detonation, $\rho_{c,9}=5.0$, $\mathbf{Z_{\rm \textbf{{init}}}}=1.0$ \\
Le18\_5.0\_2.0 & 6 & 2D Deflagration-to-Detonation, $\rho_{c,9}=5.0$, $\mathbf{Z_{\rm \textbf{{init}}}}=2.0$ \\
Le18\_5.0\_3.0 & 6 & 2D Deflagration-to-Detonation, $\rho_{c,9}=5.0$, $\mathbf{Z_{\rm \textbf{{init}}}}=3.0$ \\
Le18\_5.0\_5.0 & 6 & 2D Deflagration-to-Detonation, $\rho_{c,9}=5.0$, $\mathbf{Z_{\rm \textbf{{init}}}}=5.0$ \\[1mm]
Le20a\_1.10\_0.10\_0\_B & 7 & 2D Double-Deton., $M_\text{WD}=1.10$, $M_\text{He}=0.10$, $Z_{\rm init}=0$, one He deton. bubble \\
Le20a\_1.10\_0.10\_2E-3\_B & 7 & 2D Double-Deton., $M_\text{WD}=1.10$, $M_\text{He}=0.10$, $Z_{\rm init}=0.002$, one He deton. bubble  \\
Le20a\_1.10\_0.10\_1E-2\_B & 7 & 2D Double-Deton., $M_\text{WD}=1.10$, $M_\text{He}=0.10$, $Z_{\rm init}=0.01$, one He deton. bubble  \\
Le20a\_1.10\_0.10\_2E-2\_B & 7 & 2D Double-Deton., $M_\text{WD}=1.10$, $M_\text{He}=0.10$, $Z_{\rm init}=0.02$, one He deton. bubble  \\
Le20a\_1.10\_0.10\_4E-2\_B & 7 & 2D Double-Deton., $M_\text{WD}=1.10$, $M_\text{He}=0.10$, $Z_{\rm init}=0.04$, one He deton. bubble  \\
Le20a\_1.10\_0.10\_6E-2\_B & 7 & 2D Double-Deton., $M_\text{WD}=1.10$, $M_\text{He}=0.10$, $Z_{\rm init}=0.06$, one He deton. bubble  \\
Le20a\_1.10\_0.10\_1E-1\_B & 7 & 2D Double-Deton., $M_\text{WD}=1.10$, $M_\text{He}=0.10$, $Z_{\rm init}=0.1$, one He deton. bubble  \\
Le20a\_1.10\_0.05\_0\_R & 7 & 2D Double-Deton., $M_\text{WD}=1.10$, $M_\text{He}=0.05$, $Z_{\rm init}=0$, He deton. ring \\
Le20a\_1.10\_0.05\_2E-3\_R & 7 & 2D Double-Deton., $M_\text{WD}=1.10$, $M_\text{He}=0.05$, $Z_{\rm init}=0.002$, He deton. ring \\
Le20a\_1.10\_0.05\_1E-2\_R & 7 & 2D Double-Deton., $M_\text{WD}=1.10$, $M_\text{He}=0.05$, $Z_{\rm init}=0.01$, He deton. ring \\
Le20a\_1.10\_0.05\_2E-2\_R & 7 & 2D Double-Deton., $M_\text{WD}=1.10$, $M_\text{He}=0.05$, $Z_{\rm init}=0.02$, He deton. ring \\
Le20a\_1.10\_0.05\_4E-2\_R & 7 & 2D Double-Deton., $M_\text{WD}=1.10$, $M_\text{He}=0.05$, $Z_{\rm init}=0.04$, He deton. ring \\
Le20a\_1.10\_0.05\_6E-2\_R & 7 & 2D Double-Deton., $M_\text{WD}=1.10$, $M_\text{He}=0.05$, $Z_{\rm init}=0.06$, He deton. ring \\
Le20a\_1.10\_0.05\_1E-1\_R & 7 & 2D Double-Deton., $M_\text{WD}=1.10$, $M_\text{He}=0.05$, $Z_{\rm init}=0.1$, He deton. ring \\
Le20a\_1.00\_0.05\_0\_S & 7 & 2D Double-Deton., $M_\text{WD}=1.00$, $M_\text{He}=0.05$, $Z_{\rm init}=0$, spherical He deton. \\
Le20a\_1.00\_0.05\_2E-3\_S & 7 & 2D Double-Deton., $M_\text{WD}=1.00$, $M_\text{He}=0.05$, $Z_{\rm init}=0.002$, spherical He deton. \\
Le20a\_1.00\_0.05\_1E-2\_S & 7 & 2D Double-Deton., $M_\text{WD}=1.00$, $M_\text{He}=0.05$, $Z_{\rm init}=0.01$, spherical He deton. \\
Le20a\_1.00\_0.05\_2E-2\_S & 7 & 2D Double-Deton., $M_\text{WD}=1.00$, $M_\text{He}=0.05$, $Z_{\rm init}=0.02$, spherical He deton. \\
Le20a\_1.00\_0.05\_4E-2\_S & 7 & 2D Double-Deton., $M_\text{WD}=1.00$, $M_\text{He}=0.05$, $Z_{\rm init}=0.04$, spherical He deton. \\
Le20a\_1.00\_0.05\_6E-2\_S & 7 & 2D Double-Deton., $M_\text{WD}=1.00$, $M_\text{He}=0.05$, $Z_{\rm init}=0.06$, spherical He deton. \\
Le20a\_1.00\_0.05\_1E-1\_S & 7 & 2D Double-Deton., $M_\text{WD}=1.00$, $M_\text{He}=0.05$, $Z_{\rm init}=0.1$, spherical He deton. \\
Le20a\_0.90\_0.15\_2E-2\_B & 7 & 2D Double-Deton., $M_\text{WD}=0.90$, $M_\text{He}=0.15$, $Z_{\rm init}=0.02$, one He deton. bubble \\
Le20a\_0.95\_0.15\_2E-2\_B & 7 & 2D Double-Deton., $M_\text{WD}=0.95$, $M_\text{He}=0.15$, $Z_{\rm init}=0.02$, one He deton. bubble \\
Le20a\_1.00\_0.10\_2E-2\_B & 7 & 2D Double-Deton., $M_\text{WD}=1.00$, $M_\text{He}=0.10$, $Z_{\rm init}=0.02$, one He deton. bubble \\
Le20a\_1.05\_0.10\_2E-2\_B & 7 & 2D Double-Deton., $M_\text{WD}=1.05$, $M_\text{He}=0.10$, $Z_{\rm init}=0.02$, one He deton. bubble \\
Le20a\_1.15\_0.10\_2E-2\_B & 7 & 2D Double-Deton., $M_\text{WD}=1.15$, $M_\text{He}=0.10$, $Z_{\rm init}=0.02$, one He deton. bubble \\
Le20a\_1.20\_0.05\_2E-2\_B & 7 & 2D Double-Deton., $M_\text{WD}=1.20$, $M_\text{He}=0.05$, $Z_{\rm init}=0.02$, one He deton. bubble \\
Le20a\_0.90\_0.05\_2E-2\_R & 7 & 2D Double-Deton., $M_\text{WD}=0.90$, $M_\text{He}=0.05$, $Z_{\rm init}=0.02$, He deton. ring \\
Le20a\_0.95\_0.05\_2E-2\_R & 7 & 2D Double-Deton., $M_\text{WD}=0.95$, $M_\text{He}=0.05$, $Z_{\rm init}=0.02$, He deton. ring \\
Le20a\_1.00\_0.05\_2E-2\_R & 7 & 2D Double-Deton., $M_\text{WD}=1.00$, $M_\text{He}=0.05$, $Z_{\rm init}=0.02$, He deton. ring \\
Le20a\_1.05\_0.05\_2E-2\_R & 7 & 2D Double-Deton., $M_\text{WD}=1.05$, $M_\text{He}=0.05$, $Z_{\rm init}=0.02$, He deton. ring \\
Le20a\_1.20\_0.05\_2E-2\_R & 7 & 2D Double-Deton., $M_\text{WD}=1.20$, $M_\text{He}=0.05$, $Z_{\rm init}=0.02$, He deton. ring \\
Le20a\_0.90\_0.05\_2E-2\_S & 7 & 2D Double-Deton., $M_\text{WD}=0.90$, $M_\text{He}=0.05$, $Z_{\rm init}=0.02$, spherical He deton. \\
Le20a\_0.95\_0.05\_2E-2\_S & 7 & 2D Double-Deton., $M_\text{WD}=0.95$, $M_\text{He}=0.05$, $Z_{\rm init}=0.02$, spherical He deton. \\
Le20a\_1.05\_0.05\_2E-2\_S & 7 & 2D Double-Deton., $M_\text{WD}=1.05$, $M_\text{He}=0.05$, $Z_{\rm init}=0.02$, spherical He deton. \\
Le20a\_1.10\_0.05\_2E-2\_S & 7 & 2D Double-Deton., $M_\text{WD}=1.10$, $M_\text{He}=0.05$, $Z_{\rm init}=0.02$, spherical He deton. \\
Le20a\_1.20\_0.05\_2E-2\_S & 7 & 2D Double-Deton., $M_\text{WD}=1.20$, $M_\text{He}=0.05$, $Z_{\rm init}=0.02$, spherical He deton. \\[1mm]
Le20b\_1.0\_1.33 & 8 & 2D Turbulent Deflagration, $\rho_{c,9}=1.0$, $M_{\rm C+O}=1.33$, C/O=1 \\
Le20b\_2.0\_1.35 & 8 & 2D Turbulent Deflagration, $\rho_{c,9}=2.0$, $M_{\rm C+O}=1.35$, C/O=1 \\
Le20b\_3.0\_1.37 & 8 & 2D Turbulent Deflagration, $\rho_{c,9}=3.0$, $M_{\rm C+O}=1.37$, C/O=1 \\
Le20b\_5.0\_1.38 & 8 & 2D Turbulent Deflagration, $\rho_{c,9}=5.0$, $M_{\rm C+O}=1.38$, C/O=1 \\
Le20b\_5.5\_1.38 & 8 & 2D Turbulent Deflagration, $\rho_{c,9}=5.5$, $M_{\rm C+O}=1.38$, C/O=1 \\
Le20b\_7.5\_1.39 & 8 & 2D Turbulent Deflagration, $\rho_{c,9}=7.0$, $M_{\rm C+O}=1.39$, C/O=1 \\
Le20b\_9.0\_1.40 & 8 & 2D Turbulent Deflagration, $\rho_{c,9}=9.0$, $M_{\rm C+O}=1.4$, C/O=1 \\
Le20b\_hybrid\_1.0\_0.43 & 8 & 2D Hybrid WD Deflagr. (C+O core w/ O+Ne+Mg layer), $\rho_{c,9}=1.0$, $M_{\rm C+O}=0.43$ \\
Le20b\_hybrid\_2.0\_0.45 & 8 & 2D Hybrid WD Deflagr. (C+O core w/ O+Ne+Mg layer), $\rho_{c,9}=2.0$, $M_{\rm C+O}=0.45$ \\
Le20b\_hybrid\_3.0\_0.47 & 8 & 2D Hybrid WD Deflagr. (C+O core w/ O+Ne+Mg layer), $\rho_{c,9}=3.0$, $M_{\rm C+O}=0.47$ \\
Le20b\_hybrid\_5.0\_0.48 & 8 & 2D Hybrid WD Deflagr. (C+O core w/ O+Ne+Mg layer), $\rho_{c,9}=5.0$, $M_{\rm C+O}=0.48$ \\
Le20b\_hybrid\_5.5\_0.48 & 8 & 2D Hybrid WD Deflagr. (C+O core w/ O+Ne+Mg layer), $\rho_{c,9}=5.5$, $M_{\rm C+O}=0.48$ \\
Le20b\_hybrid\_7.5\_0.49 & 8 & 2D Hybrid WD Deflagr. (C+O core w/ O+Ne+Mg layer), $\rho_{c,9}=7.5$, $M_{\rm C+O}=0.49$ \\
Le20b\_hybrid\_9.0\_0.50 & 8 & 2D Hybrid WD Deflagr. (C+O core w/ O+Ne+Mg layer), $\rho_{c,9}=9.0$, $M_{\rm C+O}=0.5$ \\[1mm]
Ma10\_ctr\_DF & 9 & 2D Deflagration, $\rho_{c,9}=2.9$, centred ignition \\
Ma10\_ctr\_DD & 9 & 2D Delayed-Detonation, $\rho_{c,9}=2.9$, $\rho_\text{T,7} \le 1.0$, centred ignition \\
Ma10\_off\_DD& 9 & 2D Delayed-Detonation, $\rho_{c,9}=2.9$, $\rho_\text{T,7} \le 1.0$, off-centre ignition \\ [2mm]
Mk15\_DD\_1.23\_0.15\_CO & 10 & 3D C+O WD Delayed-Detonation, $M_\text{WD}=1.23$, $\rho_{c,9}=0.15$, C/O=1 \\
Mk15\_DD\_1.18\_0.1\_ONe & 10 & 3D O+Ne WD Delayed-Detonation, $M_\text{WD}=1.18$, $\rho_{c,9}=0.1$ \\
Mk15\_DD\_1.21\_0.13\_ONe & 10 & 3D O+Ne WD Delayed-Detonation, $M_\text{WD}=1.21$, $\rho_{c,9}=0.13$ \\
Mk15\_DD\_1.23\_0.15\_ONe & 10 & 3D O+Ne WD Delayed-Detonation, $M_\text{WD}=1.23$, $\rho_{c,9}=0.15$ \\
Mk15\_DD\_1.24\_0.17\_ONe & 10 & 3D O+Ne WD Delayed-Detonation, $M_\text{WD}=1.24$, $\rho_{c,9}=0.17$ \\
Mk15\_DD\_1.25\_0.2\_ONe & 10 & 3D O+Ne WD Delayed-Detonation, $M_\text{WD}=1.25$, $\rho_{c,9}=0.2$ \\ [2mm]
Oh14\_DD\_50 & 11 & Se13\_100\_2.9, Homogeneous WD with 50\% carbon mass fraction \\
Oh14\_DD\_20 & 11 & Se13\_100\_2.9, Carbon depleted core WD with 20\% carbon mass fraction \\
Oh14\_DD\_32 & 11 & Se13\_100\_2.9, Carbon depleted core WD with 32\% carbon mass fraction \\
Oh14\_DD\_40 & 11 & Se13\_100\_2.9, Carbon depleted core WD with 40\% carbon mass fraction \\ [2mm]
Pr10\_0.9\_0.9 & 12 & 3D WDs violent merger; $(M_{1,_\text{WD}}, M_{2,_\text{WD}})=(0.9,\,0.9)$ \\
Pr12\_1.1\_0.9 & 13 & 3D WDs violent merger; $(M_{1,_\text{WD}}, M_{2,_\text{WD}})=(1.1,\,0.9)$ \\ [2mm]
Pa16\_1A & 14 & 3D WDs collision, $(M_{1,_\text{WD}}, M_{2,_\text{WD}})=(0.6,\,0.6)$, $\rho_{c,9}=3.4\times 10^{-3}$ \\
Pa16\_1C\_He & 14 & 3D WDs collision, $(M_{1,_\text{WD}}, M_{2,_\text{WD}})=(0.6,\,0.6)$, $\rho_{c,9}=3.4\times 10^{-3}$, both w/ $M_\text{He}=0.01$ \\ [2mm]
Se13\_1\_2.9 & 15 & 3D Delayed-Detonation, $\rho_{c,9}=2.9$, 1 off-centred ignition spot \\
Se13\_3\_2.9 & 15 & 3D Delayed-Detonation, $\rho_{c,9}=2.9$, 3 near-center ignition spots \\
Se13\_5\_2.9 & 15 & 3D Delayed-Detonation, $\rho_{c,9}=2.9$, 5 near-center ignition spots \\
Se13\_10\_2.9 & 15 & 3D Delayed-Detonation, $\rho_{c,9}=2.9$, 10 near-center ignition spots \\
Se13\_20\_2.9 & 15 & 3D Delayed-Detonation, $\rho_{c,9}=2.9$, 20 near-center ignition spots \\
Se13\_40\_2.9 & 15 & 3D Delayed-Detonation, $\rho_{c,9}=2.9$, 40 near-center ignition spots \\
Se13\_100\_5.5 & 15 & 3D Delayed-Detonation, $\rho_{c,9}=5.5$, 100 near-center ignition spots \\
Se13\_100\_2.9 & 15 & 3D Delayed-Detonation, $\rho_{c,9}=2.9$, 100 near-center ignition spots \\
Se13\_100\_1.0 & 15 & 3D Delayed-Detonation, $\rho_{c,9}=1.0$, 100 near-center ignition spots \\
Se13\_150\_2.9 & 15 & 3D Delayed-Detonation, $\rho_{c,9}=2.9$, 150 near-center ignition spots \\
Se13\_200\_2.9 & 15 & 3D Delayed-Detonation, $\rho_{c,9}=2.9$, 200 near-center ignition spots \\
Se13\_300\_2.9\_c & 15 & 3D Delayed-Detonation, $\rho_{c,9}=2.9$, 300 compact centred ignition spots, sph. symmetry \\
Se13\_1600\_2.9 & 15 & 3D Delayed-Detonation, $\rho_{c,9}=2.9$, 1600 near-center ignition spots \\
Se13\_1600\_2.9\_c & 15 & 3D Delayed-Detonation, $\rho_{c,9}=2.9$, 1600 compact centred ignition spots, sph. symmetry \\
Se13\_100\_2.9\_Z5E-1 & 15 & 3D Delayed-Detonation, $\rho_{c,9}=2.9$, 100 near-center ignition spots, $\mathbf{Z_{\rm \textbf{{init}}}}=0.5$ \\
Se13\_100\_2.9\_Z1E-1 & 15 & 3D Delayed-Detonation, $\rho_{c,9}=2.9$, 100 near-center ignition spots, $\mathbf{Z_{\rm \textbf{{init}}}}=0.1$ \\
Se13\_100\_2.9\_Z1E-2 & 15 & 3D Delayed-Detonation, $\rho_{c,9}=2.9$, 100 near-center ignition spots, $\mathbf{Z_{\rm \textbf{{init}}}}=0.01$ \\[1mm]
Se16\_GCD & 16 & 3D Gravitationally confined detonation, $\rho_{c,9}=2.9$, one off-centre ignition spot \\
            &&with distance from the center of the ignition kernel to the center of WD of 200km \\[1mm]
Sh18\_0.8\_30/70\_5E-3\_0.1 & 17 & 1D DDDDDD, $M_\text{WD}=0.8$, C/O=30/70, $Z_{\rm init}=0.005$, $\text{f}_{\rm ^{12}C+^{16}O}=0.1$ \\
Sh18\_0.8\_30/70\_5E-3\_1 & 17 & 1D DDDDDD, $M_\text{WD}=0.8$, C/O=30/70, $Z_{\rm init}=0.005$, $\text{f}_{\rm ^{12}C+^{16}O}=1.0$ \\
Sh18\_0.8\_30/70\_1E-2\_0.1 & 17 & 1D DDDDDD, $M_\text{WD}=0.8$, C/O=30/70, $Z_{\rm init}=0.01$, $\text{f}_{\rm ^{12}C+^{16}O}=0.1$ \\
Sh18\_0.8\_30/70\_1E-2\_1 & 17 & 1D DDDDDD, $M_\text{WD}=0.8$, C/O=30/70, $Z_{\rm init}=0.01$, $\text{f}_{\rm ^{12}C+^{16}O}=1.0$ \\
Sh18\_0.8\_30/70\_0\_0.1 & 17 & 1D DDDDDD, $M_\text{WD}=0.8$, C/O=30/70, $Z_{\rm init}=0.00$, $\text{f}_{\rm ^{12}C+^{16}O}=0.1$ \\
Sh18\_0.8\_30/70\_2E-2\_0.1 & 17 & 1D DDDDDD, $M_\text{WD}=0.8$, C/O=30/70, $Z_{\rm init}=0.02$, $\text{f}_{\rm ^{12}C+^{16}O}=1.0$ \\
Sh18\_0.8\_30/70\_2E-2\_1 & 17 & 1D DDDDDD, $M_\text{WD}=0.8$, C/O=30/70, $Z_{\rm init}=0.02$, $\text{f}_{\rm ^{12}C+^{16}O}=0.1$ \\
Sh18\_0.8\_30/70\_0\_1 & 17 & 1D DDDDDD, $M_\text{WD}=0.8$, C/O=30/70, $Z_{\rm init}=0.00$, $\text{f}_{\rm ^{12}C+^{16}O}=1.0$ \\
Sh18\_0.8\_50/50\_5E-3\_0.1 & 17 & 1D DDDDDD, $M_\text{WD}=0.8$, C/O=50/50, $Z_{\rm init}=0.005$, $\text{f}_{\rm ^{12}C+^{16}O}=0.1$ \\
Sh18\_0.8\_50/50\_5E-3\_1 & 17 & 1D DDDDDD, $M_\text{WD}=0.8$, C/O=50/50, $Z_{\rm init}=0.005$, $\text{f}_{\rm ^{12}C+^{16}O}=1.0$ \\
Sh18\_0.8\_50/50\_1E-2\_0.1 & 17 & 1D DDDDDD, $M_\text{WD}=0.8$, C/O=50/50, $Z_{\rm init}=0.01$, $\text{f}_{\rm ^{12}C+^{16}O}=0.1$ \\
Sh18\_0.8\_50/50\_1E-2\_1 & 17 & 1D DDDDDD, $M_\text{WD}=0.8$, C/O=50/50, $Z_{\rm init}=0.01$, $\text{f}_{\rm ^{12}C+^{16}O}=1.0$ \\
Sh18\_0.8\_50/50\_0\_0.1 & 17 & 1D DDDDDD, $M_\text{WD}=0.8$, C/O=50/50, $Z_{\rm init}=0.00$, $\text{f}_{\rm ^{12}C+^{16}O}=0.1$ \\
Sh18\_0.8\_50/50\_2E-2\_0.1 & 17 & 1D DDDDDD, $M_\text{WD}=0.8$, C/O=50/50, $Z_{\rm init}=0.02$, $\text{f}_{\rm ^{12}C+^{16}O}=1.0$ \\
Sh18\_0.8\_50/50\_2E-2\_1 & 17 & 1D DDDDDD, $M_\text{WD}=0.8$, C/O=50/50, $Z_{\rm init}=0.02$, $\text{f}_{\rm ^{12}C+^{16}O}=0.1$ \\
Sh18\_0.8\_50/50\_0\_1 & 17 & 1D DDDDDD, $M_\text{WD}=0.8$, C/O=50/50, $Z_{\rm init}=0.00$, $\text{f}_{\rm ^{12}C+^{16}O}=1.0$ \\
Sh18\_0.85\_30/70\_5E-3\_0.1 & 17 & 1D DDDDDD, $M_\text{WD}=0.85$, C/O=30/70, $Z_{\rm init}=0.005$, $\text{f}_{\rm ^{12}C+^{16}O}=0.1$ \\
Sh18\_0.85\_30/70\_5E-3\_1 & 17 & 1D DDDDDD, $M_\text{WD}=0.85$, C/O=30/70, $Z_{\rm init}=0.005$, $\text{f}_{\rm ^{12}C+^{16}O}=1.0$ \\
Sh18\_0.85\_30/70\_1E-2\_0.1 & 17 & 1D DDDDDD, $M_\text{WD}=0.85$, C/O=30/70, $Z_{\rm init}=0.01$, $\text{f}_{\rm ^{12}C+^{16}O}=0.1$ \\
Sh18\_0.85\_30/70\_1E-2\_1 & 17 & 1D DDDDDD, $M_\text{WD}=0.85$, C/O=30/70, $Z_{\rm init}=0.01$, $\text{f}_{\rm ^{12}C+^{16}O}=1.0$ \\
Sh18\_0.85\_30/70\_0\_0.1 & 17 & 1D DDDDDD, $M_\text{WD}=0.85$, C/O=30/70, $Z_{\rm init}=0.00$, $\text{f}_{\rm ^{12}C+^{16}O}=0.1$ \\
Sh18\_0.85\_30/70\_2E-2\_0.1 & 17 & 1D DDDDDD, $M_\text{WD}=0.85$, C/O=30/70, $Z_{\rm init}=0.02$, $\text{f}_{\rm ^{12}C+^{16}O}=1.0$ \\
Sh18\_0.85\_30/70\_2E-2\_1 & 17 & 1D DDDDDD, $M_\text{WD}=0.85$, C/O=30/70, $Z_{\rm init}=0.02$, $\text{f}_{\rm ^{12}C+^{16}O}=0.1$ \\
Sh18\_0.85\_30/70\_0\_1 & 17 & 1D DDDDDD, $M_\text{WD}=0.85$, C/O=30/70, $Z_{\rm init}=0.00$, $\text{f}_{\rm ^{12}C+^{16}O}=1.0$ \\
Sh18\_0.85\_50/50\_5E-3\_0.1 & 17 & 1D DDDDDD, $M_\text{WD}=0.85$, C/O=50/50, $Z_{\rm init}=0.005$, $\text{f}_{\rm ^{12}C+^{16}O}=0.1$ \\
Sh18\_0.85\_50/50\_5E-3\_1 & 17 & 1D DDDDDD, $M_\text{WD}=0.85$, C/O=50/50, $Z_{\rm init}=0.005$, $\text{f}_{\rm ^{12}C+^{16}O}=1.0$ \\
Sh18\_0.85\_50/50\_1E-2\_0.1 & 17 & 1D DDDDDD, $M_\text{WD}=0.85$, C/O=50/50, $Z_{\rm init}=0.01$, $\text{f}_{\rm ^{12}C+^{16}O}=0.1$ \\
Sh18\_0.85\_50/50\_1E-2\_1 & 17 & 1D DDDDDD, $M_\text{WD}=0.85$, C/O=50/50, $Z_{\rm init}=0.01$, $\text{f}_{\rm ^{12}C+^{16}O}=1.0$ \\
Sh18\_0.85\_50/50\_0\_0.1 & 17 & 1D DDDDDD, $M_\text{WD}=0.85$, C/O=50/50, $Z_{\rm init}=0.00$, $\text{f}_{\rm ^{12}C+^{16}O}=0.1$ \\
Sh18\_0.85\_50/50\_2E-2\_0.1 & 17 & 1D DDDDDD, $M_\text{WD}=0.85$, C/O=50/50, $Z_{\rm init}=0.02$, $\text{f}_{\rm ^{12}C+^{16}O}=1.0$ \\
Sh18\_0.85\_50/50\_2E-2\_1 & 17 & 1D DDDDDD, $M_\text{WD}=0.85$, C/O=50/50, $Z_{\rm init}=0.02$, $\text{f}_{\rm ^{12}C+^{16}O}=0.1$ \\
Sh18\_0.85\_50/50\_0\_1 & 17 & 1D DDDDDD, $M_\text{WD}=0.85$, C/O=50/50, $Z_{\rm init}=0.00$, $\text{f}_{\rm ^{12}C+^{16}O}=1.0$ \\
Sh18\_0.9\_30/70\_5E-3\_0.1 & 17 & 1D DDDDDD, $M_\text{WD}=0.90$, C/O=30/70, $Z_{\rm init}=0.005$, $\text{f}_{\rm ^{12}C+^{16}O}=0.1$ \\
Sh18\_0.9\_30/70\_5E-3\_1 & 17 & 1D DDDDDD, $M_\text{WD}=0.90$, C/O=30/70, $Z_{\rm init}=0.005$, $\text{f}_{\rm ^{12}C+^{16}O}=1.0$ \\
Sh18\_0.9\_30/70\_1E-2\_0.1 & 17 & 1D DDDDDD, $M_\text{WD}=0.90$, C/O=30/70, $Z_{\rm init}=0.01$, $\text{f}_{\rm ^{12}C+^{16}O}=0.1$ \\
Sh18\_0.9\_30/70\_1E-2\_1 & 17 & 1D DDDDDD, $M_\text{WD}=0.90$, C/O=30/70, $Z_{\rm init}=0.01$, $\text{f}_{\rm ^{12}C+^{16}O}=1.0$ \\
Sh18\_0.9\_30/70\_0\_0.1 & 17 & 1D DDDDDD, $M_\text{WD}=0.90$, C/O=30/70, $Z_{\rm init}=0.00$, $\text{f}_{\rm ^{12}C+^{16}O}=0.1$ \\
Sh18\_0.9\_30/70\_2E-2\_0.1 & 17 & 1D DDDDDD, $M_\text{WD}=0.90$, C/O=30/70, $Z_{\rm init}=0.02$, $\text{f}_{\rm ^{12}C+^{16}O}=1.0$ \\
Sh18\_0.9\_30/70\_2E-2\_1 & 17 & 1D DDDDDD, $M_\text{WD}=0.90$, C/O=30/70, $Z_{\rm init}=0.02$, $\text{f}_{\rm ^{12}C+^{16}O}=0.1$ \\
Sh18\_0.9\_30/70\_0\_1 & 17 & 1D DDDDDD, $M_\text{WD}=0.90$, C/O=30/70, $Z_{\rm init}=0.00$, $\text{f}_{\rm ^{12}C+^{16}O}=1.0$ \\
Sh18\_0.9\_50/50\_5E-3\_0.1 & 17 & 1D DDDDDD, $M_\text{WD}=0.90$, C/O=50/50, $Z_{\rm init}=0.005$, $\text{f}_{\rm ^{12}C+^{16}O}=0.1$ \\
Sh18\_0.9\_50/50\_5E-3\_1 & 17 & 1D DDDDDD, $M_\text{WD}=0.90$, C/O=50/50, $Z_{\rm init}=0.005$, $\text{f}_{\rm ^{12}C+^{16}O}=1.0$ \\
Sh18\_0.9\_50/50\_1E-2\_0.1 & 17 & 1D DDDDDD, $M_\text{WD}=0.90$, C/O=50/50, $Z_{\rm init}=0.01$, $\text{f}_{\rm ^{12}C+^{16}O}=0.1$ \\
Sh18\_0.9\_50/50\_1E-2\_1 & 17 & 1D DDDDDD, $M_\text{WD}=0.90$, C/O=50/50, $Z_{\rm init}=0.01$, $\text{f}_{\rm ^{12}C+^{16}O}=1.0$ \\
Sh18\_0.9\_50/50\_0\_0.1 & 17 & 1D DDDDDD, $M_\text{WD}=0.90$, C/O=50/50, $Z_{\rm init}=0.00$, $\text{f}_{\rm ^{12}C+^{16}O}=0.1$ \\
Sh18\_0.9\_50/50\_2E-2\_0.1 & 17 & 1D DDDDDD, $M_\text{WD}=0.90$, C/O=50/50, $Z_{\rm init}=0.02$, $\text{f}_{\rm ^{12}C+^{16}O}=1.0$ \\
Sh18\_0.9\_50/50\_2E-2\_1 & 17 & 1D DDDDDD, $M_\text{WD}=0.90$, C/O=50/50, $Z_{\rm init}=0.02$, $\text{f}_{\rm ^{12}C+^{16}O}=0.1$ \\
Sh18\_0.9\_50/50\_0\_1 & 17 & 1D DDDDDD, $M_\text{WD}=0.90$, C/O=50/50, $Z_{\rm init}=0.00$, $\text{f}_{\rm ^{12}C+^{16}O}=1.0$ \\
Sh18\_1.0\_30/70\_5E-3\_0.1 & 17 & 1D DDDDDD, $M_\text{WD}=1.0$, C/O=30/70, $Z_{\rm init}=0.005$, $\text{f}_{\rm ^{12}C+^{16}O}=0.1$ \\
Sh18\_1.0\_30/70\_5E-3\_1 & 17 & 1D DDDDDD, $M_\text{WD}=1.0$, C/O=30/70, $Z_{\rm init}=0.005$, $\text{f}_{\rm ^{12}C+^{16}O}=1.0$ \\
Sh18\_1.0\_30/70\_1E-2\_0.1 & 17 & 1D DDDDDD, $M_\text{WD}=1.0$, C/O=30/70, $Z_{\rm init}=0.01$, $\text{f}_{\rm ^{12}C+^{16}O}=0.1$ \\
Sh18\_1.0\_30/70\_1E-2\_1 & 17 & 1D DDDDDD, $M_\text{WD}=1.0$, C/O=30/70, $Z_{\rm init}=0.01$, $\text{f}_{\rm ^{12}C+^{16}O}=1.0$ \\
Sh18\_1.0\_30/70\_0\_0.1 & 17 & 1D DDDDDD, $M_\text{WD}=1.0$, C/O=30/70, $Z_{\rm init}=0.00$, $\text{f}_{\rm ^{12}C+^{16}O}=0.1$ \\
Sh18\_1.0\_30/70\_2E-2\_0.1 & 17 & 1D DDDDDD, $M_\text{WD}=1.0$, C/O=30/70, $Z_{\rm init}=0.02$, $\text{f}_{\rm ^{12}C+^{16}O}=1.0$ \\
Sh18\_1.0\_30/70\_2E-2\_1 & 17 & 1D DDDDDD, $M_\text{WD}=1.0$, C/O=30/70, $Z_{\rm init}=0.02$, $\text{f}_{\rm ^{12}C+^{16}O}=0.1$ \\
Sh18\_1.0\_30/70\_0\_1 & 17 & 1D DDDDDD, $M_\text{WD}=1.0$, C/O=30/70, $Z_{\rm init}=0.00$, $\text{f}_{\rm ^{12}C+^{16}O}=1.0$ \\
Sh18\_1.0\_50/50\_5E-3\_0.1 & 17 & 1D DDDDDD, $M_\text{WD}=1.0$, C/O=50/50, $Z_{\rm init}=0.005$, $\text{f}_{\rm ^{12}C+^{16}O}=0.1$ \\
Sh18\_1.0\_50/50\_5E-3\_1 & 17 & 1D DDDDDD, $M_\text{WD}=1.0$, C/O=50/50, $Z_{\rm init}=0.005$, $\text{f}_{\rm ^{12}C+^{16}O}=1.0$ \\
Sh18\_1.0\_50/50\_1E-2\_0.1 & 17 & 1D DDDDDD, $M_\text{WD}=1.0$, C/O=50/50, $Z_{\rm init}=0.01$, $\text{f}_{\rm ^{12}C+^{16}O}=0.1$ \\
Sh18\_1.0\_50/50\_1E-2\_1 & 17 & 1D DDDDDD, $M_\text{WD}=1.0$, C/O=50/50, $Z_{\rm init}=0.01$, $\text{f}_{\rm ^{12}C+^{16}O}=1.0$ \\
Sh18\_1.0\_50/50\_0\_0.1 & 17 & 1D DDDDDD, $M_\text{WD}=1.0$, C/O=50/50, $Z_{\rm init}=0.00$, $\text{f}_{\rm ^{12}C+^{16}O}=0.1$ \\
Sh18\_1.0\_50/50\_2E-2\_0.1 & 17 & 1D DDDDDD, $M_\text{WD}=1.0$, C/O=50/50, $Z_{\rm init}=0.02$, $\text{f}_{\rm ^{12}C+^{16}O}=1.0$ \\
Sh18\_1.0\_50/50\_2E-2\_1 & 17 & 1D DDDDDD, $M_\text{WD}=1.0$, C/O=50/50, $Z_{\rm init}=0.02$, $\text{f}_{\rm ^{12}C+^{16}O}=0.1$ \\
Sh18\_1.0\_50/50\_0\_1 & 17 & 1D DDDDDD, $M_\text{WD}=1.0$, C/O=50/50, $Z_{\rm init}=0.00$, $\text{f}_{\rm ^{12}C+^{16}O}=1.0$ \\
Sh18\_1.1\_30/70\_5E-3\_0.1 & 17 & 1D DDDDDD, $M_\text{WD}=1.10$, C/O=30/70, $Z_{\rm init}=0.005$, $\text{f}_{\rm ^{12}C+^{16}O}=0.1$ \\
Sh18\_1.1\_30/70\_5E-3\_1 & 17 & 1D DDDDDD, $M_\text{WD}=1.10$, C/O=30/70, $Z_{\rm init}=0.005$, $\text{f}_{\rm ^{12}C+^{16}O}=1.0$ \\
Sh18\_1.1\_30/70\_1E-2\_0.1 & 17 & 1D DDDDDD, $M_\text{WD}=1.10$, C/O=30/70, $Z_{\rm init}=0.01$, $\text{f}_{\rm ^{12}C+^{16}O}=0.1$ \\
Sh18\_1.1\_30/70\_1E-2\_1 & 17 & 1D DDDDDD, $M_\text{WD}=1.10$, C/O=30/70, $Z_{\rm init}=0.01$, $\text{f}_{\rm ^{12}C+^{16}O}=1.0$ \\
Sh18\_1.1\_30/70\_0\_0.1 & 17 & 1D DDDDDD, $M_\text{WD}=1.10$, C/O=30/70, $Z_{\rm init}=0.00$, $\text{f}_{\rm ^{12}C+^{16}O}=0.1$ \\
Sh18\_1.1\_30/70\_2E-2\_0.1 & 17 & 1D DDDDDD, $M_\text{WD}=1.10$, C/O=30/70, $Z_{\rm init}=0.02$, $\text{f}_{\rm ^{12}C+^{16}O}=1.0$ \\
Sh18\_1.1\_30/70\_2E-2\_1 & 17 & 1D DDDDDD, $M_\text{WD}=1.10$, C/O=30/70, $Z_{\rm init}=0.02$, $\text{f}_{\rm ^{12}C+^{16}O}=0.1$ \\
Sh18\_1.1\_30/70\_0\_1 & 17 & 1D DDDDDD, $M_\text{WD}=1.10$, C/O=30/70, $Z_{\rm init}=0.00$, $\text{f}_{\rm ^{12}C+^{16}O}=1.0$ \\
Sh18\_1.1\_50/50\_5E-3\_0.1 & 17 & 1D DDDDDD, $M_\text{WD}=1.10$, C/O=50/50, $Z_{\rm init}=0.005$, $\text{f}_{\rm ^{12}C+^{16}O}=0.1$ \\
Sh18\_1.1\_50/50\_5E-3\_1 & 17 & 1D DDDDDD, $M_\text{WD}=1.10$, C/O=50/50, $Z_{\rm init}=0.005$, $\text{f}_{\rm ^{12}C+^{16}O}=1.0$ \\
Sh18\_1.1\_50/50\_1E-2\_0.1 & 17 & 1D DDDDDD, $M_\text{WD}=1.10$, C/O=50/50, $Z_{\rm init}=0.01$, $\text{f}_{\rm ^{12}C+^{16}O}=0.1$ \\
Sh18\_1.1\_50/50\_1E-2\_1 & 17 & 1D DDDDDD, $M_\text{WD}=1.10$, C/O=50/50, $Z_{\rm init}=0.01$, $\text{f}_{\rm ^{12}C+^{16}O}=1.0$ \\
Sh18\_1.1\_50/50\_0\_0.1 & 17 & 1D DDDDDD, $M_\text{WD}=1.10$, C/O=50/50, $Z_{\rm init}=0.00$, $\text{f}_{\rm ^{12}C+^{16}O}=0.1$ \\
Sh18\_1.1\_50/50\_2E-2\_0.1 & 17 & 1D DDDDDD, $M_\text{WD}=1.10$, C/O=50/50, $Z_{\rm init}=0.02$, $\text{f}_{\rm ^{12}C+^{16}O}=1.0$ \\
Sh18\_1.1\_50/50\_2E-2\_1 & 17 & 1D DDDDDD, $M_\text{WD}=1.10$, C/O=50/50, $Z_{\rm init}=0.02$, $\text{f}_{\rm ^{12}C+^{16}O}=0.1$ \\
Sh18\_1.1\_50/50\_0\_1 & 17 & 1D DDDDDD, $M_\text{WD}=1.10$, C/O=50/50, $Z_{\rm init}=0.00$, $\text{f}_{\rm ^{12}C+^{16}O}=1.0$ \\[1mm]
Si10\_Det\_0.81\_1.0 & 18 & 1D sub-$M_\text{Ch}$ C+O WD Detonation, $M_\text{WD}=0.81$, $\rho_{c,7}=1.0$, C/O/Ne=50/50/0 \\
Si10\_Det\_0.88\_1.45 & 18 & 1D sub-$M_\text{Ch}$ C+O WD Detonation, $M_\text{WD}=0.88$, $\rho_{c,7}=1.45$, C/O/Ne=50/50/0 \\
Si10\_Det\_0.97\_2.4 & 18 & 1D sub-$M_\text{Ch}$ C+O WD Detonation, $M_\text{WD}=0.97$, $\rho_{c,7}=2.4$, C/O/Ne=50/50/0 \\
Si10\_Det\_1.06\_4.15 & 18 & 1D sub-$M_\text{Ch}$ C+O WD Detonation, $M_\text{WD}=1.06$, $\rho_{c,7}=4.15$, C/O/Ne=50/50/0 \\
Si10\_Det\_1.06\_4.15\_Ne & 18 & 1D sub-$M_\text{Ch}$ C+O WD Detonation, $M_\text{WD}=1.06$, $\rho_{c,7}=4.15$, C/O/Ne=42.5/50/7.5 \\
Si10\_Det\_1.15\_7.9 & 18 & 1D sub-$M_\text{Ch}$ C+O WD Detonation, $M_\text{WD}=1.15$, $\rho_{c,7}=7.9$, C/O/Ne=50/50/0 \\[1mm]
Si12\_DDet\_0.66\_0.38\_CS & 19 & 2D Converging-shock Double-Deton. (low-mass), $M_{\rm WD}=0.66$, $\rho_{c,7}=3.81\times 10^{-1}$ \\
Si12\_DDet\_0.79\_0.85\_CS & 19 & 2D Converging-shock Double-Deton. (standard), $M_{\rm WD}=0.79$, $\rho_{c,7}=8.5\times 10^{-1}$ \\
Si12\_DDet\_0.66\_0.38\_EL & 19 & 2D Edge-lit double-detonation (low-mass), $M_{\rm WD}=0.66 $, $\rho_{c,7}=3.81\times 10^{-1}$ \\
Si12\_DDet\_0.79\_0.85\_EL & 19 & 2D Edge-lit double-detonation (standard), $M_{\rm WD}=0.79$, $\rho_{c,7}=8.5\times 10^{-1}$ \\
Si12\_DDet\_0.66\_0.38\_He & 19 & 2D He-only Detonation (low-mass), $M_{\rm WD}=0.66$, $\rho_{c,7}=3.81\times 10^{-1}$ \\
Si12\_DDet\_0.79\_0.85\_He & 19 & 2D He-only Detonation (standard), $M_{\rm WD}=0.79$, $\rho_{c,7}=8.5\times 10^{-1}$ \\[1mm]
Tr04\_2D\_512 & 20 & 2D central ignition, gridsize=$512^2$, $\rho_{c,9}=2.9$  \\
Tr04\_3D\_256 & 20 & 3D central ignition, gridsize=$256^3$, $\rho_{c,9}=2.9$  \\
Tr04\_3D\_256\_T$^a$ & 20 & 3D central ignition, gridsize=$256^3$, $\rho_{c,9}=2.9$  \\
Tr04\_3D\_5\_256$^a$ & 20 & 3D multi-point ignition, 5 bubbles, gridsize=$256^3$, $\rho_{c,9}=2.9$  \\
Tr04\_3D\_30\_768$^a$ & 20 & 3D multi-point ignition, 30 bubbles, gridsize=$768^3$, $\rho_{c,9}=2.9$  \\
\enddata
\tablecomments{The central density of the white dwarf ($\rho_{c,7}$ and $\rho_{c,9}$) is given in units of $10^7$ and $10^9 \mathrm{g\,cm^{-3}}$, respectively. The density of the deflagration-to-detonation transition ($\rho_{T,7}$) is given in units of $10^7 \mathrm{g\,cm^{-3}}$. The mass fraction of the carbon/oxygen ratio (C/O) of the white dwarf and the ${\rm ^{12}C+^{16}O}$ reaction rate ($\text{f}_{\rm ^{12}C+^{16}O}$) are dimensionless. The initial mass of the white dwarf ($M_\text{WD}$), the masses of CO core ($M_{\rm C+O}$) and of the helium envelope ($M_\text{He}$), and the initial masses of the primary and secondary white dwarfs in collision and mergers ($M_{1,_\text{WD}}$ and $M_{2,_\text{WD}}$, respectively) are all given in units of $M_\odot$. The initial mass fraction composition of WD (C/O/Ne) are given by mass fraction of ${\rm ^{12}C/^{16}O/^{22}Ne}$. The progenitor initial metallicity is denoted by $Z_{\rm init}$, while bold ($\mathbf{Z_{\rm \textbf{{init}}}}$) is given in terms of solar metallicity units, $Z_\odot$; each SNcc model can assume different solar values to define their initial metallicity composition (see references for details). We denote DDDDDD for the Dynamically-Driven Double-Degenerate Double-Detonation models from \citet{Shen2018}. $^a$Only those tracer particles that reach NSE conditions starting 90\% of the peak temperature ($\sim 8.5\ \mathrm{x}\ 10^9$ K) are considered in the nucleosynthesis calculations.}
\tablereferences{(1) \citet{Fink2014}; (2) \citet{Iwamoto1999}; 
(3) \citet{Kromer2013}; (4) \citet{Kromer2015}; (5) \citet{Kromer2016}; 
(6) \citet{Leung2018}; (7) \citet{Leung2020a}; (8) \citet{Leung2020b}; 
(9) \citet{Maeda2010}; (10) \citet{Marquardt2015}; (11) \citet{Ohlmann2014}; 
(12) \citet{Pakmor2010}; (13) \citet{Pakmor2012}; (14) \citet{Papish2016}
(15) \citet{Seitenzahl2013}; (16) \citet{Seitenzahl2016}; (17) \citet{Shen2018};
(18) \citet{Sim2010}; (19) \citet{Sim2012}; (20) \citet{Travaglio_2004}.}
\end{deluxetable*}

\begin{deluxetable*}{lll}
\tablecaption{List of 31 SNcc explosion models tested in this study.\label{tab:list_SNcc_models}}
\tablehead{
\colhead{Model Name} & \colhead{Ref.} & \colhead{Basic Proprieties}
}
\startdata
Ch04\_SNII\_Z0 & a & Type II Supernova, $Z_{\rm init}=0$, $m=13,15,20,25,30,35$\\
Ch04\_SNII\_Z1E-6 & a & Type II Supernova, $Z_{\rm init}=1\text{E-}06$, $m=13,15,20,25,30,35$\\
Ch04\_SNII\_Z1E-4 & a & Type II Supernova, $Z_{\rm init}=1\text{E-}04$, $m=13,15,20,25,30,35$\\
Ch04\_SNII\_Z1E-3 & a & Type II Supernova, $Z_{\rm init}=1\text{E-}03$, $m=13,15,20,25,30,35$\\
Ch04\_SNII\_Z6E-3 & a & Type II Supernova, $Z_{\rm init}=6\text{E-}03$, $m=13,15,20,25,30,35$\\
Ch04\_SNII\_Z2E-2 & a & Type II Supernova, $Z_{\rm init}=2\text{E-}02$, $m=13,15,20,25,30,35$\\[1mm]
No97\_SNII\_Z1 & b & Type II Supernova, $\mathbf{Z_{\rm \textbf{{init}}}}=1$, $(m_{\text{low}}, m_{\text{up}})=(11,50)$\\[1mm]
No06\_SNII\_Z4E-3 & c & Type II Supernova, $Z_{\rm init}=0.004$, $m=13, 15, 18, 20, 25, 30, 40$ \\
No06\_SNII\_Z2E-2 & c & Type II Supernova, $Z_{\rm init}=0.02$, $m=13, 15, 18, 20, 25, 30, 40$ \\
No06\_HN\_Z2E-2 & c & Hypernovae, $Z_{\rm init}=0.02$, $m=20, 25, 30, 40$ \\[1mm]
No13\_SNII\_Z0 & c,d & Type II Supernova, $Z_{\rm init}=0$, $m=11, 13, 15, 18, 20, 25, 30, 40$ \\
No13\_SNII\_Z0\_ext & c,d & Type II Supernova, $Z_{\rm init}=0$, $m=11, 13, 15, 18, 20, 25, 30, 40, 100, 140$\\
No13\_SNII\_Z1E-3 & c,d & Type II Supernova, $Z_{\rm init}=0.001$, $m=13, 15, 18, 20, 25, 30, 40$\\
No13\_SNII\_Z4E-3 & c,d,e & Type II Supernova, $Z_{\rm init}=0.004$, $m=13, 15, 18, 20, 25, 30, 40$\\
No13\_SNII\_Z8E-3 & c,d & Type II Supernova, $Z_{\rm init}=0.008$, $m=13, 15, 18, 20, 25, 30, 40$\\
No13\_SNII\_Z2E-2 & c,d,e & Type II Supernova, $Z_{\rm init}=0.02$, $m=13, 15, 18, 20, 25, 30, 40$\\
No13\_SNII\_Z5E-2 & c,d & Type II Supernova, $Z_{\rm init}=0.05$, $m=13, 15, 18, 20, 25, 30, 40$\\
No13\_HN\_Z0 & c,d & Hypernovae, $Z_{\rm init}=0$, $m=20, 25, 30, 40$\\
No13\_HN\_Z0\_ext & c,d & Hypernovae, $Z_{\rm init}=0$, $m=20, 25, 30, 40, 100, 140$ \\
No13\_HN\_Z1E-3 & c,d & Hypernovae, $Z_{\rm init}=0.001$, $m=20, 25, 30, 40$\\
No13\_HN\_Z4E-3 & c,d & Hypernovae, $Z_{\rm init}=0.004$, $m=20, 25, 30, 40$\\
No13\_HN\_Z8E-3 & c,d & Hypernovae, $Z_{\rm init}=0.008$, $m=20, 25, 30, 40$\\
No13\_HN\_Z2E-2 & c,d,e & Hypernovae, $Z_{\rm init}=0.02$, $m=20, 25, 30, 40$\\
No13\_HN\_Z5E-2 & c,d & Hypernovae, $Z_{\rm init}=0.05$, $m=20, 25, 30, 40$\\
No13\_PISN\_Z0 & c,d & Pair-instability, $Z_{\rm init}=0$, $m=140, 150, 170, 200, 270, 300 $\\
No13\_SNII\&PISN\_Z0 & c,d & Type II Supernova and Pair-instability, $Z_{\rm init}=0$, \\
                        & & $m=11, 13, 15, 18, 20, 25, 30, 40, 100, 140, 150, 170, 200, 270, 300$\\[1mm]
He10\_SNII\_Z0 & f,g & Type II Supernova, $Z_{\rm init}=0$, $m=10, 12, 15, 20, 25, 35, 50, 75, 100$\\
He02\_PISN\_Z0 & f,g & Type II Supernova, $Z_{\rm init}=0$, $(m_{\text{low}}, m_{\text{up}})=(140,260)$\\
He0210\_SNII\&PISN\_Z0 & f,g & Type II SN and Pair-instability, $Z_{\rm init}=0$, $(m_{\text{low}}, m_{\text{up}})=(10,260)$\\[1mm]
Su16\_N20 & h & Nonrotating SN1987A model calibrated by \citet{Nomoto1988},\\
&&$\mathbf{Z_{\rm \textbf{{init}}}}=1$, $(m_{\text{low}}, m_{\text{up}})=(12.25, 120.0)$ \\
Su16\_W18 & h & Rotating SN1987A model calibrated by \citet{Utrobin2015}, \\
&& $\mathbf{Z_{\rm \textbf{{init}}}}=1$, $(m_{\text{low}}, m_{\text{up}})=(12.25, 120.0)$ \\
\enddata
\tablecomments{Respectively, the $m$, $m_{\text{low}}$ and $m_{\text{up}}$ are the initial, lowest, and greatest progenitor stellar mass simulated given in units of M$_\odot$ for a SNcc model. The progenitor initial metallicity is denoted by $Z_{\rm init}$, while bold ($\mathbf{Z_{\rm \textbf{{init}}}}$) is given in terms of solar metallicity units, $Z_\odot$; each SNcc model can assume different solar values to define their initial metallicity composition (see references for details).}
\tablereferences{(a) \citet{Chieffi2004}; (b) \citet{Nomoto1997}; (c) \citet{Nomoto2006};
(d) \citet{Nomoto2013}; (e) \citet{Kobayashi2011}; (f) \citet{Heger2002}; 
(g) \citet{Heger2010}; (h) \citet{Sukhbold2016}.}
\end{deluxetable*}

\end{document}